\documentclass[twocolumn,tighten]{aastex631}
\usepackage{amsmath, mathrsfs}

\shorttitle{Spin Evolution and Mass Distribution of BNSs}
\shortauthors{Chu, Lu, \& Yu}

\begin{document}

\title{Spin evolution and mass distribution of the Galactic Binary Neutron Stars}

\author{Qingbo Chu}
\affiliation{National Astronomical Observatories, Chinese Academy of Sciences, 
20A Datun Road, Beijing 100101, China}
\affiliation{School of Astronomy and Space Sciences, University of Chinese Academy of Sciences, 
19A Yuquan Road, Beijing 100049, China}
\author[0000-0002-1310-4664]{Youjun Lu}\thanks{luyj@nao.cas.cn}
\affiliation{National Astronomical Observatories, Chinese Academy of Sciences, 
20A Datun Road, Beijing 100101, China}
\affiliation{School of Astronomy and Space Sciences, University of Chinese Academy of Sciences, 
19A Yuquan Road, Beijing 100049, China}
\author{Shenghua Yu}
\affiliation{National Astronomical Observatories, Chinese Academy of Sciences, 
20A Datun Road, Beijing 100101, China}

\correspondingauthor{Youjun Lu}

\begin{abstract}
Binary neutron stars (BNSs) detected in the Milky Way have the total masses distributing narrowly around $\sim2.6-2.7M_\odot$, while the BNS merger GW190425 detected via gravitational wave has a significantly larger mass ($\sim3.4M_\odot$). This difference is not well understood, yet. In this paper, we investigate the BNS spin evolution via an improved binary star evolution model and its effects on the BNS observability, with implementation of various relevant astrophysical processes. We find that the first-born neutron star component in low-mass BNSs can be spun up to millisecond pulsars by the accretion of Roche-lobe overflow from its companion and its radio lifetime can be comparable to the Hubble time. However, most high-mass BNSs have substantially shorter radio lifetime than the low-mass BNSs, and thus smaller probability being detected via radio emission. 
Adopting the star formation and metal enrichment history of the Milky Way given by observations,
we obtain the survived Galactic BNSs with pulsar components from our population synthesis model and find that their distributions on the diagrams of spin period versus spin-period-time-derivative 
($P-\dot{P}$) and orbital period versus eccentricity ($P_{\rm orb}-e$) can well match those of the observed Galactic BNSs. The total mass distribution of the observed Galactic BNSs can also be matched by the model. A significant fraction 
($\sim19\%-22\%$) 
of merging BNSs at redshift $z\sim0$ have masses $\gtrsim3M_\odot$, which seems compatible with the GW observations. Future radio observations may detect many more Galactic BNSs, which will put strong constraint on the spin evolution of BNSs during their formation processes.
\end{abstract}

\keywords{Accretion(14) -- Compact binary stars(283) -- Gravitational wave sources(677) -- Neutron stars(1108) -- Stellar evolution(1599)	-- X-ray binary stars(1811)}

\section{Introduction} 
\label{sec:intro}

The mass distribution of binary neutron stars (BNSs) may encode critical information about the formation and evolution of BNSs. More than $20$ BNSs have been found in the Milky Way, and $19$ BNSs among them have well measured total masses [see Table 1 in \citet{2024arXiv240204658D}]. For these Galactic BNSs, their total mass ($M_{\rm tot}$) distribution seems to follow a two-component Gaussian distribution with the mean ($\bar{M}_{\rm tot}$) and the scatter ($\sigma_{\bar{M}_{\rm tot}}$) of $(\bar{M}_{\rm tot},\sigma_{\bar{M}_{\rm tot}})=(2.58M_\odot, 0.01M_\odot)$ and $(2.72M_\odot, 0.08M_\odot)$, respectively, according to \citet[][see also other studies by \citep{2012ApJ...757...55O, 2013ApJ...778...66K, 2016ARA&A..54..401O, 2018arXiv180403101H, 2019ApJ...876...18F, 2019MNRAS.485.1665K} in the literature]{2018arXiv180403101H}. Two BNS mergers, i.e., GW170817 and GW190425, have been detected by the Laser Interferometer Gravitational wave Observatories (LIGO) and Virgo \citep{2017PhRvL.119p1101A, 2017ApJ...848L..12A, 2020ApJ...892L...3A}, of which the total mass $M_{\rm tot} = 2.74^{+0.04}_{-0.01} M_\odot$ and $3.4^{+0.3}_{-0.1} M_\odot$, respectively \citep{2017PhRvL.119p1101A, 2020ApJ...892L...3A}. The total mass of GW170817 is well consistent with those of the Galactic BNSs. However, the total mass of GW190425 is significantly larger than that of any known Galactic BNS systems \citep{2020ApJ...892L...3A}, which may suggest a tension between the mass distribution of the gravitational wave (GW) detected BNSs and that of the Galactic BNSs detected by radio emission. \citet{2020A&A...639A.123K} discussed that this discrepancy may be naturally produced in binary evolution, though without detailed consideration of the selection effects in both radio and GW detection.  

Heavy BNSs like GW190425 may be formed via a channel different from that for observed Galactic BNSs. \citet{2020MNRAS.496L..64R} suggested that GW190425-like BNSs were formed via the unstable case BB mass transfer, which quickly merged within a time of $\lesssim10$\,Myr, and such a short delay time may explain why similar systems are not observed via radio emission. However, such a fast-merging channel may be not efficient enough to form heavy BNSs with abundance consistent with GW observations \citep[see][]{2020ApJ...900...13S}. It is also possible to form heavy BNSs like GW190425 by revising the recipes for several physical processes involved in the evolution of binary stars, such as different common-envelope (CE) evolution channels, neutron star masses and kicks \citep{2021ApJ...909L..19G, 2021ApJ...920L..17V, 2021MNRAS.500.1380M}. However, it is not clear whether these heavy BNSs can be detected by radio emission and the total mass distributions of the Galactic BNSs and GW detected BNSs can be well explained.

One should note that the selection effects may play important roles in detecting BNSs. For the search of Galactic BNSs by radio emission, at least one pulsar is required to be in an observed system. In contrast, the GW detection only depends on the BNS chirp mass, luminosity distance, orientation and sky localization of the source, regardless of whether it has a pulsar component or not. One simple idea is that neutron stars in the high-mass BNSs (e.g., GW190425) have much shorter lifetime in the radio band than those in the low-mass ones (e.g., Galactic BNSs), and thus the  GW190425-like heavy BNSs can hardly be observed via radio emission in our Galaxy while they can be easier detected comparing with those lower mass BNS mergers by GW detectors due to their higher chirp masses.

The spin evolution of neutron stars may be closely related to their observabilities in the radio band and thus could be crucial for the understanding of the BNS total mass distributions obtained for Galactic BNSs and those detected via GWs. Such spin evolution has not been thoroughly considered in most previous works for the formation of BNSs via rapid binary population synthesis codes \citep[e.g.,][]{2012ApJ...759...52D, 2013ApJ...779...72D, 2016ApJ...819..108B, 2017ApJ...846..170T, 2018A&A...615A..91B, 2018MNRAS.474.2937C, 2018MNRAS.480.2011G, 2018MNRAS.481.1908K, 2018MNRAS.481.4009V, 2022ApJS..258...34R}, though it has been investigated in details for the high mass X-ray binary (HMXB) systems \citep[e.g.,][]{1981MNRAS.196..209D, 1991PhR...203....1B, 2004ChJAA...4..320Z, 2008ApJ...683.1031B,2010MNRAS.407.1090B,2012MNRAS.421L.127P,2016ApJ...824..143L, 2020PASJ...72...95K}. In this paper, we further develop the binary star evolution (BSE) code presented in \citet{2022MNRAS.509.1557C} by including the detailed consideration of the neutron star spin evolution, and illustrate whether high-mass BNSs can be detected by their radio emission in the Milky Way and how the difference of the total mass distribution of Galactic BNSs from that of GW detected BNSs may be explained by the selection effects. 

This paper is organized as follows. In Section~\ref{sec:method}, we introduce the population synthesis model and the spin evolution model for binary neutron stars, the settings of the initial binary star population, and the settings for various parameters of the relevant physical processes involved in the binary evolution. According to these models and the parameter settings, we perform Monte Carlo simulations of binary star evolution and obtain evolution tracks for a large amount of BNSs. In Section~\ref{sec:results}, we show several spin evolution tracks of the typical BNSs, as examples, and explain the behaviour of these tracked by the underlying physics. Using the obtained BNS evolution tracks as the templates, we further produce mock survived BNS samples by assuming the star formation history in the Milky Way.  We compare the distribution of these mock samples in the spin period versus the time derivative of spin period ($P-\dot{P}$) diagram and the orbital period-eccentricity ($P_{\rm orb}-e$) diagram with those of the observed BNSs, and find that the observations can be well matched. We also obtain the total mass distributions of survived BNSs with pulsar components and compare them with observational ones. Discussions and conclusions are given in Sections~\ref{sec:discussion} and \ref{sec:conclusion}, respectively.

\section{Methods} 
\label{sec:method}

We adopt a modified version of binary star evolution (BSE) code \citep{2000MNRAS.315..543H, 2002MNRAS.329..897H} to perform BSE calculations [see more descriptions in \citet{2022MNRAS.509.1557C}, for a study of the formation and evolution of BNSs and Galactic binary compact objects, see also \citet{2010A&A...521A..85Y, 2011MNRAS.417.1392Y, 2015MNRAS.448.1078Y}]. In this section, we first briefly introduce the basic model settings for the initial properties of binary stars, the consideration of the magnetic field evolution, and the treatments of the stellar wind velocity, the remnant compact object masses, supernova and natal kicks in Section~\ref{subsec:bse}. Then we introduce the implementation of various processes that affect the spin evolution in Section~\ref{subsec:spin_e}. Finally, we briefly describe the orbital evolution of BNSs under the GW radiation after their formation in Section~\ref{subsec:orbevol}.

\subsection{BSE model settings}
\label{subsec:bse}

We set the initial condition for the binary stars at their formation time as follows. The initial mass function (IMF) for the primary components ($M_1$) of the binary stars is assumed to follow $p(M_1)\propto M_1^{-2.3}$ for $M_1>1\ M_{\odot}$ \citep{2001MNRAS.322..231K}, where $p(M_1)$ denotes the probability that $M_1$ is in the range from $M_1$ to $M_1+dM_1$. The mass ratio $q$ of the secondary component ($M_2$) to the primary one is assumed to follow a uniform distribution $\sim U(0.01,1)$ \citep{1989ApJ...347..998E}. The initial semi-major axis ($a$) of binary stars is assumed to follow the distribution as \citep[see][]{1998MNRAS.296.1019H, 2003MNRAS.341..669H}
\begin{equation}
an(a)=\left\{
\begin{array}{c}
\alpha_{\rm sep}(\frac{a}{a_{0}})^{k},\ a\leqslant a_{0},\\
\alpha_{\rm sep},\ a_{0}<a<a_{1}.
\end{array}
\right.
\label{eq_a}
\end{equation}	
where $n(a)$ denotes the probability of binaries with semimajor axis in the range from $a$ to $a+da$, $\alpha_{\rm sep}\approx0.070$, $a_{0}=10 R_{\odot}$, $a_{1}=5.75\times10^{6} R_{\odot}=0.13$\,pc, and $k\approx1.2$. All binary systems are assumed to be on circular orbits at the birth time and their initial eccentricities $e_{\rm i}=0$.

Massive stars may have substantial wind ejection during its evolution. In this paper, we consider the wind velocity with a $\beta$-type velocity law for the accelerating part of the wind as \citep{2001A&A...369..574V}
\begin{equation}
v_{\rm w}(r)=v_\infty \left( 1-\frac{R_{i}}{r}  \right)^{\beta},
\label{eq:vw}
\end{equation}
where $v_\infty=k v_{\rm esc}$ is the terminal velocity. The escape velocity is calculated as $v_{\rm esc}=(2GM_i/R_i)^{1/2}$, where $M_i$ and $R_i$ denoting the mass and radius of the $i$-component of the binary, and $i=1$ and $2$. The numerical factor $k=2.6$ or $1.3$ for hot stars or less hot stars due to so-called ``bi-stability'' jump (mass loss rate $\dot{M}$ of the nuclear burning star via stellar wind decreases with decreasing star temperature and drops sharply at the so-called ``jump temperature''). The jump temperature between hot stars and less hot stars can be estimated as \citep{2001A&A...369..574V}
\begin{equation}
T_{\rm eff}^{\rm jump}=61.2(\pm 4.0)+2.59(\pm 0.28){\rm log}\langle \rho \rangle \  {\rm kK},
\end{equation}
where the characteristic density $\langle \rho \rangle$, defined as the wind density at $0.5v_\infty$ of the wind, is given by \citep{2001A&A...369..574V}
\begin{equation}
\log \langle \rho \rangle=-13.636(\pm 0.029)+0.889(\pm 0.026){\rm log}(Z/Z_{\odot}).
\end{equation}
The parameter $\beta$ in Equation~\eqref{eq:vw} governs the degree of acceleration and the wind velocity can be controlled by changing $\beta$. Here, we set $\beta=1$, which is corresponding to the fast wind model in \citet{2020PASJ...72...95K}, and we also discuss the possible effects of the wind velocity setting on the neutron star spin evolution by choosing different $\beta$ in Section~\ref{sec:discussion}. The stellar wind loss rate $\dot{M}_{\rm d}$ follows the formulas given in \citet{2000MNRAS.315..543H}, and one can find that summarized in \citet{2022ApJS..258...34R}. The mass accretion rate on to the neutron star is given as
\begin{equation}
\dot{M}_{\rm acc}=\frac{G^2M_{\rm NS}^2}{v_{\rm w}^{4}a^{2}(1-e^{2})^{1/2}}\dot{M}_{\rm d},
\end{equation}
where $a$ and $e$ denote the semi-major axis and eccentricity of the binary orbit, respectively. Besides, this mass accretion rate is limited by the Eddington limit \citep{1967Natur.215..464C,2002MNRAS.329..897H}
\begin{equation}
\label{eq:dmedd}
\dot{M}_{\rm Edd}=2.08\times10^{-3}(1+X)^{-1}\frac{R_{\rm NS}}{{\rm R}_{\odot}}{\rm M}_{\odot}{\rm yr}^{-1},
\end{equation}
where $X$ represents the hydrogen mass fraction. Once the radius of one of the two stars becomes equal to (or larger than) the Roche lobe (RL) radius \citep{1983ApJ...268..368E}
\begin{equation}
r_{\rm L}=a\frac{0.49q^{2/3}}{0.6q^{2/3}+{\rm ln}(1+q^{1/3})},
\end{equation}
the Roche lobe overflow (RLOF) begins and mass transfer occurs in close binary systems. Here, we adopt the mass transfer stability options used in \citet{2002MNRAS.329..897H} and enforce the Eddington limit if the accretor is a compact object \citep{2023MNRAS.524..426I}.

The common-envelope (CE) evolution can be simplified by using the description of the $\alpha$-formalism  \citep{1984ApJ...277..355W}, i.e.,
\begin{equation}
\frac{G(M_1-m_{1\rm c})M_1}{\lambda r_{\rm L}} = \alpha_{\rm CE} \left( \frac{Gm_{1\rm c}M_2}{2a_{\rm f}} - \frac{GM_1 M_2}{2a_{\rm i}}\right),
\end{equation}
where $\alpha_{\rm CE}$ represents the CE ejection efficiency, $\lambda$ is a structure parameter which depends on the evolutionary state of the donor, $r_{\rm L}$ is the Roche lobe radius, $a_{\rm i}$ and $a_{\rm f}$ are the semi-major axis of the binary before and after the CE stage, $M_1$ and $m_{\rm 1c}$ are the masses of the star and its core, $M_2$ is the companion mass. The structure parameter $\lambda$ is given in \citet{2014A&A...563A..83C} and \citet{2021A&A...650A.107M}. The value of $\alpha_{\rm CE}$ is highly uncertain. According to its definition, $\alpha_{\rm CE}$ should range from $0$ to $1$. However, \citet{2023MNRAS.524..426I} found that for the cases with $\alpha_{\rm CE}\leq1$, the formation of BNSs is highly suppressed, and consequently the merger rate. We also check the cases with $\alpha_{\rm CE}<1$ and find that the distribution of observed Galactic BNSs cannot be reproduced. Recent works have explored the cases with $\alpha_{\rm CE}>1$, corresponding to the assumption that the orbital energy is not the only source of energy contributing to unbind the CE and  the ejection of CE is more efficient than that for the cases with $\alpha_{\rm CE}<1$ \citep[e.g.,][and references therein]{2023LRCA....9....2R}. Adopting a larger $\alpha_{\rm CE}$ ($\alpha_{\rm CE}>1$) also leads to the formation of spiraling-in BNSs at relatively larger radii. In this paper, we adopt $\alpha_{\rm CE}=3$ according to \citet{2023MNRAS.526.2210S} because adopting this value the population synthesis model can best reproduces the BNS merger rate in our Galaxy. Note also that the $\alpha_{\rm CE}\lambda$ parameterization might be too simplistic to deal with the complex CE evolution, especially assuming a constant value of $\alpha_{\rm CE}$ for the full stellar mass range. The more complicated formalism for the CE phase \citep[e.g.,][]{2011MNRAS.411.2277D,2022ApJ...937L..42H,2023ApJ...944...87D} may be alternative adopted in future work.

We consider several different types of supernovae in the model. For stars with helium core masses in the range of $1.6$-$2.25M_{\odot}$ \citep{2002MNRAS.329..897H} and $>2.25M_{\odot}$, we assume that they end up as the electron-capture supernovae \citep[ECSNe;][]{1980PASJ...32..303M, 1984ApJ...277..791N, 1987ApJ...322..206N} and the core-collapse supernovae \citep[CCSNe;][]{2002MNRAS.329..897H}, respectively. In a short-orbital-period binary with a compact companion, mass transfer may occur when the helium star is re-expanding after core helium burning and the donor is severely stripped by the companion, leaving behind a helium envelope with mass less than $0.1M_{\odot}$ \citep[so-called ``case BB mass transfer'';][]{2013ApJ...778L..23T, 2015MNRAS.451.2123T}. If the remaining stellar core is massive enough to undergo core collapse, we assume that it ends up as an ultra-stripped supernova \citep[USSNe;][]{2013ApJ...778L..23T, 2015MNRAS.451.2123T}. In addition, we assume that the case BB mass transfer is always stable \citep{2018MNRAS.481.4009V} and it removes the entire helium envelope \citep{2022ApJS..258...34R}. 

The mass of the remnant leaving by a supernova depends on the type of the supernova, the core mass, and the neutron star equation of state (EOS). For stars undergoing ECSNe, we set the remnant mass as $1.26M_{\odot}$ \citep{1996ApJ...457..834T}. For stars undergoing CCSNe or USSNe, we calculate the remnant mass following the \textit{delayed} supernova remnant mass function (RMF) given by  \citet{2012ApJ...749...91F}. We adopt two types of EOSs for neutron stars, i.e., one is the DD2 EOS with the maximum mass for nonrotating neutron stars of $M_{\rm TOV}\sim2.42 M_{\odot}$ \citep[see][]{2014ApJS..214...22B}, and the other is the SLy4 EOS with $M_{\rm TOV}\sim2.05 M_{\odot}$ \citep[e.g.,][]{2001A&A...380..151D}. For a neutron star with given mass ($M_{\rm NS}$) and EOS, its radius ($R_{\rm NS}$) can be estimated from its mass [see Fig.~6 in \citet{2001A&A...380..151D} and Fig.~2 in \citet{2014ApJS..214...22B}]. The moment of inertia for the neutron star can be estimated as $I=0.4M_{\rm NS}R_{\rm NS}^{3}$ under the assumption of the rigid body, for simplicity. Note that a fast spinning neutron star can be substantially heavier than $M_{\rm TOV}$ \citep{1992ApJ...398..203C,1994ApJ...424..823C, 2000ApJ...528L..29B}. However, these fast spinning neutron stars may collapse to black holes when they evolve to slower spinning ones with longer spin periods. In our simulations, fast spinning neutron stars indeed evolve to slower spinning ones with longer spin period (see details in Section~\ref{sec:results}). Therefore, we do not consider the neutron stars with masses larger than $M_{\rm TOV}$ for simplicity. Note that the mass-radius relation may be dependent on the spin of the neutron star, and the spinning ones have the relation slightly different from that of non-rotating ones. For simplicity, we also ignore this difference in this paper for calculating the spin evolution.

Different type of supernovae may result in neutron stars with different natal-kick velocities ($v_{\rm k}$). We draw the natal-kick velocities from a Maxwellian distribution
\begin{equation}
\label{eq:kickv}
\frac{dN}{Ndv_{\rm k}}=\left( \frac{2}{\pi} \right)^{1/2}\frac{v_{\rm k}^{2}}{\sigma_{\rm k}^{3}}
\exp\left(-\frac{v_{\rm k}^{2}}{2\sigma_{\rm k}^{2}}\right),
\end{equation}
where $\sigma_{\rm k}$ is its dispersion, $dN/N$ is the normalized number in a kick velocity bin $dv_{\rm k}$. For ECSNe and USSNe, we set $\sigma_{\rm k}=30\ {\rm km\,s}^{-1}$ \citep{2018MNRAS.481.4009V} and for CCSNe, we set $\sigma_{\rm k}=265\ {\rm km}\,s^{-1}$ \citep{2005MNRAS.360..974H}.

The survived number of BNSs depends on the star formation history of the Milky Way as well as the metallicities of the stars. In this paper, we assume 
that the total mass of the stars formed in the disc and bulge amount to $5.2\times10^{10}\ M_{\odot}$, $0.9\times10^{10}\ M_{\odot}$, respectively \citep{2015ApJ...806...96L}. We adopt the star formation and metallicity enrichment histories from \citet{2003MNRAS.345.1381F} (see Fig.~1 therein) and \citet{2014ApJ...781L..31S} (see Fig.~2 therein) to reproduce the star formation in the Galactic bulge and disc, respectively. The grid of metallicity is set to be $Z=0.1 Z_\odot$, $0.3 Z_\odot$, $0.5 Z_\odot$, $0.7 Z_\odot$, and $1 Z_\odot$, to cope with metallicity bins of zero-age main sequence (ZAMS) binaries $Z/Z_\odot \in (0,0.2]$, ($0.2, 0.4$], ($0.4, 0.6$], ($0.6, 0.85$], and $(0.85, 5)$, separately. Here, we take the solar metallicity as $Z_\odot=0.02$. For simplicity, we assume that half of the massive stars are in binaries.

\subsection{Spin evolution of neutron stars}
\label{subsec:spin_e}

The spin of a neutron star can be characterized by its spin period, and the neutron star may spin up or spin down due to various processes. Here spin-up and spin-down mean that the spin period goes down and up, respectively. For isolated neutron stars, the spin declines continuously and the spin-down rate highly depends on the magnetic field $B$, the braking index $n$ (a dimensionless parameter characterizing how the pulsar spin-down rate varies with the rotation frequency), the alignment angle $\chi$ (the angle between the rotational and magnetic axes of the pulsar), and the detailed spin-down model. For simplicity, we treat all pulsars as orthogonal rotators (i.e., $\chi=90^{\circ}$) with a fixed index \citep[e.g., $n=3$, see ][]{2010MNRAS.404.1081R} and set the initial spin period as $P_{0}=0.01$\,s \citep[see][a different set of the initial spin has little effect on our final results]{2020PASJ...72...95K}. The spin-down rate is given by
\begin{equation}
\dot{P}= 9.73\times10^{-16}\mu_{30}^{2}I_{45}^{-1}P^{-1} {\rm s\,s}^{-1},
\label{eq:isolate}
\end{equation}
where $\mu_{30}$ ($\mu=BR_{\rm NS}^{3}$) and $I_{45}$ are the magnetic dipole moment in units of $10^{30}$\,G\,cm$^3$ and the moment of inertia in units of $10^{45}$\,g\,cm$^2$, respectively \citep{2010MNRAS.404.1081R}. The so-called ``death line" which separate the radio-quiet pulsars from others is given by
\begin{equation}
P_{\rm death}= \sqrt{\frac{B}{1.7\times10^{11}\,{\rm  G} } }\, {\rm s},
\label{eq:pdeath}
\end{equation}
and pulsars with spin period $P>P_{\rm death}$ are radio-quiet \citep{1991PhR...203....1B}.\footnote{One may note that a small fraction of pulsars \citep[e.g., J2144-3933;][]{1999Natur.400..848Y} with $P>P_{\rm death}$ are still observable at radio band.}

The spin evolution of pulsars during the BNS formation processes may be vastly different from the cases for isolated pulsars. In the binary cases, mass accretion onto the first-born neutron star may occur due to a massive donor, which leads to significant evolution of the neutron star spin. At this stage, the system can be treated as a high-mass X-ray binary (HMXB). According to different types of the donor, HMXBs can be divided into roughly two groups, i.e., OB-type and Be-type \citep{1984A&A...141...91C, 1986MNRAS.220.1047C, 1997ApJS..113..367B}. For an OB-type HMXB, the donor fills its Roche lobe (RL) and the neutron star accretes matter overflown from the RL. In contrast, a neutron star in a Be-type HMXB accretes the stellar wind from the donor.

Below, we first summarize the definitions for different radii related to various processes used in our following analysis and calculations, for convenience.
\begin{itemize}
\item $R_{\rm g}$: the Bondi radius, defined as $R_{\rm g}=2GM_{\rm NS}/v_{\rm w}^2$ \citep{1952MNRAS.112..195B}. Material inside $R_{\rm g}$ can be captured by the neutron star gravity.
\item $R_{\rm lc}$: the light cylinder radius, defined as $R_{\rm lc}=cP/2\pi$ \citep{1981MNRAS.196..209D}.
\item $R_{\rm mag,1}$: the magnetic field radius when $R_{\rm mag,1}<R_{\rm g}$, defined as $R_{\rm mag,1}=(\mu^2/2\dot{M}\sqrt{2GM_{\rm NS}})^{2/7}$ \citep{2010MNRAS.407.1090B}.
\item $R_{\rm mag,2}$: the magnetic field radius when $R_{\rm mag,2}>R_{\rm g}$, defined as $R_{\rm mag,2}=(2\mu^2G^2M_{\rm NS}^2/\dot{M}v_{\rm w}^5)^{1/6}$ \citep{2010MNRAS.407.1090B}.
\item $R_{\rm a}$: the inner boundary radius (with respect to the neutron star) of the wind flow, defined as $R_{\rm a}=R_{\rm mag,1}^{7/9}R_{\rm g}^{2/9}$ \citep{1981MNRAS.196..209D}. A pressure balance is achieved between the magnetosphere and the wind material at $R_{\rm a}$.
\item $R_{\rm b}$: the outer boundary radius (with respect to the neutron star) of the wind flow, defined as $R_{\rm b}=R_{\rm g}^3 \Omega^2 /v_{\rm w}^2$ \citep{1981MNRAS.196..209D} with $\Omega$ representing the rotation velocity of the pulsar. For the wind material outside $R_{\rm b}$, the pressure and density are approximately constant.
\item $R_{\rm cor}$: the corotation radius, defined as $R_{\rm cor}=(GM_{\rm NS}P^2/4\pi^2)^{1/3}$ \citep{2007rapp.book.....G}.
\end{itemize}

\subsubsection{Wind-fed accretion}

The donor star in a binary may not expand right after the formation of the first born neutron star due to different lifetimes of massive main sequence stars, and thus the accretion onto the neutron star may be first fed by wind only as the Roche Lobe is not filled, yet. In this case, the system appears as a wind-fed HXMB. The newly formed neutron star may have strong magnetic field and rapid spin, and it emits strong electro-magnetic radiation and relativistic particles. As a result, the mass accretion is inhibited until the outgoing flux is less than the accretion ram pressure \citep{2012MNRAS.421L.127P, 2020PASJ...72...95K}. This period is denoted as the \textit{ejector phase} (phase a). In this phase, the spin evolution of the neutron star follows the spin-down model for isolated pulsars as above (Eq.~\ref{eq:isolate}). 
For the weak stellar wind case, $\dot{M}<\dot{M}_{\rm c1}$, where \citep{1975A&A....39..185I}
\begin{equation}
\label{eq:dmc1}
\dot{M}_{\rm c1}=10^{10} B_{11}^{2} v_{8}^{7} / M_{\rm NS}^{4},
\end{equation}
here $B_{11}$ represents the magnetic field in unit of $10^{11}$\,G, and $v_8$ denotes the wind velocity $v_{\rm w}$ in unit of $10^{8}$\,cm\,s$^{-1}$,
this phase ends at \citep{1981MNRAS.196..209D}
\begin{equation}
P_{\rm ab}=0.8\mu_{30}^{1/3}\dot{M}_{15}^{-1/6}v_{8}^{-5/6}\left( \frac{M_{\rm NS}}{M_{\odot}} \right)^{1/3}\ \rm s.
\label{eq:Pab}
\end{equation}
For the strong stellar wind case ($\dot{M}>\dot{M}_{\rm c1}$), this phase ends at \citep{1981MNRAS.196..209D}
\begin{equation}
P_{\rm ac}=1.2\mu_{30}^{1/2}\dot{M}_{15}^{-1/4}v_{8}^{-1/2}\ \rm s.
\label{eq:Pac}
\end{equation}

The neutron star may evolve into the \textit{rapid rotator phase} (phase b) after the ejector phase in the weak stellar wind case ($P>P_{\rm ab}$). In the rapid rotator phase, the wind material from the donor enters the light cylinder of the neutron star before it can enter into the magnetic field radius and the Bondi radius, i.e., $R_{\rm g}<R_{\rm mag,2}<R_{\rm lc}$. It can hardly penetrate deep into the region with radius $\lesssim R_{\rm mag,2}$ due to the strong magnetic field inside the light cylinder and thus the rotational energy of the neutron star will be dissipated at $R_{\rm mag,2}$. The spin-down rate in this phase can be estimated as \citep{2004ChJAA...4..320Z}
\begin{equation}
\dot{P}=8.59\times10^{-16}\mu_{30}\dot{M}_{15}^{1/2}v_{8}^{5/2}P^{2}\ \rm s\ s^{-1},
\label{eq:dp_phaseb}
\end{equation}
and this phase ends when the outer boundary of the wind plasma $R_{\rm b}$ decreases to the Bondi radius $R_{\rm g}$ \citep{1981MNRAS.196..209D}, at which the neutron star period is
\begin{equation}
P_{\rm r}=2.2\mu_{30}\dot{M}_{15}^{-1/2}v_{8}^{1/2}\left( \frac{M_{\rm NS}}{M_{\odot}} \right)^{-1}\ \rm s.
\label{eq:Pbc}
\end{equation}

The neutron star may enter the \textit{propeller phase} (phase c) after the ending of the rapid rotator phase in the weak stellar wind case with $P=P_{\rm r}$ or the ejector phase in the strong stellar wind case with $P=P_{\rm ac}$. In this propeller phase, the inner boundary of the wind matter $R_{\rm a}$ is larger than the corotation radius $R_{\rm cor}$ but smaller than the Bondi radius $R_{\rm g}$, i.e., $R_{\rm g}>R_{\rm a}>R_{\rm cor}$. The spin angular momentum is transferred to the ejected wind matter through the connection with the rotating magnetic field. Hence, the neutron star spin declines rapidly and the spin-down torque in this phase is \citep{2007rapp.book.....G, 2020PASJ...72...95K}
\begin{equation}
N_{\rm c}=-I\dot{\Omega}= \frac{2\pi I\dot{P}}{P^2}
=\left( \frac{\pi^2 \mu^4}{9GM_{\rm NS}P^2 R_{\rm mag,1}^3} \right)^{1/2}.
\label{eq:dp_phasec}
\end{equation}
This propeller phase ends at the quasi-equilibrium rate when the inner boundary of the wind matter $R_{\rm a}$  reaches the corotation radius $R_{\rm cor}$ and the resulting spin of the neutron star is \citep{1981MNRAS.196..209D}
\begin{equation}
P_{\rm eq}=23 \mu_{30}^{2/3}\dot{M}_{15}^{-1/3}v_{8}^{-2/3}\ \rm s.
\label{eq:Peq1}
\end{equation}
However, if the magnetic field remains at high value at the end of the ejector phase, this field may inhibit the accretion onto the neutron star. For strong stellar wind case, when $R_{\rm mag,2}>R_{\rm cor}$, the wind material can be blown off via a strong centrifugal force provided by the rapid magnetospheric rotation, and this state corresponds to the \textit{super-Keplerian magnetic inhibition phase} \citep[phase b2;][]{2008ApJ...683.1031B}. During this phase, the spin declines rapidly and the spin-down torque can be estimated as \citep{2020PASJ...72...95K}
\begin{equation}
N_{\rm b2}=-I\dot{\Omega}=R_{\rm mag,2}(\dot{M}_{\rm mag}v_{\rm spin}),
\end{equation}
where $\dot{M}_{\rm mag}$ denotes the rate of material entering into the magnetic radius and $v_{\rm spin}$ is the rotational velocity of the magnetic field lines at $R_{\rm mag,2}$. Consequently, the corotation radius $R_{\rm cor}$ continues to expand until it reaches the magnetic radius $R_{\rm mag,2}$. Once $R_{\rm cor}>R_{\rm mag,2}$, the wind matter will fall on to the neutron star due to the Kelvin-Helmholtz instability instead of being blown off \citep{1992Natur.357..388H}. This phase (phase b3) is subject to the \textit{sub-Keplerian magnetic inhibition phase} \citep{2008ApJ...683.1031B} or the \textit{georotator phase} \citep{1992ans..book.....L}. During this phase, a quasi-equilibrium state with $P=P_{\rm eq}$ (Eq.~\ref{eq:Peq1}) is achieved ($R_{\rm cor}=R_{\rm mag,2}>R_{\rm g}$). The spin period of neutron star declines gradually  considering the slow decrease of $P_{\rm eq}$ due to the decay of the magnetic field \citep{2010MNRAS.407.1090B,2020PASJ...72...95K}.

A quasi-spherical wind accretion can occur at the end of the propeller phase if the corotation radius $R_{\rm cor}$ becomes larger than the inner boundary of the wind plasma $R_{\rm a}$, or occur at the end of the sub-Keplerian magnetic inhibition phase if the Bondi radius $R_{\rm g}$ becomes larger than the magnetic field radius $R_{\rm mag,2}$. Therefore, a thick settling shell can be formed around the neutron star \citep{2012MNRAS.420..216S}. According to the value of the accretion rate, the accretion phases can be divided into two cases. If $\dot{M}>\dot{M}_{\rm c2}$, where \citep{1984ApJ...278..326E}
\begin{equation}
\dot{M}_{\rm c2}=1.6\times10^{16}\mu_{30}^{1/4}\left( \frac{M_{\rm NS}}{M_{\odot}} \right)^{-1/2}\left( \frac{R_{\rm NS}}{10\ \rm km}\right)^{7/8}\ \rm g\ s^{-1},
\label{eq:dmcrit}
\end{equation}
the cooling of the shell is efficient due to the inverse-Compton mechanism and the matter falls toward the neutron star at a supersonic speed, corresponding to the \textit{Bondi-Hoyle-Littleton (BHL) accretion phase} \citep[phase d1;][]{1952MNRAS.112..195B, 1939PCPS...35..405H}. Otherwise, the shell is cooled inefficiently. Here, the angular momentum of neutron star can be removed from the magnetosphere in two ways. One is that the material falls on to neutron star at a  subsonic speed, corresponding to the \textit{subsonic settling accretion phase} \citep[phase d2;][]{2012MNRAS.420..216S}. The other is that the material is expelled from the magnetospheric boundary without accretion, corresponding to the \textit{subsonic propeller phase} \citep[phase d3;][]{2008ApJ...683.1031B}. In the phase d1, the spin-up torque can be estimated as
\begin{equation}
N_{\rm d1}=I\dot{\Omega}=\dot{M}R_{\rm mag,1}^{2}\Omega.
\end{equation}
This phase ends at the spin equilibrium state $P_{\rm eq}$ (see Eq.~\ref{eq:Peq1}). In the phase d2, the settling accretion shell can inhibit the accretion, and the interaction between the magnetic field and the settling shell material exerts an additional torque on the neutron star. The total torque can be estimated as \citep{2020PASJ...72...95K}
\begin{equation}
N_{\rm d2}=I\dot{\Omega}=A\dot{M}_{16}^{7/11}-B\dot{M}_{16}^{3/11},
\end{equation}
where $A$ and $B$ are two functions given by
\begin{equation}
A=4.6\times10^{31}K\mu_{30}^{1/11}v_{8}^{-4}\left( \frac{P_{\rm orb}}{10\ {\rm day}} \right)^{-1},
\end{equation}
\begin{equation}
B=5.5\times10^{32}K\mu_{30}^{13/11}\left( \frac{P}{100\ {\rm s}} \right)^{-1}.
\end{equation}
Here $K\sim 40$ is a non-dimensional factor \citep{2012MNRAS.421L.127P}. This phase ends at the equilibrium spin period \citep{2016ApJ...824..143L}
\begin{equation}
P_{\rm eq,d2}=1193\mu_{30}^{12/11}v_{8}^{4}\dot{M}_{16}^{-4/11}\left( \frac{P_{\rm orb}}{10\ {\rm day}} \right) \ \rm s.
\label{eq:Peq2}
\end{equation}
In the phase d3, the spin-down rate is \citep{2010MNRAS.407.1090B} 
\begin{equation}
\dot{P}=2.4\times10^{-11}\mu_{30}^{2}\left( \frac{M_{\rm NS}}{M_{\odot}} \right)^{-1}\ \rm s\ s^{-1},
\end{equation}
and this phase ends at the critical spin period \citep{2001A&A...368L...5I}
\begin{equation}
P_{\rm eq,d3}=450\mu_{30}^{16/21}\dot{M}_{15}^{-5/7}\left( \frac{M_{\rm NS}}{M_{\odot}} \right)^{-4/21}\ \rm s.
\end{equation}
According to \citet{2016ApJ...824..143L}, the subsonic settling accretion model can well reconstruct the spin orbital-period distribution of HMXBs and the measured wind velocities from several HMXBs seem to favour this viewpoint. Therefore, we mainly adopt the subsonic settling accretion model for cases with $\dot{M}<\dot{M}_{\rm c2}$ to estimate the spin evolution and show the results in Section~\ref{sec:results}. For comparison, the results of subsonic propeller model are discussed in Section~\ref{sec:discussion}.

\subsubsection{Roche Lobe overflow}

Second, we consider the spin evolution of a neutron star due to Roche-Lobe overflow. The systems with RLOF (Cen X-3, Her X-1, etc.) show that once a permanent accretion disc forms, the neutron star is rapidly spun up back to a short pulse period and the spin-up rate is estimated as \citep{1972A&A....21....1P}
\begin{eqnarray}
\frac{\dot{P}}{P}& = & -7.2\times10^{-7}\dot{M}_{16}\left( \frac{M_{\rm NS}}{M_{\odot}}\right)^{-1/2} \times \\ \nonumber
& & \left( \frac{R_{\rm mag,1}}{10\ \rm km}\right)^{1/2}\left( \frac{R_{\rm NS}}{10\ \rm km}\right)^{-2}P \ {\rm yr^{-1}},
\end{eqnarray}
The spin-up process ends when the neutron star reaches an equilibrium spin period ($R_{\rm cor}=R_{\rm mag,1}$) given by \citep{1991PhR...203....1B} 
\begin{eqnarray}
P_{\rm eq,RL}& = & 1.6\left( \frac{B}{10^{12}\ {\rm G}} \right)^{6/7}\left( \frac{R_{\rm NS}}{10\ {\rm km}} \right)^{18/7} \times \\ \nonumber 
& & \left( \frac{M_{\odot}}{M_{\rm NS}} \right)^{5/7}\left( \frac{10^{17}\ {\rm g\ s^{-1}}}{\dot{M}} \right)^{3/7}\ \rm s.
\label{eq:peqrl}
\end{eqnarray}

After the birth of the second neutron star, we assume that both neutron stars spin down following Equation~\eqref{eq:isolate} and no further accretion occurs before the coalescence of the two neutron stars.

\begin{figure*}
\centering
\includegraphics[width=0.95\textwidth]{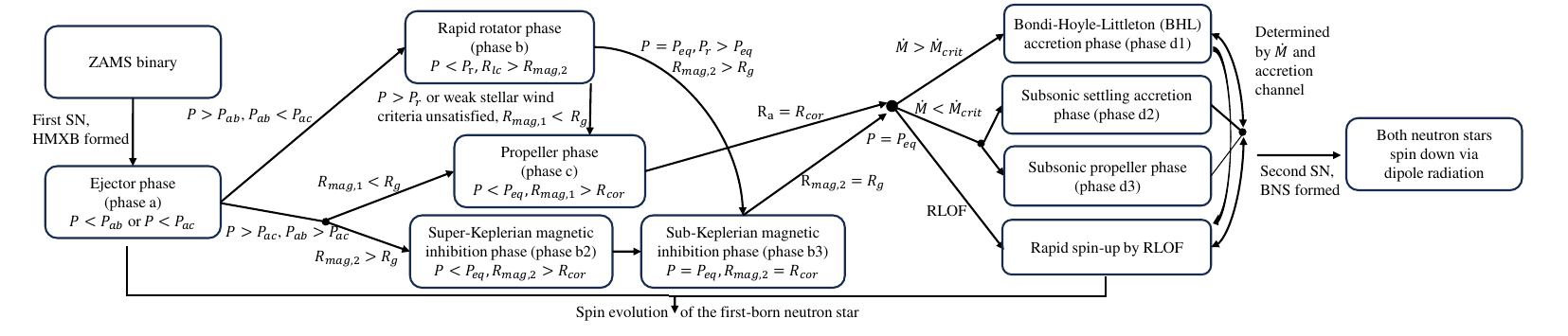}
\caption{A flow chart for the spin evolution of BNSs.}
\label{fig:f1}
\end{figure*}

A flow chart is shown in Figure~\ref{fig:f1} to exemplify the spin evolution of two neutron star components in the formation processes of a BNS system, by including various astrophysical processes as described above that affect the spin evolution of the first-born neutron star component before the formation of the second one.

\subsubsection{Magnetic field evolution}

It is widely accepted that a neutron star has a strong magnetic field right after its birth and this initial strong magnetic field declines with time. Here we assume the magnetic field decays in an exponential form
\begin{equation}
B(t)=(B_{\rm ini}-B_{\rm fin})\times{\rm exp}(-t/t_{\rm d})+B_{\rm fin},
\end{equation}
where $t_{\rm d}$ is the magnetic field decay timescale, and $B_{\rm fin}=10^{8}\ {\rm G}$ is the final field corresponding to the typical values observed in neutron stars in old systems (e.g., low mass X-ray binaries; LMXBs) \citep{2011MNRAS.413..461O}. The magnitude of initial field $B_{\rm ini}$ is still under debate. In our calculations, we assign the initial magnetic field for a neutron star formed from stars with mass $<20\ M_\odot$ by assuming it follows a log-normal distribution with the mean of $\langle\log (B_{\rm ini}/\rm G)\rangle \sim13$ and the scatter of $\sim0.55$, according to the constraints on the distribution obtained in the literature \citep[see,][]{2006ApJ...643..332F, 2013MNRAS.432..967I}. However, for those neutron stars descend from particularly massive stars ($M_{\rm MS}\gtrsim20\ M_{\odot}$), magnetars can be formed via the fossil field scenario or the dynamo hypothesis and we assume that their initial magnetic field is uniformly distributed in the range of $\sim 10^{14}-10^{15}$\,G \citep{2021ASSL..461...97E}, according to the magnetar catalog from \citet{2014ApJS..212....6O}. There is a large uncertainty associated with the magnetic field decay timescale. It has been proposed that the accretion material from the companion star onto the pulsar will bury the pulsar magnetic field at an accelerated rate \citep{2006MNRAS.366..137Z}. Therefore, we set $t_{\rm d}=10^6$\,yr when the neutron star accretes from the donor \citep{2020PASJ...72...95K}. Otherwise, we set $t_{\rm d}=10^9$\,yr \citep{2020MNRAS.494.1587C}.

\subsection{Orbital evolution of BNSs after their formation}
\label{subsec:orbevol}

BNSs continue to evolve under the gravitational wave (GW) radiation after their formation. In this stage, the evolution of the BNS semimajor axes ($a$) and eccentricities ($e$) are given by \citep{Peters1964}
\begin{eqnarray}
\left< \frac{da}{dt} \right>& = & -\frac{64}{5}\frac{G^{3}m_{1}m_{2}(m_{1}+m_{2})}{c^{5}a^{3}(1-e^{2})^{7/2}}
\left(1+\frac{73}{24}e^{2}+\frac{37}{96}e^{4}\right), \nonumber \\ \\
\left< \frac{de}{dt} \right>& = & -\frac{304}{15}e\frac{G^{3}m_{1}m_{2}(m_{1}+m_{2})}{c^{5}a^{4}(1-e^{2})^{5/2}}
\left(1+\frac{121}{304}e^{2}\right), 
\label{eq:aeevol}
\end{eqnarray}
where $m_{1}$ and $m_2$ denotes the primary and secondary component masses of a BNS. In this way, we can obtain the merger timescale for each BNS. 

\section{Results} 
\label{sec:results}

For each metallicity,
we perform Monte Carlo simulations for the evolution of $10^7$ binaries by using the above settings for the BSE model. The total mass of the evolved binaries is $1.97\times10^{8}\ M_{\odot}$, representing $7.65\times10^8\ M_{\odot}$ of total star forming mass under the assumed initial mass distribution. According to these Monte Carlo simulations, we record the evolution track for each binary, especially those that finally form BNSs, and take them as the templates to further investigate the survived BNSs in the Milky Way.

\subsection{Evolution tracks of BNSs}
\label{sub:tracks}

According to our model and simulations, BNSs are mainly formed from two channels, i.e., the wind-fed channel and the wind-fed $+$ RLOF channel. For the wind-fed $+$ RLOF formation channel, the first-born neutron star can be spun up to have a relatively short spin period due to the prolonged RLOF accretion. In contrast, for the wind-fed channel, HMXBs go through the subsonic settling accretion phase when the donor evolves in the main sequence (MS) stage or the naked helium star MS (HeMS) stage, and the BHL accretion phase when the donor enters the Hertzsprung gap (HG) stage or the naked helium star Hertzsprung gap (HeHG) stage. In the subsonic settling accretion phase, the neutron star spins down to the quasi-equilibrium state with spin period of $P_{\rm eq,d2}$ (Eq.~\ref{eq:Peq2}) and then its spin period decreases from this quasi-equilibrium value gradually because of the decay of the magnetic field. Although a neutron star can be spun up during the BHL accretion phase, the spin period is still substantially larger than that formed from the wind-fed $+$ RLOF formation channel due to the limited timescale of the HG stage or the HeHG stage and the spin-up rate. 

The age of HMXBs, $t_{\rm e}$, is another parameter that affects the spin evolution. Here $t_{\rm e}$ is defined as the time since the formation of the first neutron star component, i.e., $t_{\rm e}=0$\,yr when this component is born. If the initial mass ratio $q\sim1$ or the initial binary system is massive enough, the lifetime of the HMXB is normally short. Hence the magnetic field is always strong and the quasi-equilibrium spin periods, e.g., $P_{\rm eq}$, $P_{\rm eq,d2}$, and $P_{\rm eq,RL}$, are relatively large in value, resulting in large spin period of the neutron star. If there is a significant difference in mass between the input binary components (i.e., a smaller mass ratio) and the total mass is relatively low, the HMXB exists for a longer time duration and at the end of the HMXB state, the magnetic field declines significantly. In this way, the first-born neutron star can be recycled as a millisecond pulsar through RLOF. 

\begin{figure*}[!htb]
\centering
\includegraphics[width=0.9\textwidth]{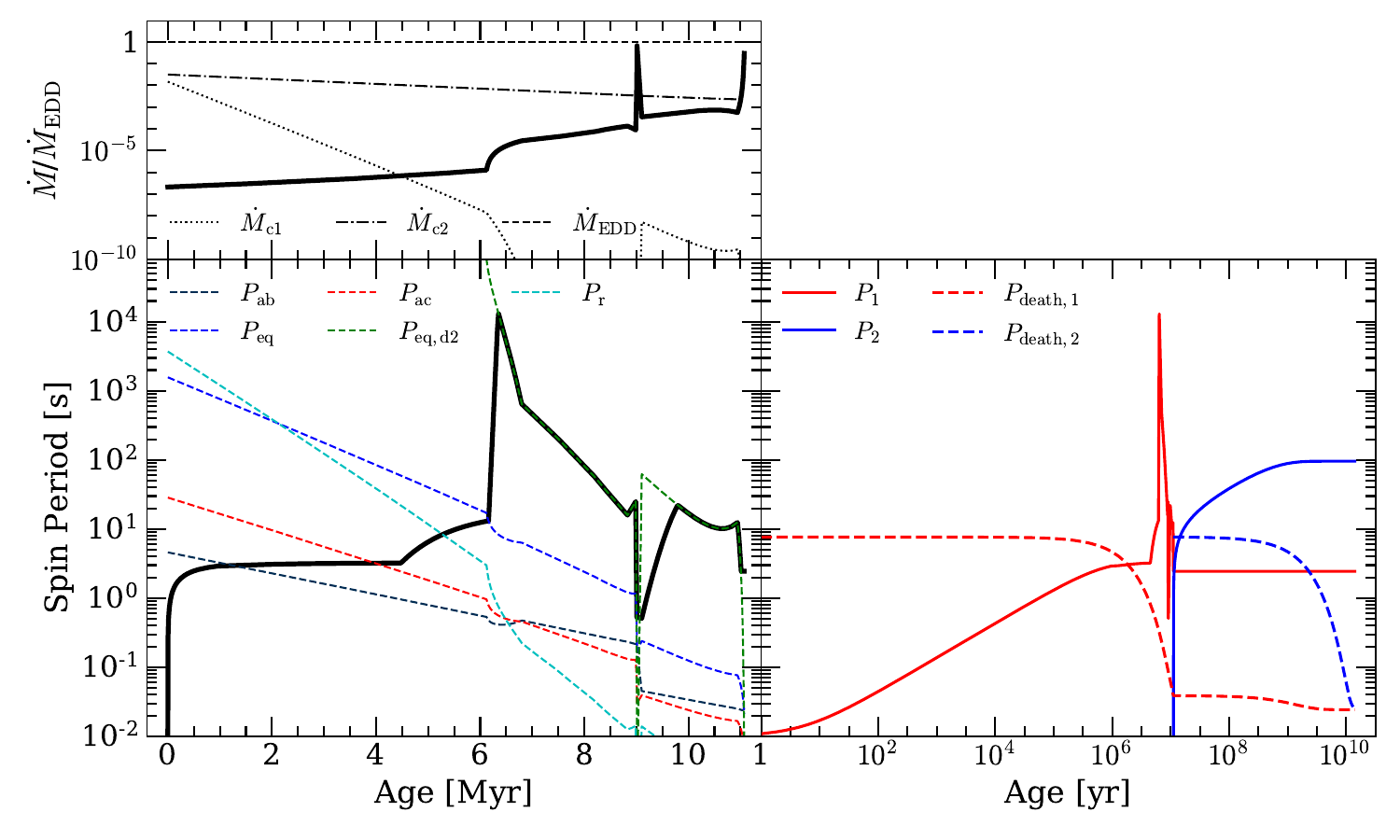}
\caption{
Spin evolution of the two neutron star components of a binary system, in which the first-born neutron star experienced wind-fed accretion during the HMXB stage. The finally formed BNS has a total mass of $M_{\rm tot}\sim2.6\ M_{\odot}$, similar to those Galactic BNSs. The total evolution period of the HMXB (from first SN to second SN) is $\sim11$\,Myr, and the initial and the final orbital periods of the HMXB are $P_{\rm orb,i}\sim53.3$\,day and $P_{\rm orb,f}\sim1$\,day at $t_{\rm e}=0$ and $11$\,Myr, respectively. 
The bottom left panel focuses on the spin evolution of the first-born neutron star during the HMXB stage, and the bottom right panel shows the spin evolution of both neutron stars since their formation. In the left panel, the black solid line shows the spin period evolution of the first-born neutron star during the HMXB stage; the black and the red dashed lines show the values of $P_{\rm ab}$ and $P_{\rm ac}$, the spin period for the first-born neutron star at the end of the ejector phase, in the weak and strong stellar wind cases, estimated according to Eqs.~\eqref{eq:Pab} and ~\eqref{eq:Pac}, respectively; the cyan dashed line shows the value of $P_{\rm r}$, the final spin period of the rapid rotator phase in the weak stellar wind case, estimated by Eq.~\eqref{eq:Pbc}; the blue dashed line shows the value of $P_{\rm eq}$, the quasi-equilibrium spin period of the sub-Keplerian magnetic inhibition phase, the minimum spin period of the BHL accretion phase, or the final spin period of the super-Kelperian magnetic inhibition phase and the propeller phase, estimated from Eq.~\eqref{eq:Peq1}; the green dashed line shows $P_{\rm eq,d2}$, the maximum spin period during the subsonic settling accretion phase, estimated from Eq.~\eqref{eq:Peq2}. In the right panel, the red and the blue dashed lines show the death line for the first-born and the second-born neutron stars, respectively. The red and the blue solid lines show the evolution of the spin period of the first and the second-born neutron star components, respectively.
The top panel shows the accretion rate (in unit of the Eddington rate) of the first-born neutron star during the HMXB stage. The dotted and dash-dot lines show the criteria for the weak or strong stellar wind case, subsonic or BHL accretion phase, estimated according to Eqs.~\eqref{eq:dmc1} and \eqref{eq:dmcrit}, respectively. 
}
\label{fig:f2}
\end{figure*}

\begin{figure*}
\centering
\includegraphics[width=0.9\textwidth]{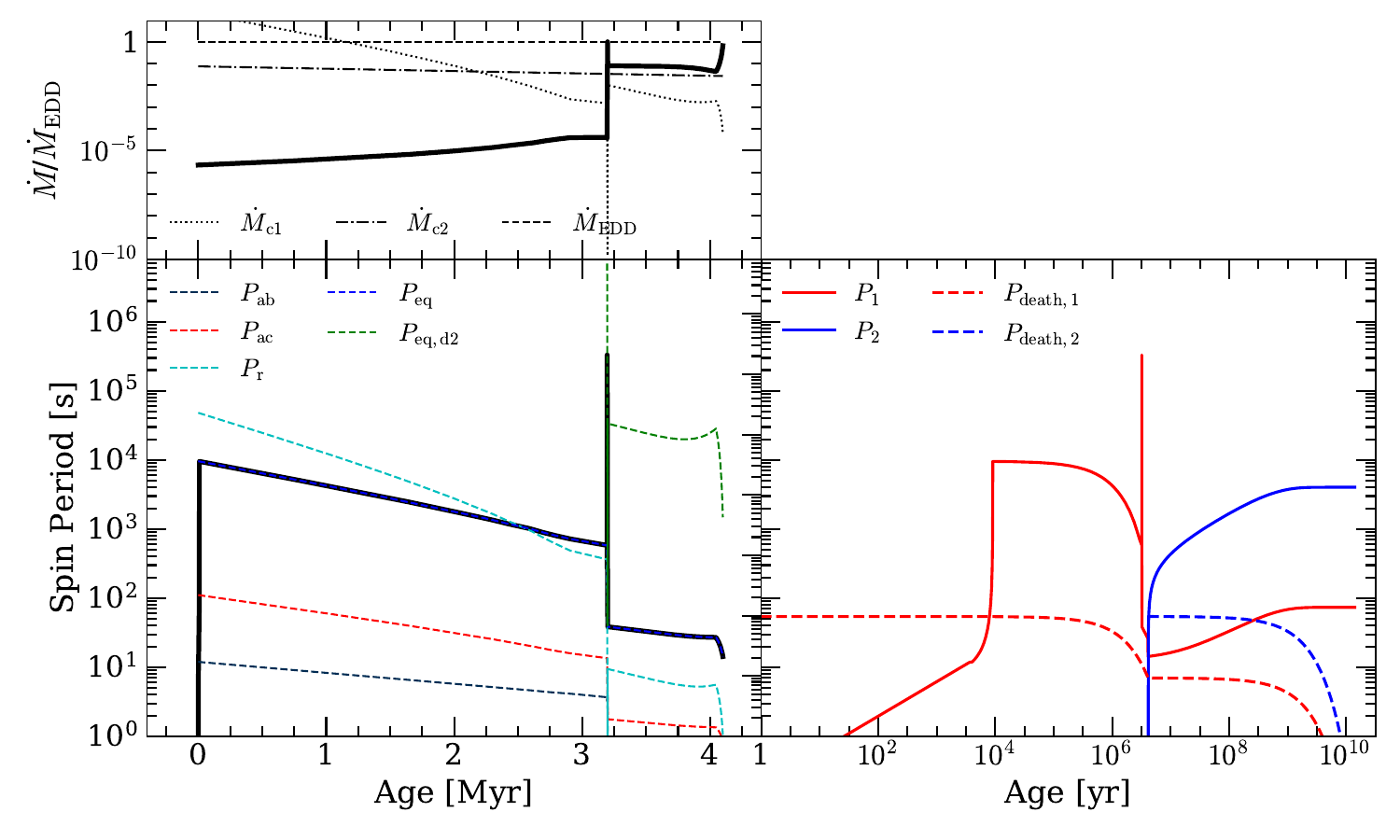}
\caption{
Legend is similar to Fig.~\ref{fig:f2}, except that the BNS resulting from the binary has a total mass of $M_{\rm tot}\sim3.4\ M_{\odot}$, similar to that of GW190425. The duration of the HMXB stage in this case is $\sim4.1$\,Myr, and the initial and the final orbital periods of the HMXB are $P_{\rm orb,i}\sim80.9$\,day and $P_{\rm orb,f}\sim1.3$\,day, respectively.
}
\label{fig:f3}
\end{figure*}

\begin{figure*}
\includegraphics[width=0.9\textwidth]{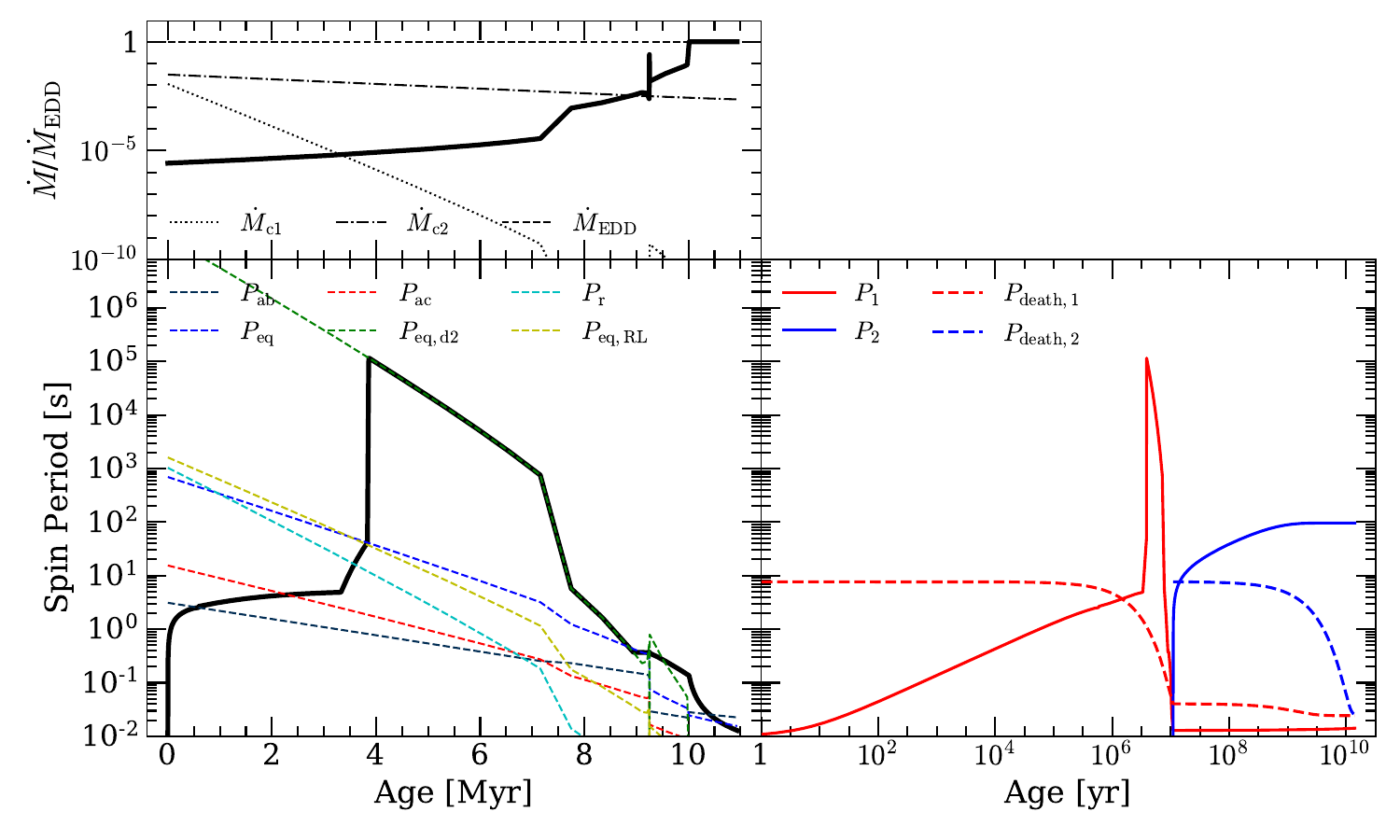}
\centering
\caption{
Legend is the same as that for Fig.~\ref{fig:f2}, except that the system experienced both wind-fed and RLOF accretion. The duration of the HMXB stage is $\sim11$\,Myr, and the initial and the final orbital periods of the HXMB are 
\textcolor{black}{ \bf $P_{\rm orb,i}\sim10.2$\,day, $P_{\rm orb,f}\sim0.11$\,day, respectively.} 
}
\label{fig:f4}
\end{figure*}

\begin{figure*}
\centering
\includegraphics[width=0.9\textwidth]{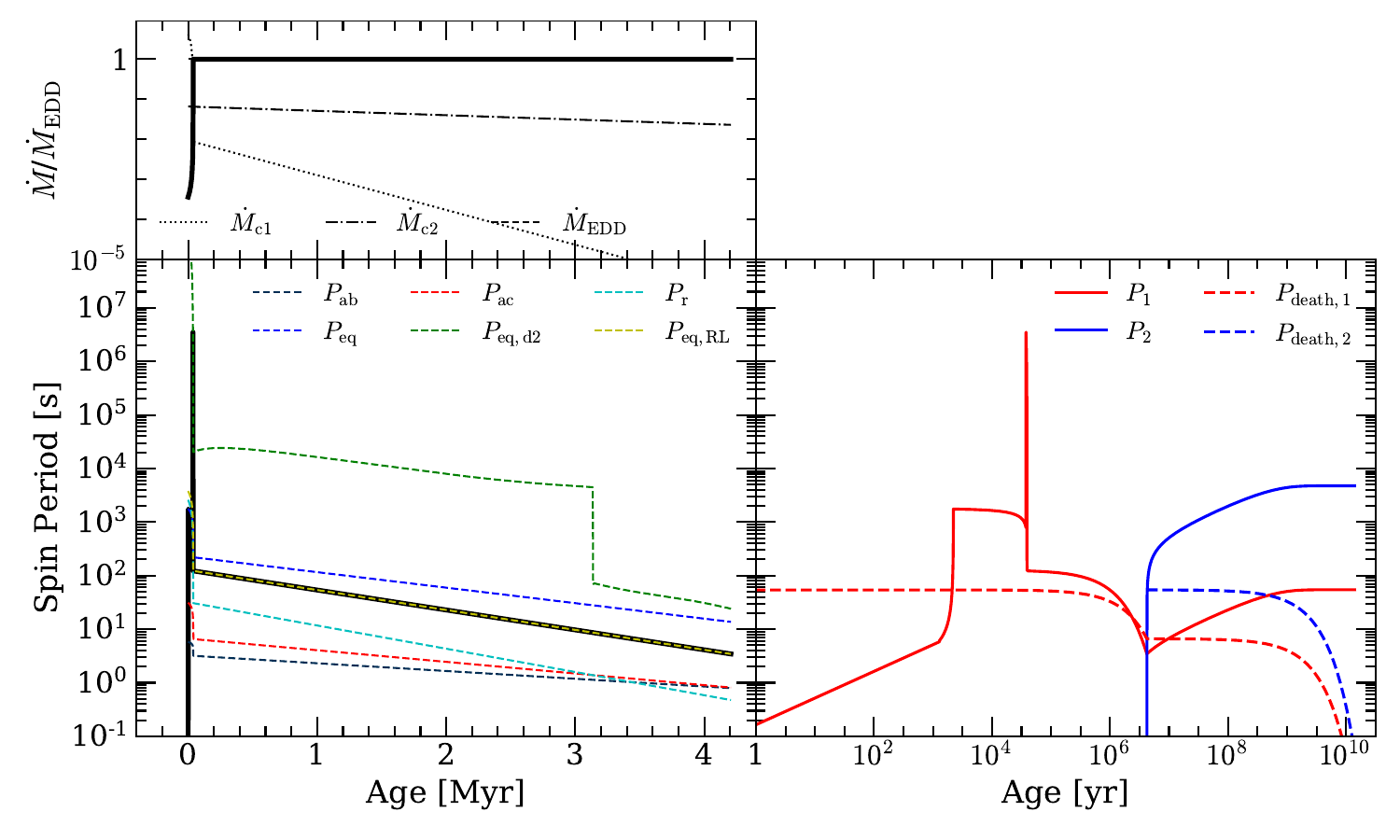}
\caption{
Legend is the same as that for Fig.~\ref{fig:f3}, except that the system experienced both wind-fed accretion and RLOF. The duration of the HMXB stage is $\sim4.2$\,Myr, and the initial and the final orbital periods of the HXMB are $P_{\rm orb,i}\sim10.5$\,day and $P_{\rm orb,f}\sim0.1$\,day, respectively.
}
\label{fig:f5}
\end{figure*}

Figures~\ref{fig:f2}-\ref{fig:f5} show several examples of the obtained spin evolution tracks for BNSs. Figure~\ref{fig:f2} shows the spin evolution of both neutron star components in a binary system, in which the first-born neutron star only experienced the wind-fed accretion during its evolution in the HXMB stage (hereafter case A). This binary star evolved to a BNS with mass $M_{\rm tot}\sim2.6 M_{\odot}$, similar to the typical mass of Galactic BNSs, at $t_{\rm e}\sim11$\,Myr after the formation of the first-born neutron star. The orbital period of the binary just after the first SN ($t_{\rm e}=0$) and just before the second SN ($t_{\rm e}=11$\,Myr) in the HMXB stage are 
 $P_{\rm orb,i}\sim53.3$\,day and $P_{\rm orb,f}\sim1$\,day, respectively.  
In this case, both progenitors of the binary have masses $<20\ M_{\odot}$ and the initial magnetic field of both neutron star components is set as $B_{\rm ini}=10^{13}$\,G.

The left panel of Figure~\ref{fig:f2} shows the spin evolution of the first-born neutron star during the HMXB stage. For case A, as seen from the figure, the first-born neutron star enters the ejector phase and spins down to $P_{\rm ab}$ continuously until $t_{\rm e}\sim0.9$\,Myr. After that, the orbital period is large and the donor is still at the MS stage, the weak stellar wind criterion is satisfied ($\dot{M}<\dot{M}_{\rm c1}$) and the neutron star then evolves into the rapid rotator phase.  At $t_{\rm e}\sim4.5$\,Myr, the weak stellar wind criterion is violated ($\dot{M}>\dot{M}_{\rm c1}$) due to the decay of the magnetic field and the expansion of the donor. The propeller spin-down phase occurs since then and the spin period grows up to $P_{\rm eq}$ at $t_{\rm e}\sim6.2$\,Myr. It takes about $0.2$\,Myr for the neutron star to spin down to the quasi-equilibrium spin period $P_{\rm eq,d2}$ in the subsonic settling accretion phase, and then the same track of spin evolution as this quasi-equilibrium state is adopted (see Eq.~\ref{eq:Peq2}). When the donor runs out of the MS state and changes into the HG stage, the wind velocity decreases and the accretion rate increases due to the expansion of the donor, resulting in the spin period slipping dramatically at  $t_{\rm e}\sim9$\,Myr. Once the increasing accretion rate reaches $\dot{M}_{\rm c2}$ (see Eq.~\ref{eq:dmcrit}), the BHL accretion phase happens and the neutron star can be spun up. Because of the weak magnetic field and low magnitude of $R_{\rm mag,1}$, the spin-up is not efficient and the neutron star can hardly be spun up to $P_{\rm eq}$ (see Eq~\ref{eq:Peq1}). After that, the donor becomes a HeMS star and the accretion rate decreases below $\dot{M}_{\rm c2}$. The neutron star spends about $0.7$\,Myr to spin down to $P_{\rm eq,d2}$ again in the subsonic settling accretion phase, and then stays at this quasi-equilibrium state in the following $\sim1.2$\,Myr. The BHL accretion phase occurs for a limited time in the late stage of HeHG stage but the spin-up is not efficient. Finally, the HMXB ends up as a BNS, with the first-born and the second-born neutron stars having  $(P,B)=(2.45$\,s, $2.6\times10^8$\,G) and $(0.01$\,s, $10^{13}$\,G) at $t_{\rm e}\sim 11$\,Myr. The right panel of Figure~\ref{fig:f2} shows the spin evolution of both neutron stars. After the second SN, both neutron stars spin down through the dipole radiation (see Eq.~\ref{eq:isolate}). The first-born neutron star is radio-quiet in the BNS stage. This BNS system can be observed through the radio emission from the second-born neutron star (i.e., observed as a young pulsar) but only within a limited period of $\sim3.3$\,Myr.

Figure~\ref{fig:f3} shows the spin evolution of both neutron star components in another binary system in which the first-born neutron star only experienced wind-fed accretion during its evolution in the HXMB stage (hereafter case B). The BNS formed from this binary has a total mass of $\sim3.4\ M_{\odot}$, similar as that of GW190425. In this case, the duration of the HMXB stage is $\sim4.1$\,Myr, and the initial and the final orbital periods of the HMXB are $P_{\rm orb,i}\sim80.9$\,day and $P_{\rm orb,f}\sim1.3$\,day, respectively. The progenitors of both neutron stars are sufficiently massive ($>20\ M_{\odot}$) and thus we assume that they are initially magnetars. We specifically assume that the initial magnetic fields of these two magnetars are $B_{\rm ini}=5\times10^{14}$\,G. 

The evolution tracks shown for the case B (Fig.~\ref{fig:f3}) are quite different from that of the case A (Fig.~\ref{fig:f2}). As seen from this figure, the ejector phase becomes shorter (at $t_{\rm e}\sim0-3.6$\,kyr)  and ends at higher spin period due to the stronger magnetic field. After that, the neutron star rapidly spins down to $P_{\rm eq}$ during the rapid rotator phase (at $t_{\rm e}\sim3.6-9.1$\,kyr). At the moment $P=P_{\rm eq}$, the magnetic radius is equal to the corotation radius and larger than the Bondi radius, i.e., $R_{\rm mag,2}=R_{\rm cor}>R_{\rm g}$. A quasi-equilibrium state is achieved and the spin period decreases gradually with the decaying magnetic field until $R_{\rm mag,2}=R_{\rm g}$ in the sub-Keplerian magnetic inhibition phase (at $t_{\rm e}\sim9.1$\,kyr-$3.2$\,Myr). Then, the neutron star quickly evolves to another quasi-equilibrium state with $P_{\rm eq,d2}$ during the short-lived subsonic settling accretion phase $\sim3$\,kyr. When the donor changes into the HG, HeMS or HeHG stages, the accretion rate of the neutron star is above $\dot{M}_{\rm c2}$ and the BHL accretion phase dominates the spin evolution. Notably, because the lifetime of the HMXB is much shorter, the magnetic field does not decrease significantly. Therefore, the magnetic radius $R_{\rm mag,1}$ is sufficiently large and the spin-up rate in the BHL accretion phase is remarkable compared with that in the case A, resulting in $P=P_{\rm eq}$ at the end of the HMXB stage. Finally, the HMXB leaves the first-born neutron star with $(P, B)=(14.3$\,s, $8.3\times10^{12}$\,G) at the formation time of the BNS. As seen from the right panel, the first-born neutron star is radio-quiet in the BNS stage and the BNS system can be detected through the radio emission from the second-born neutron star (i.e., observed as a magnetar) in $9.3\times10^4$\,yr. 

Figure~\ref{fig:f4} shows the spin evolution of both neutron star components in a binary system, in which the first-born neutron star experienced both wind-fed accretion and RLOF (hereafter case C). The BNS formed from this binary at $t_{\rm e}\sim11$\,Myr has a total mass of $M_{\rm tot}\sim2.6\ M_{\odot}$, similar to the typical mass of Galactic BNSs. The initial and the final orbital period of the HMXB at $t_{\rm e}=0$ and $11$\,Myr are  $P_{\rm orb,i}\sim10.2$\,day and $P_{\rm orb,f}\sim0.11$\,day, respectively. Similar to the case A, the initial magnetic fields of both neutron stars are set as $B_{\rm ini}=10^{13}$\,G for illustration, as the masses of both progenitors are small ($<20M_\odot$).

The evolution track of the first-born neutron star in case C is similar to that of the case A for the first half part. The neutron star spins down continuously in the ejector phase  ($t_{\rm e}\sim0-0.6$\,Myr), rapid rotator phase ($t_{\rm e}\sim0.6-3.3$\,Myr), and propeller phase ($t_{\rm e}\sim3.3-3.8$\,Myr). When  $t_{\rm e}\sim3.8-8.3$\,Myr, the subsonic settling accretion phase occurs and the spin period decreases gradually due to the decay of the magnetic field after it increases to the quasi-equilibrium state $P_{\rm eq,d2}$. In the HMXB stage, the orbital period is shorter compared with that in the case A, and the accretion rate can overtake $\dot{M}_{\rm c2}$ at the end of the MS stage of the donor. Therefore, the BHL accretion phase happens much sooner ($t_{\rm e}\sim8.3-9.2$\,Myr). When the donor becomes a HeMS star, it fills the RL and case BB mass transfer happens due to the short orbital period and small semimajor axis. Here, we re-declare that the case BB mass transfer is always stable \citep{2018MNRAS.481.4009V} and removes the entire helium envelope \citep{2022ApJS..258...34R}. In the long term of the RLOF ($t_{\rm e}\sim9.25-11$\,Myr), the neutron star can be spun up to a millisecond pulsar. Note that the magnetic field declines significantly at the end of the HMXB stage and it is hard for the neutron star to be spun up to $P_{\rm eq,RL}$. Finally, the HMXB ends to form a BNS, with the first-born neutron star having  $(P, B)=(0.0129$\,s, $2.8\times10^8$\,G) at the BNS formation time. The right panel shows the spin evolution for both neutron stars. As clearly seen from the figure, the first-born neutron star becomes a millisecond pulsar through the recycling process and its lifetime in the radio band can be as long as or even exceed the Hubble time. After the formation of the BNS, both components may be detected as radio pulsars (i.e., a recycled pulsar and a young pulsar) over a period of $3.3$\,Myr, and only the recycled neutron star component can be observed as a pulsar after that.

Figure~\ref{fig:f5} shows the spin evolution of both neutron star components in another binary, in which the first neutron star component experienced wind-fed accretion and RLOF (hereafter case D). The BNS formed from this binary has a total mass of $\sim3.4\ M_{\odot}$, similar to that of GW190425. The total lifetime of the HMXB is $\sim4.2$\,Myr, the initial and the final orbital periods of the HMXB are $P_{\rm orb,i}=10.5$\,day and $P_{\rm orb,f}=0.1$\,day at $t_{\rm e}=0$ and $4.2$\,Myr, respectively. Similar to case B, the neutron stars are the outcomes of massive MS stars ($>20 M_{\odot}$), and the initial magnetic field of both neutron star components in this case are set as $B_{\rm ini}=5\times10^{14}$\,G.

The accretion rates in relevant accretion phases in the case D are relatively larger than those in the other three cases (Figs.~\ref{fig:f2}-\ref{fig:f4}) due to the heavier donor and the shorter orbital period of the binary system. Comparing with the case B, the first-born neutron star quickly goes through the ejector phase ($t_{\rm e}\sim0-1.2$\,kyr), rapid rotator phase ($t_{\rm e}\sim1.2-2.2$\,kyr), sub-Keplerian magnetic inhibition phase ($t_{\rm e}\sim2.2-37.5$\,kyr), subsonic settling accretion phase ($t_{\rm e}\sim37.5-38.9$\,kyr), and BHL accretion phase ($t_{\rm e}\sim38.9-39.0$\,kyr). In addition, the donor fills its RL during the MS stage and the RLOF dominates the spin evolution for the rest of time ($t_{\rm e}\sim39.0\ {\rm kyr}-4.2$\,Myr). Finally, the HMXB ends to form a BNS, with the first-born neutron star having $(P, B)=(3.45$\,s, $7.4\times10^{12}$\,G) at the BNS formation time. As seen from the right panel of the figure, the first-born neutron star may be detected as a pulsar for $\sim 5.5$\,Myr after the formation of the BNS, while the second-born neutron star only has the lifetime $6.9\times10^4$\,yr in the radio band.

\subsection{Survived BNSs in the Milky Way}
\label{subsec:MWBNS}

\begin{figure*}
\centering
\includegraphics[width=0.95\textwidth]{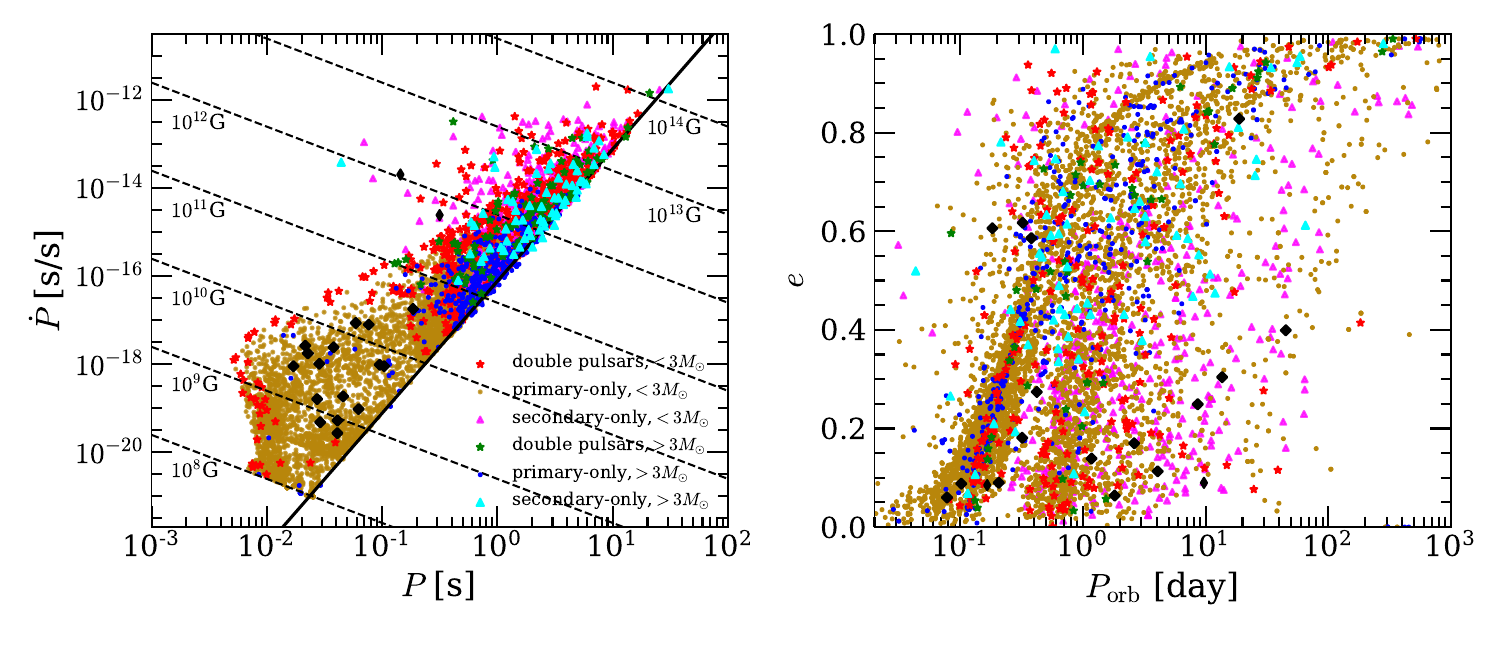}
\caption{
The $P-\dot{P}$ diagram (left panel) and the orbital period-eccentricity ($P_{\rm orb}-e$) diagram (right panel) for survived BNSs with pulsar components in the Milky Way at the present time obtained from our population synthesis model. The red and green stars represent the pulsar-pulsar binaries with total mass $M_{\rm tot}<3 M_\odot$ and $\ge 3 M_\odot$, respectively, and both pulsar components of these binaries are shown in the $P-\dot{P}$ diagram. The dark yellow and blue dots show the ``primary-only'' binaries (only the first-born neutron star appearing as a radio pulsar but the secondary one not) with $M_{\rm tot}<3 M_\odot$ and $\ge3 M_\odot$, respectively. The magenta and cyan triangles denote the ``secondary-only'' binaries (only the second-born neutron star appearing as a radio pulsar but the primary one not) with $M_{\rm tot}<3 M_\odot$ and $\ge3 M_\odot$, respectively. The black diamonds show the properties of the observed Galactic BNSs (excluding those in Globular clusters; thick diamonds for Galactic BNSs with recycled pulsars and thin diamonds for Galactic BNSs with young pulsars). The black solid line shows the ``death line" described by Eq.~\eqref{eq:pdeath} in Section~\ref{subsec:spin_e} by assuming typical neutron star mass $1.3\ M_{\odot}$ and radius $13.19$\,km. The black dashed lines indicate the magnetic field $B=10^{8},\cdots,10^{14}$\,G from bottom to top, respectively.
}
\label{fig:f6}
\end{figure*}

\begin{figure}
\includegraphics[width=0.45\textwidth]{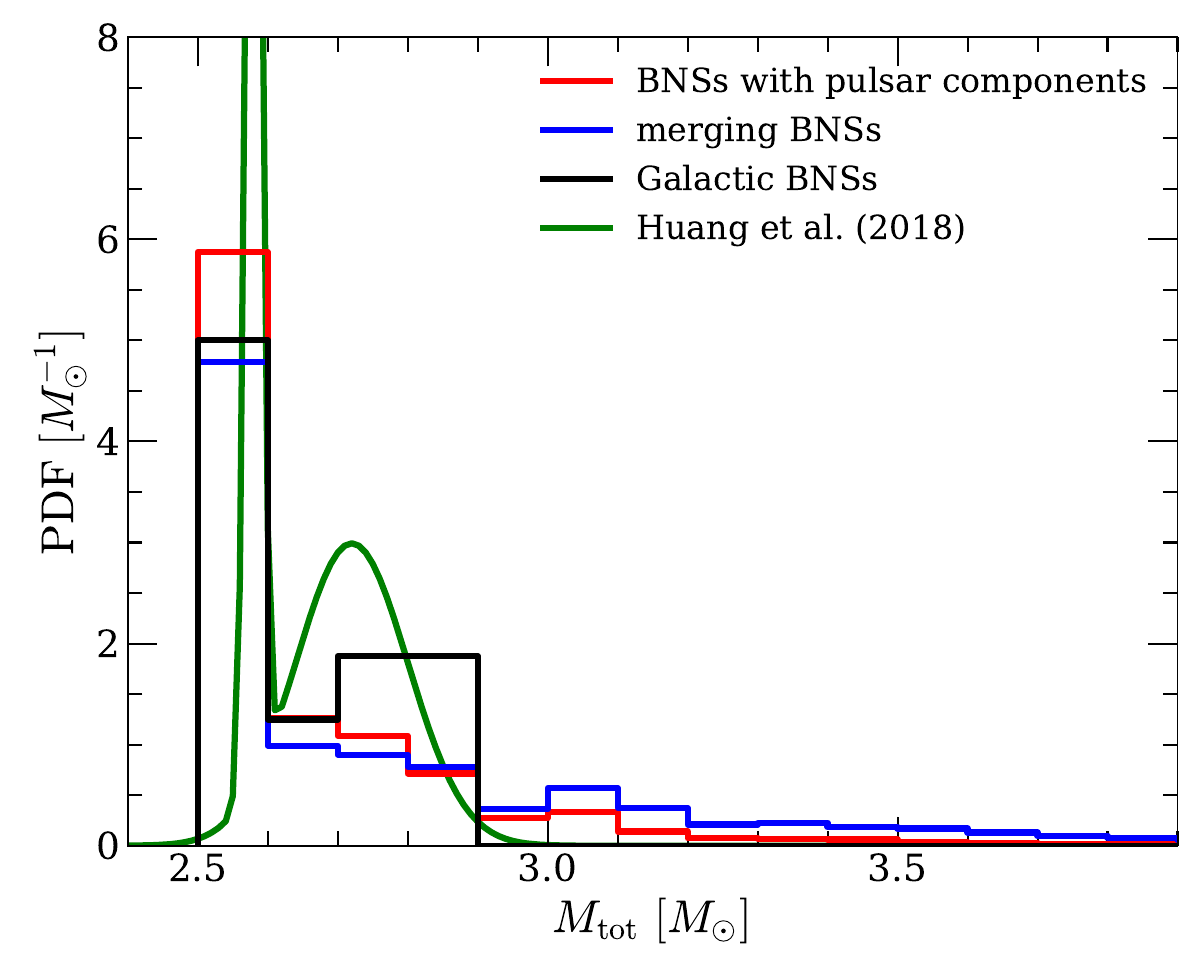}
\caption{
The probability density functions (PDFs) of the BNS total mass for the survived BNSs with pulsar components at $z=0$ in the Milky Way obtained from the population synthesis model, and those from observations. The red histogram shows the results of the survived BNSs with pulsar components obtained from our simulation. The black histogram represents the observed one. The green line shows the bimodal Gaussian distribution of the Galactic BNSs obtained by \citet{2018arXiv180403101H}. For comparison, the blue histogram shows the total mass distribution of merging BNSs at $z=0$ in the Milky Way or Milky Way like galaxies.
}
\label{fig:f7}
\end{figure}

Given the templates for the orbital evolution and spin evolution tracks of a large amount of BNSs (e.g., those shown in Figs.~\ref{fig:f2}-\ref{fig:f5}), the orbital and spin information for each BNS system can be obtained at any given time. Considering the Milky Way, some BNSs formed in the past may still survive at the present time. Among these survived BNSs, some may be detectable as pulsars, but others may not be detectable as they are radio quiet. Considering BNSs in distant universe, some of them may be detectable by GW detectors when they merge.

We simulate the survived BNSs in the Milky Way by the convolution of the evolution tracks obtained from our BSE models and the star formation history of the Milky Way. We assume that the star formation history (SFH) and metallicity evolution of the Galactic bulge and disc follow the distributions in \citet{2003MNRAS.345.1381F} and \citet{2014ApJ...781L..31S} respectively, and randomly assign the ZAMS binaries over the star-forming period. We perform Monte Carlo simulations and obtain $100$ realizations of the survived BNSs and those BNSs with radio pulsars ($P>P_{\rm death}$) in the Milky Way. Note here that we do not consider the beaming factor and observational selection effects for the survived BNSs with pulsar components, as the beaming factor ($\gtrsim 20\%$ \citealt{2008LRR....11....8L}) is quite uncertain and dependent on pulsar properties (e.g., spin period) and the selection effects are different for different telescopes. We find that the total number of survived BNSs resulting from the model is  $1.335^{+0.001}_{-0.001}\times10^6$, with the errors quote the scatter ($5\%$ to $95\%$) among different realizations. The total number of survived BNSs with pulsar components is $7383^{+129}_{-123}$. Among them,  $224^{+23}_{-22}$, $47^{+9}_{-10}$ are pulsar-pulsar binaries with $M_{\rm tot}<3 M_{\odot}$ and $\ge3 M_{\odot}$, respectively. Pulsar-pulsar BNSs are only a fraction $\sim3.7\% $ of the survived BNSs with pulsar components. The ``primary-only'' binaries occupy the majority  ($\sim 89\%$) of the survived BNSs, and  $6129^{+111}_{-108}$ and  $446^{+32}_{-23}$ of them have $M_{\rm tot}<3 M_\odot$ and  $\ge3 M_\odot$ (a fraction  $\sim83\%$  and $\sim6.0\%$ of all the survived BNSs with pulsar components), respectively. The rest  ($\sim7.3\%$) are the ``secondary-only'' binaries, where $473^{+32}_{-33}$ and $64^{+14}_{-12}$ of them have total mass $M_{\rm tot}<3 M_\odot$ and $\geq 3 M_\odot$, respectively. Those BNSs with mass $\geq 3M_\odot$ is only a fraction of  $\sim 7.5\%$ of the survived BNSs with pulsar components. Here ``primary only'' binaries and ``secondary only'' binaries mean that only the first-born neutron star component can be detected in radio and the secondary one is a non-radio neutron star, and only the second-born neutron star component can be detected and the primary one is a non-radio neutron star.

Figure~\ref{fig:f6} shows the $P-\dot{P}$ diagram and the $P_{\rm orb}-e$ diagram of the survived BNSs that have pulsar components at the present time obtained from our model. As seen from the left panel of this figure, the distribution of the recycled pulsars in Galactic BNSs (thick diamonds) on the $P-\dot{P}$ plane can be well matched by the primary-only binaries with $M_{\rm tot}<3 M_{\odot}$. In addition, the two Galactic BNSs with young pulsar properties (thin diamonds), i.e., $(P, \dot{P}) =(0.1441$\,s, $2.03\times 10^{-14}$\,s\,s$^{-1})$ and $(0.3152$\,s, $2.43\times 10^{-15}$\,s\,s$^{-1})$, are consistent with those secondary-only binaries with $M_{\rm tot}<3 M_\odot$ produced from the model. For the survived primary-only binaries with $M_{\rm tot}\ge3 M_\odot$ or $<3 M_\odot$, the spin periods and magnetic fields are in the ranges of $0.10-2.33$\,s and $5.6\times10^{9}-1.5\times10^{12}$\,G (or $0.01-1.01$\,s and $1.2\times10^8-3.1\times10^{11}$\,G) at the $90\%$ confidence interval, respectively. The reasons for these distributions are two folds. First, the initial magnetic field of the primary-only binaries with $M_{\rm tot}<3 M_\odot$ is weaker than that in the high mass ones. Second, the low mass primary-only binaries may go through the HMXB stage over an extended period and the decline of the magnetic field is more significant compared with the high mass ones. Hence, the recycled pulsars in the low mass binaries can be spun up to shorter spin periods compared with those in high mass binaries. In addition, there are considerable secondary-only binaries survived at the present time in our simulation. These young pulsars have the spin period and the magnetic field in the ranges of $0.59-7.03$\,s, $2.8\times10^{11}-1.9\times10^{13}$\,G, respectively. It is rare to detect the secondary as pulsars in high mass BNSs as they are mainly magnetars and short-lived in the radio band in our simulation. The right panel of Figure~\ref{fig:f6} shows that the $P_{\rm orb}-e$ distribution of the observed Galactic BNSs can also be well matched by the simulated BNSs from our model. 

Figure~\ref{fig:f7} shows the total mass distributions of both the survived BNSs with pulsar components obtained from our population synthesis model and the observations of the Galactic BNSs. Among the observed Galactic BNSs, the BNS total mass has been precisely measured for fifteen BNSs with recycled pulsar components and one young pulsar-NS binary, and they all have $M_{\rm tot}<3 M_\odot$. According to our simulations, the survived BNSs with pulsar components and $M_{\rm tot}<3 M_\odot$ account for $92.5\%$ of the total number of the survived BNSs with pulsar components, and the other $7.5\%$ are heavier ones with total mass $\ge 3 M_\odot$. This result shows that none of the radio observed Galactic BNSs having a mass similar to GW190425 ($\sim 3.4M_\odot$) is reasonable. In addition, we also obtain the total mass distribution of those BNSs merging at redshift $z\sim 0$ in the Milky Way and Milky way like galaxies, and find that $78\%$ of them have total mass less than $3 M_\odot$ and the other $22\%$  are heavier than $3 M_\odot$, which seems to be consistent with the GW detections of GW190425 and GW170817.

\section{Discussions} 
\label{sec:discussion}

\begin{figure*}
\centering
\includegraphics[width=0.95\textwidth]{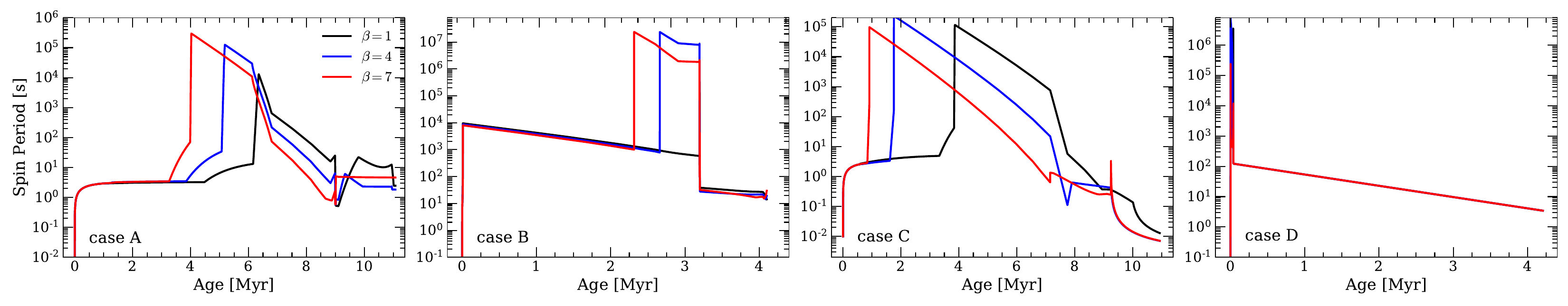}
\caption{
The spin evolution tracks of the first-born neutron star in the HMXB stage for four example BNSs with different wind velocity prescriptions. The black, blue, and red lines represent the results obtained by choosing the wind velocity prescription described by Eq.~\eqref{eq:vw} with $\beta=1,\,4$, and $7$, respectively. Panels from left to right shows the results for the cases A to D, respectively. For details, see Section~\ref{sec:discussion}.
}
\label{fig:f8}
\end{figure*}

There are a number of uncertainties in the model settings that may affect the resulting spin evolution tracks of the pulsar components and the final results on survived BNSs. These include the stellar wind velocity, the accretion models, the neutron star EOS, and also the selection effects for searching pulsars. In this section, we discuss these in further details.

The stellar wind velocity plays an important role in the spin evolution of neutron stars and thus different choices of the wind velocity prescription may have a significant effect on the spin evolution of neutron stars. For example, if the wind velocity is chosen to follow Equation~\ref{eq:vw} in Section~\ref{sec:method} but with a different  slope, e.g., $\beta=4$ or $7$ \citep[see][]{2020PASJ...72...95K} rather than $\beta=1$ for the analysis presented in Section~\ref{sec:results}, the Bondi radius $R_{\rm g}$ ($R_{\rm g}\propto v_{\rm w}^{-2}$) increases, and thus the accretion rate $\dot{M}$ ($\dot{M}\propto v_{\rm w}^{-4}$) also increases. For the cases A and C, the weak stellar wind criterion may be violated more easily and the propeller phase can happen earlier. In addition, the spin-down rates in the rapid rotator phase (Eq.~\ref{eq:dp_phaseb}) and the propeller phase (Eq.~\ref{eq:dp_phasec}) become higher. It takes a shorter time for the spin period to grow up to $P_{\rm eq}$ and the subsonic settling accretion phase happens earlier. For the case B, the magnetic field radius $R_{\rm mag,1}$ ($R_{\rm mag,1}\propto v_{\rm w}^{8/7}$) decreases with decreasing wind velocity and $R_{\rm mag,1}< R_{\rm g}$ happens earlier. Therefore, the sub-Keplerian magnetic inhibition phase ends faster as well as the subsonic settling accretion phase happens earlier. During the subsonic settling accretion phase in the cases A, B, and C, we find that the quasi-equilibrium spin period $P_{\rm eq,d2}$ ($P_{\rm eq,d2}\propto v_{\rm w}^{60/11}$) decreases in condition of low wind velocity. In addition, the criterion $\dot{M}>\dot{M}_{\rm c2}$ for the BHL accretion phase may be satisfied easier, and the BHL accretion phase occurs prematurely. For the case D, the changes in the wind velocity have a weak effect on the spin evolution track. In general, it is hard to determine whether the final spin period is higher or lower at the end of the HMXB stage when changing the wind velocity and a detailed calculation is needed. Such differences mentioned above are illustrated in Figure~\ref{fig:f8}.

\begin{figure*}
\centering
\includegraphics[width=0.95\textwidth]{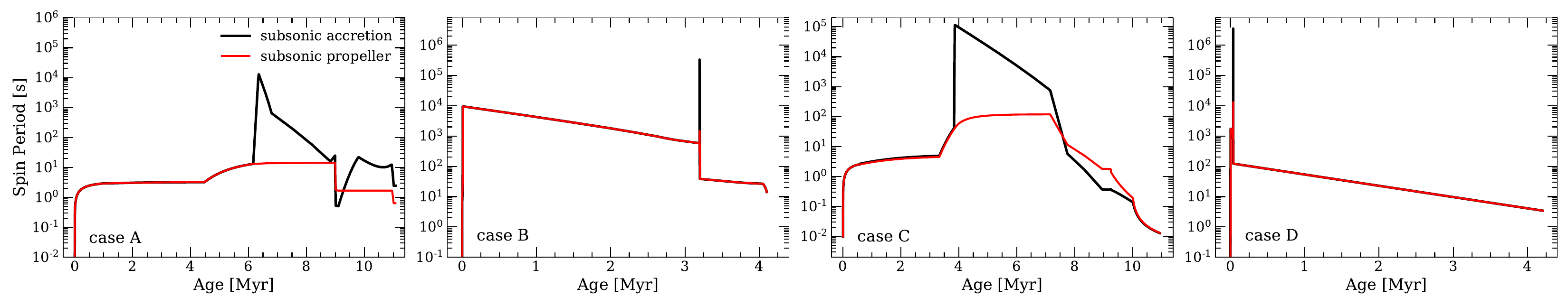}
\caption{
The spin evolution tracks of the first-born neutron star in the HMXB stage for four example BNSs obtained by choosing two different accretion mechanisms at $\dot{M}<\dot{M}_{\rm c2}$. The black and red lines represent the results obtained by adopting the subsonic settling accretion mechanism and the subsonic propeller mechanism, respectively, when $\dot{M}<\dot{M}_{\rm c2}$. Panels from left to right show the results obtained for the cases A to D, respectively. For details, see Section~\ref{sec:discussion}.
}
\label{fig:f9}
\end{figure*}

Note that the subsonic propeller model is also possible when $\dot{M}<\dot{M}_{\rm c2}$, though we adopt the subsonic settling accretion mechanism in this circumstances in Section~\ref{subsec:spin_e} in Section~\ref{sec:results}. Consider such subsonic propeller model would also lead some changes to the resulting spin evolution tracks. Figure~\ref{fig:f9} shows the differences in the spin evolution tracks between these two different choices. For most cases with $\dot{M}<\dot{M}_{\rm c2}$, we find that the subsonic propeller model results in lower spin period than the subsonic accretion model, unless the initial magnetic field is weak and declines significantly (at the end of MS stage of donor in the cases A and C). However, if the following spin-up is efficient (strong magnetic field in the case B; spin-up by the RLOF in the the cases C and D), the final spin periods at the end of HMXB stage are consistent with each other.

\begin{figure*}
\centering
\includegraphics[width=0.95\textwidth]{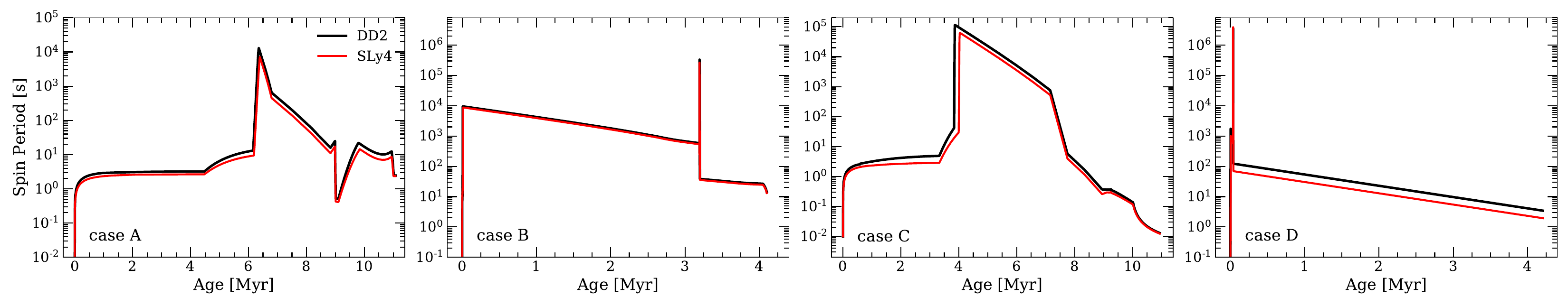}
\caption{
The spin evolution tracks of the first-born neutron star in the HMXB stage for four example BNSs by using different EOSs. The black and red lines represent the results obtained by adopting the DD2 EOS and the SLy4 EOS, respectively. Panels from left to right show the results obtained from the cases A to D, respectively. For details, see Section~\ref{sec:discussion}.
}
\label{fig:f10}
\end{figure*}

The spin evolution of neutron stars may depend on the choice of the EOS in the model, as the radius and thus the moment of inertia of a neutron star with given mass depends on the EOS. We adopt a stiff EOS (DD2) for the calculations described in Section~\ref{sec:results}. Note that a softer EOS may be preferred as suggested by the observations of GW170817 \citep{2018PhRvL.121p1101A}. To check the effects due to the choice of EOS, here we also adopt the SLy4 EOS to perform simulations as that done above and obtain the spin evolution tracks for the first-born neutron stars in the HMXB stages. Figure~\ref{fig:f10} shows the comparison of the spin evolution tracks for several example cases (corresponding to the cases A, B, C, and D shown in Figs.~\ref{fig:f2}-\ref{fig:f5}) obtained by either adopting the DD2 EOS (black lines) or the SLy4 EOS (red lines). Compared with those adopting the DD2 EOS, the neutron star radius is significantly smaller and the magnetic dipole moment $\mu$ ($\mu\propto R_{\rm NS}^3$) is also significantly smaller when assuming the SLy4 EOS. For the cases A and C, when adopting the SLy4 EOS, the spin-down rates in the rapid rotator phase (Eq.~\ref{eq:dp_phaseb}) and the propeller phase (Eq.~\ref{eq:dp_phasec}) become lower and the quasi-equilibrium spin period in the subsonic accretion phase $P_{\rm eq,d2}$ ($P_{\rm eq,d2}\propto \mu^{12/11}$) also becomes smaller compared with those obtained by adopting the DD2 EOS. Therefore, the spin evolution tracks are below those inferred from the DD2 EOS. For the case D, the quasi-equilibrium spin period during the RLOF $P_{\rm eq,RL}$ ($P_{\rm eq,RL}\propto R_{\rm NS}^{18/7}$) becomes smaller and the recycled neutron star can be spun up to a shorter spin period if using the SLy4 EOS compared with those using the DD2 EOS.

Similarly, we also obtain the survived BNSs and those with pulsar components in the Milky Way via the population synthesis model in Section~\ref{subsec:MWBNS} by assuming the SLy4 EOS. We find that the total number of survived BNSs is $1.318^{+0.001}_{-0.001} \times10^6$. The total number of the survived BNSs with pulsar component is  $9979^{+168}_{-137}$,  somewhat higher than that obtained by adopting the DD2 EOS. Among them, $343^{+32}_{-29}$ and $91^{+12}_{-11}$  BNSs are pulsar-pulsar binaries with $M_{\rm tot}<3 M_\odot$ and $M_{\rm tot}\ge3 M_\odot$, respectively. The primary-only binaries occupy the majority of the Galactic BNSs, $8045^{+117}_{-123}$ and $638^{+36}_{-34}$ of them have $M_{\rm tot}<3 M_\odot$ and $\ge3 M_\odot$, respectively. The rest are the secondary-only binaries, where $713^{+44}_{-35}$ and $89^{+13}_{-16}$ of them have $M_{\rm tot}<3 M_\odot$ and  $\ge 3 M_\odot$, respectively. According to our simulations, the survived BNSs with pulsar components and $M_{\rm tot}<3 M_\odot$ account for $91\%$ and the other $9\%$ BNSs are heavier with $M_{\rm tot} \ge 3 M_\odot$. In addition, we obtain that $81\%$ of merging BNSs at $z\sim 0$ have total mass $M_{\rm tot}<3M_{\odot}$ and the other
$19\%$ have mass $\ge 3 M_\odot$ for the case adopting the SLy4 EOS. 

Note that we do not consider the selection effects of pulsars for specific radio telescopes and adopt equal weight for the observabilities of all survived pulsar-NS binaries. In reality, the position and movement of a pulsar, as well as its radio luminosity, the sensitivity of radio telescopes, the beaming effect, and the Doppler shifting effect can significantly influence the observabilities of the pulsars, and consequently the total number of BNSs that can be detected and properties of these BNSs may depend on the selection effects in a profound way, which certainly deserves further detailed consideration for given survey radio telescopes, such as FAST, MeerKAT, and SKA, etc.

\begin{figure}
\centering
\includegraphics[width=0.45\textwidth]{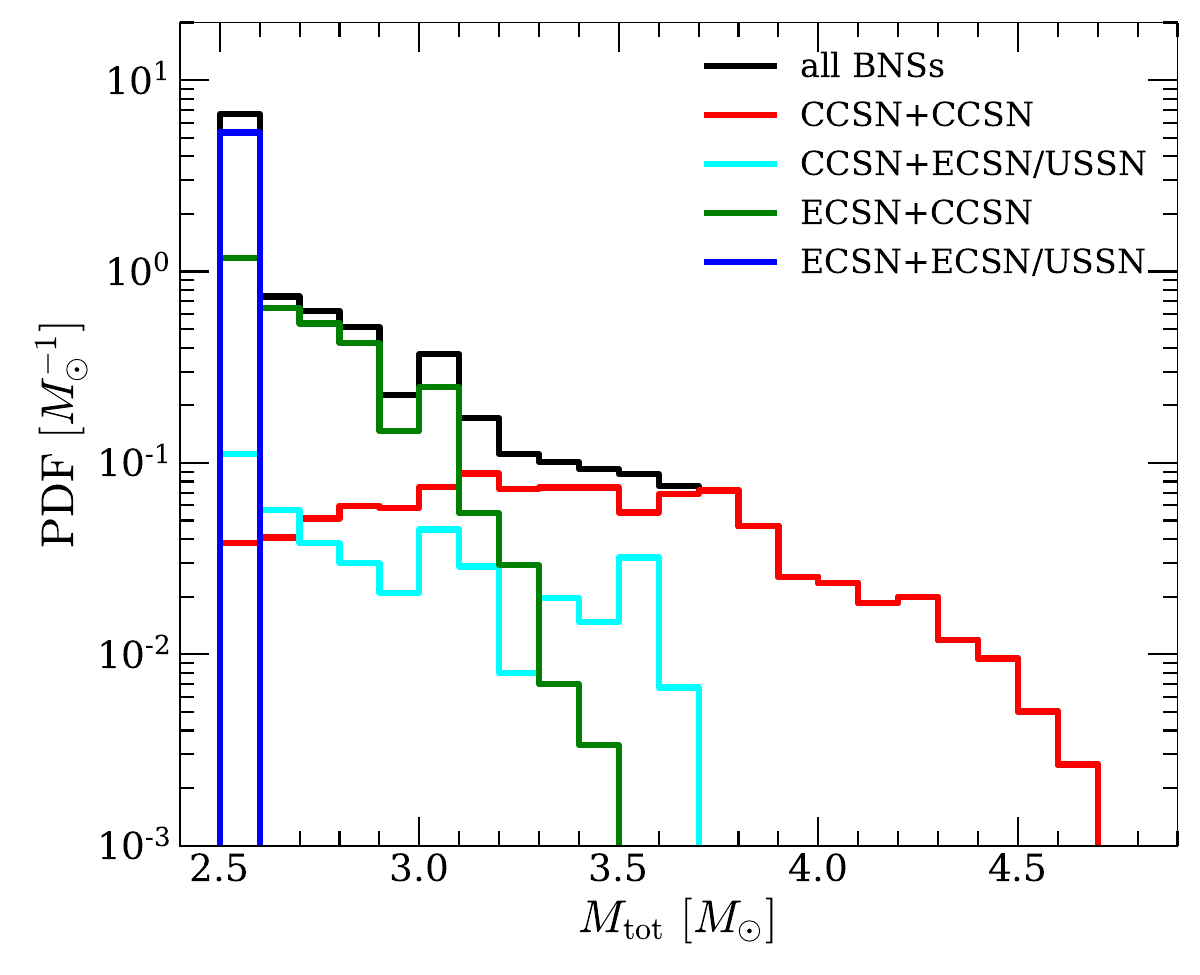}
\caption{
The probability density functions of the BNS total mass of different formation channels. The black line shows the result of all BNSs formed in the population synthesis model. The red, cyan, green, and blue lines show the results obtained from the CCSN + CCSN, CCSN + ECSN/USSN, ECSN + CCSN, and ECSN + ECSN/USSN formation channels, respectively.
}
\label{fig:f11}
\end{figure}

The observed BNSs by radio emission still suffer from small number statistics, and it is insufficient to draw firm conclusions on the bimodal Gaussian distribution of Galactic BNSs (see Fig.~\ref{fig:f7}). \citet{2019ApJ...876...18F} and \citet{2018arXiv180403101H} pointed out that approximately $60$-$100$ additional BNSs are needed to determine the detailed shape of the distribution. According to the remnant mass functions (RMFs) and supernova assumptions in our BSE model, BNSs are mainly formed via the the channels of ECSN + ECSN/USSN, ECSN + CCSN, CCSN + ECSN/USSN, and CCSN + CCSN. Figure~\ref{fig:f11} shows the total mass distributions of those BNSs formed form these different channels, respectively, as well as the distribution of all BNSs. As seen from this figure, the ECSN + ECSN/USSN channel contributes about $\sim 53\%$ of the total BNSs and the total masses $M_{\rm tot}$ of the BNSs formed from this channel distribute in a narrow range around $\sim2.54 M_\odot$; the ECSN + CCSN channel contributes  $\sim 33\%$ of all BNSs and the total masses $M_{\rm tot}$ of these BNSs distribute in $2.54-3.51 M_\odot$; the CCSN + ECSN/USSN channel contributes $\sim4\%$ of all BNSs and the resulting $M_{\rm tot}$  distribute in the range of $2.54 M_\odot-3.67 M_\odot$; the CCSN + CCSN channel contributes $\sim9.9\%$  of all BNSs, and the $M_{\rm tot}$ from this channel is in the range of $2.54-4.71 M_\odot$. It is obvious that the ECSN + ECSN/USSN channel dominates the formation of BNSs, especially at $M_{\rm tot}\sim2.5-2.6 M_\odot$. The ECSN + CCSN channel also contributes to the formation of BNSs with $M_{\rm tot}\sim2.5-2.6 M_\odot$ significantly and dominates the formation of BNSs with $M_{\rm tot}\sim2.6-3.1 M_\odot$. Therefore, we suggest that Galactic BNSs with total mass around the first peak of the bimodal distribution are formed from the ECSN + ECSN/USSN channel or the ECSN + CCSN channel, and the others with total mass around the second peak are mainly formed from the ECSN + CCSN channel, respectively \citep[][see green line in Fig.~\ref{fig:f7}]{2018arXiv180403101H}. As a results, GW170817 and GW190425 are possibly formed from the ECSN + CCSN channel and the CCSN + CCSN channel, respectively. As also seen from the right panel of Figure~\ref{fig:f6}, we find that the high mass BNSs with pulsar components are concentrated at low orbital period regions, indicating that they will merge quickly and it plays an additional role in the lack of high mass BNSs in the Galactic BNS samples \citep{2020MNRAS.496L..64R}.

The total mass distribution of merging BNSs at redshift $z=0$ is estimated and $\sim78\%-81\%$ of these merging BNSs have total mass $<3 M_\odot$ and the other $19\%-22\%$ have mass $\ge3 M_\odot$ under the assumption of SFH and metallicity evolution in \citet{2003MNRAS.345.1381F} and \citet{2014ApJ...781L..31S}, which seems compatible with the GW observations. However, one should note that GW170817 and GW190425 are located in galaxies different from the Milky Way, which have different star formation and metallicity enrichment history. Here, we point out that metallicity has a significant influence on the formation of BNSs and consequently affects the mass distribution. To explain the difference between the total-mass distribution of the GW detected BNS mergers and that of the Galactic BNSs, a more comprehensive study is needed with consideration of the effects of total mass on the signal to noise ratio (SNR) of the GW detection, the metallicity enrichment and star formation history in the local universe. as well as the detailed selection effects of the observed Galactic BNSs.

Future observations of a large amount of BNS mergers by the ground-based GW detectors may be able to reveal the total mass distribution of BNS mergers and thus put strong constraints on the formation and origin of BNSs by comparing it with the model predicted ones.

Note that \citet{2020MNRAS.494.1587C} also modelled the $P-\dot{P}$ distribution and the total mass distributions of the Galactic BNSs with pulsar component by using the population synthesis code {\bf COMPAS}. They obtained a $P-\dot{P}$ distribution consistent with the observational one with consideration of the radio selection effects. However, there are some differences of the work presented in this paper from that in \citet{2020MNRAS.494.1587C}. First, we consider the subsonic settling accretion phase or the subsonic propeller phase when $\dot{M}<\dot{M}_{\rm c2}$ and $R_{\rm mag,1}<R_{\rm cor}$, while \citet{2020MNRAS.494.1587C} ignored them and assumed that the neutron star always spins up when $R_{\rm mag,1}<R_{\rm cor}$. Second, we focus on the effects of BNS total mass on the spin evolution, and find that high mass BNSs possibly have much stronger magnetic field and longer spin period while low mass BNSs are more likely to be recycled and evolve to long-lived pulsar-neutron star binaries. Besides, we adopt different initial magnetic field models, magnetic field decay models, CE accretion models from those in \citet{2020MNRAS.494.1587C}. Nevertheless, the congruence of the $P-\dot{P}$ distributions obtained by both the work presented here and that by \citet{2020MNRAS.494.1587C} and their consistence with the observational one suggests that the spin evolution does plays a crucial role in reproducing the Galactic BNSs. Moreover, \citet{2020MNRAS.494.1587C} statistically studied the mass distributions of Galactic BNSs and found that about $70-80\%$ merging BNSs have total mass $\leq3.0\ M_{\odot}$, which is consistent with our results ($\sim78\%-81\%$ merging BNSs with $M_{\rm tot}\leq3.0\ M_{\odot}$).

\section{Conclusions} 
\label{sec:conclusion}

In this paper, we investigate the spin evolution of binary neutron stars by implementing various processes that may affect the spin evolution into the BSE code, with a focus on the spin evolution of the first-born neutron star components in the binary systems during the HMXB stages. We update the recipes for stellar wind velocity, remnant mass function for different types of supernovae and their corresponding natal kicks, and further consider the effects of the wind-fed accretion, RLOF accretion, and different initial magnetic field on the neutron star evolution. We find that the spin evolution of the first-born neutron star in binary systems that result in BNSs are significantly affected by which channel that the BNSs are produced. For the wind-fed formation channel, the primary neutron star goes through ejector phase - rapid rotator phase - propeller phase (for the low mass BNSs) or ejector phase - rapid rotator phase - sub-Keplerian magnetic inhibition phase (for high mass BNSs) before accretion phase. The final spin period is determined by the subsonic accretion phase (for the low mass BNSs; due to the weak magnetic field and limited spin-up rate) or the BHL accretion phase (for the high mass BNSs; due to the strong magnetic field and remarkable spin-up rate). For the wind-fed + RLOF formation channel, the primary neutron star goes through ejector phase - rapid rotator phase - propeller phase (for the low mass BNSs) or ejector phase - rapid rotator phase - sub-Keplerian magnetic inhibition phase (for the high mass BNSs) before the accretion phase. For low mass BNSs, the primary neutron star can be spun up to a millisecond pulsar and have the radio lifetime comparable to or even larger than the Hubble time. For the high mass BNSs, the primary neutron star components can be spun up to a short spin period but remains strong magnetic field, resulting in a limited radio lifetime and thus smaller probability to be detected as pulsars.

Furthermore, we study the properties of the survived BNSs that may be detectable in the Milky Way as radio pulsars. The distribution of the Galactic BNSs on the $P-\dot{P}$ diagram and the $P-e$ diagram can be well matched by the survived BNSs with pulsar components obtained from our population synthesis model under the assumed star formation  and metallicity enrichment history in the Milky Way \citep{2003MNRAS.345.1381F,2014ApJ...781L..31S,2015ApJ...806...96L}. We estimate that there are  $\sim 7383^{+129}_{-123}$ survived BNSs with pulsar components in the Milky Way if adopting the DD2 EOS. The majority of these BNSs are the recycled pulsar-neutron star binaries with  $\sim 6129^{+111}_{-108}$ and $446^{+32}_{-23}$ having $M_{\rm tot}<3 M_\odot$ and $\ge3 M_\odot$, respectively. The rest are mainly young pulsar-NS binaries ($473^{+32}_{-33}$) with $M_{\rm tot}<3 M_\odot$. We find that the total mass distributions of the survived BNSs with pulsar components can be well consistent with that of the observed Galactic BNSs. Heavy BNSs like GW190425 can hardly be detected in the Milky Way by radio telescopes simply due to that most survived BNSs with $M_{\rm tot}\ge 3M_\odot$ do not have pulsar components. Our model also predicts that a substantial fraction ($\sim19\%-22\%$) of the BNSs merging at redshift $z\sim0$ can have mass $\gtrsim 3.0M_\odot$ in the Milky Way and Milky Way like galaxies, which seems compatible with current GW observations of GW170817 and GW190425.

In future, observations by the Five Hundred Aperture Sphere radio Telescope (FAST) and Square Kilometer Array may also find many more BNSs in the MW \citep{2011IJMPD..20..989N, 2009IEEEP..97.1482D}, and the third generation ground-based GW detectors, the Einstein Telescope \citep{2010CQGra..27s4002P, 2020JCAP...03..050M} and the Cosmic Explorer \citep{2019BAAS...51g..35R}, will allow us to observe a huge number of BNS mergers with redshift up to $z\sim2$ \citep{2019BAAS...51c.242K}. These observations will provide better statistics on the distributions of the BNS properties, and thus enable a detailed comparison with the evolution model of BNSs (including the spin evolution) and deepen our understanding of the BNS formation and evolution.

\section*{acknowledgments}
This work is partly supported by the Strategic Priority Research Program of the Chinese Academy of Sciences (grant no. XDB0550300), the National Natural Science Foundation of China (grant nos. 12273050, 11991052), and the National Key Research and Development Program of China (grant nos. 2020YFC2201400, 2022YFC2205201). We thank the referee for her/his helpful insights and suggestions, which have improved the paper. We thank Zhiwei Chen for his helpful suggestions.

\bibliography{reference}{}

\begin{thebibliography}{}
\expandafter\ifx\csname natexlab\endcsname\relax\def\natexlab#1{#1}\fi
\providecommand{\url}[1]{\href{#1}{#1}}
\providecommand{\dodoi}[1]{doi:~\href{http://doi.org/#1}{\nolinkurl{#1}}}
\providecommand{\doeprint}[1]{\href{http://ascl.net/#1}{\nolinkurl{http://ascl.net/#1}}}
\providecommand{\doarXiv}[1]{\href{https://arxiv.org/abs/#1}{\nolinkurl{https://arxiv.org/abs/#1}}}

\bibitem[{{Abbott} {et~al.}(2017{\natexlab{a}}){Abbott}, {Abbott}, {Abbott}, {Acernese}, {Ackley}, {Adams}, {Adams}, {Addesso}, {Adhikari}, {Adya}, {Affeldt}, {Afrough}, {Agarwal}, {Agathos}, {Agatsuma}, {Aggarwal}, {Aguiar}, {Aiello}, {Ain}, {Ajith}, {Allen}, {Allen}, {Allocca}, {Altin}, {Amato}, {Ananyeva}, {Anderson}, {Anderson}, {Angelova}, {Antier}, {Appert}, {Arai}, {Araya}, {Areeda}, {Arnaud}, {Arun}, {Ascenzi}, {Ashton}, {Ast}, {Aston}, {Astone}, {Atallah}, {Aufmuth}, {Aulbert}, {AultONeal}, {Austin}, {Avila-Alvarez}, {Babak}, {Bacon}, {Bader}, {Bae}, {Bailes}, {Baker}, {Baldaccini}, {Ballardin}, {Ballmer}, {Banagiri}, {Barayoga}, {Barclay}, {Barish}, {Barker}, {Barkett}, {Barone}, {Barr}, {Barsotti}, {Barsuglia}, {Barta}, {Barthelmy}, {Bartlett}, {Bartos}, {Bassiri}, {Basti}, {Batch}, {Bawaj}, {Bayley}, {Bazzan}, {B{\'e}csy}, {Beer}, {Bejger}, {Belahcene}, {Bell}, {Berger}, {Bergmann}, {Bernuzzi}, {Bero}, {Berry}, {Bersanetti}, {Bertolini}, {Betzwieser}, {Bhagwat}, {Bhandare}, {Bilenko},
  {Billingsley}, {Billman}, {Birch}, {Birney}, {Birnholtz}, {Biscans}, {Biscoveanu}, {Bisht}, {Bitossi}, {Biwer}, {Bizouard}, {Blackburn}, {Blackman}, {Blair}, {Blair}, {Blair}, {Bloemen}, {Bock}, {Bode}, {Boer}, {Bogaert}, {Bohe}, {Bondu}, {Bonilla}, {Bonnand}, {Boom}, {Bork}, {Boschi}, {Bose}, {Bossie}, {Bouffanais}, {Bozzi}, {Bradaschia}, {Brady}, {Branchesi}, {Brau}, {Briant}, {Brillet}, {Brinkmann}, {Brisson}, {Brockill}, {Broida}, {Brooks}, {Brown}, {Brown}, {Brunett}, {Buchanan}, {Buikema}, {Bulik}, {Bulten}, {Buonanno}, {Buskulic}, {Buy}, {Byer}, {Cabero}, {Cadonati}, {Cagnoli}, {Cahillane}, {Calder{\'o}n Bustillo}, {Callister}, {Calloni}, {Camp}, {Canepa}, {Canizares}, {Cannon}, {Cao}, {Cao}, {Capano}, {Capocasa}, {Carbognani}, {Caride}, {Carney}, {Carullo}, {Casanueva Diaz}, {Casentini}, {Caudill}, {Cavagli{\`a}}, {Cavalier}, {Cavalieri}, {Cella}, {Cepeda}, {Cerd{\'a}-Dur{\'a}n}, {Cerretani}, {Cesarini}, {Chamberlin}, {Chan}, {Chao}, {Charlton}, {Chase}, {Chassande-Mottin}, {Chatterjee},
  {Chatziioannou}, {Cheeseboro}, {Chen}, {Chen}, {Chen}, {Cheng}, {Chia}, {Chincarini}, {Chiummo}, {Chmiel}, {Cho}, {Cho}, {Chow}, {Christensen}, {Chu}, {Chua}, {Chua}, {Chung}, {Chung}, {Ciani}, {Ciolfi}, {Cirelli}, {Cirone}, {Clara}, {Clark}, {Clearwater}, {Cleva}, {Cocchieri}, {Coccia}, {Cohadon}, {Cohen}, {Colla}, {Collette}, {Cominsky}, {Constancio}, {Conti}, {Cooper}, {Corban}, {Corbitt}, {Cordero-Carri{\'o}n}, {Corley}, {Cornish}, {Corsi}, {Cortese}, {Costa}, {Coughlin}, {Coughlin}, {Coulon}, {Countryman}, {Couvares}, {Covas}, {Cowan}, {Coward}, {Cowart}, {Coyne}, {Coyne}, {Creighton}, {Creighton}, {Cripe}, {Crowder}, {Cullen}, {Cumming}, {Cunningham}, {Cuoco}, {Dal Canton}, {D{\'a}lya}, {Danilishin}, {D'Antonio}, {Danzmann}, {Dasgupta}, {Da Silva Costa}, {Dattilo}, {Dave}, {Davier}, {Davis}, {Daw}, {Day}, {De}, {DeBra}, {Degallaix}, {De Laurentis}, {Del{\'e}glise}, {Del Pozzo}, {Demos}, {Denker}, {Dent}, {De Pietri}, {Dergachev}, {De Rosa}, {DeRosa}, {De Rossi}, {DeSalvo}, {de Varona}, {Devenson},
  {Dhurandhar}, {D{\'\i}az}, {Dietrich}, {Di Fiore}, {Di Giovanni}, {Di Girolamo}, {Di Lieto}, {Di Pace}, {Di Palma}, {Di Renzo}, {Doctor}, {Dolique}, {Donovan}, {Dooley}, {Doravari}, {Dorrington}, {Douglas}, {Dovale {\'A}lvarez}, {Downes}, {Drago}, {Dreissigacker}, {Driggers}, {Du}, {Ducrot}, {Dudi}, {Dupej}, {Dwyer}, {Edo}, {Edwards}, {Effler}, {Eggenstein}, {Ehrens}, {Eichholz}, {Eikenberry}, {Eisenstein}, {Essick}, {Estevez}, {Etienne}, {Etzel}, {Evans}, {Evans}, {Factourovich}, {Fafone}, {Fair}, {Fairhurst}, {Fan}, {Farinon}, {Farr}, {Farr}, {Fauchon-Jones}, {Favata}, {Fays}, {Fee}, {Fehrmann}, {Feicht}, {Fejer}, {Fernandez-Galiana}, {Ferrante}, {Ferreira}, {Ferrini}, {Fidecaro}, {Finstad}, {Fiori}, {Fiorucci}, {Fishbach}, {Fisher}, {Fitz-Axen}, {Flaminio}, {Fletcher}, {Fong}, {Font}, {Forsyth}, {Forsyth}, {Fournier}, {Frasca}, {Frasconi}, {Frei}, {Freise}, {Frey}, {Frey}, {Fries}, {Fritschel}, {Frolov}, {Fulda}, {Fyffe}, {Gabbard}, {Gadre}, {Gaebel}, {Gair}, {Gammaitoni}, {Ganija}, {Gaonkar},
  {Garcia-Quiros}, {Garufi}, {Gateley}, {Gaudio}, {Gaur}, {Gayathri}, {Gehrels}, {Gemme}, {Genin}, {Gennai}, {George}, {George}, {Gergely}, {Germain}, {Ghonge}, {Ghosh}, {Ghosh}, {Ghosh}, {Giaime}, {Giardina}, {Giazotto}, {Gill}, {Glover}, {Goetz}, {Goetz}, {Gomes}, {Goncharov}, {Gonz{\'a}lez}, {Gonzalez Castro}, {Gopakumar}, {Gorodetsky}, {Gossan}, {Gosselin}, {Gouaty}, {Grado}, {Graef}, {Granata}, {Grant}, {Gras}, {Gray}, {Greco}, {Green}, {Gretarsson}, {Groot}, {Grote}, {Grunewald}, {Gruning}, {Guidi}, {Guo}, {Gupta}, {Gupta}, {Gushwa}, {Gustafson}, {Gustafson}, {Halim}, {Hall}, {Hall}, {Hamilton}, {Hammond}, {Haney}, {Hanke}, {Hanks}, {Hanna}, {Hannam}, {Hannuksela}, {Hanson}, {Hardwick}, {Harms}, {Harry}, {Harry}, {Hart}, {Haster}, {Haughian}, {Healy}, {Heidmann}, {Heintze}, {Heitmann}, {Hello}, {Hemming}, {Hendry}, {Heng}, {Hennig}, {Heptonstall}, {Heurs}, {Hild}, {Hinderer}, {Ho}, {Hoak}, {Hofman}, {Holt}, {Holz}, {Hopkins}, {Horst}, {Hough}, {Houston}, {Howell}, {Hreibi}, {Hu}, {Huerta}, {Huet},
  {Hughey}, {Husa}, {Huttner}, {Huynh-Dinh}, {Indik}, {Inta}, {Intini}, {Isa}, {Isac}, {Isi}, {Iyer}, {Izumi}, {Jacqmin}, {Jani}, {Jaranowski}, {Jawahar}, {Jim{\'e}nez-Forteza}, {Johnson}, {Johnson-McDaniel}, {Jones}, {Jones}, {Jonker}, {Ju}, {Junker}, {Kalaghatgi}, {Kalogera}, {Kamai}, {Kandhasamy}, {Kang}, {Kanner}, {Kapadia}, {Karki}, {Karvinen}, {Kasprzack}, {Kastaun}, {Katolik}, {Katsavounidis}, {Katzman}, {Kaufer}, {Kawabe}, {K{\'e}f{\'e}lian}, {Keitel}, {Kemball}, {Kennedy}, {Kent}, {Key}, {Khalili}, {Khan}, {Khan}, {Khan}, {Khazanov}, {Kijbunchoo}, {Kim}, {Kim}, {Kim}, {Kim}, {Kim}, {Kim}, {Kimbrell}, {King}, {King}, {Kinley-Hanlon}, {Kirchhoff}, {Kissel}, {Kleybolte}, {Klimenko}, {Knowles}, {Koch}, {Koehlenbeck}, {Koley}, {Kondrashov}, {Kontos}, {Korobko}, {Korth}, {Kowalska}, {Kozak}, {Kr{\"a}mer}, {Kringel}, {Krishnan}, {Kr{\'o}lak}, {Kuehn}, {Kumar}, {Kumar}, {Kumar}, {Kuo}, {Kutynia}, {Kwang}, {Lackey}, {Lai}, {Landry}, {Lang}, {Lange}, {Lantz}, {Lanza}, {Larson}, {Lartaux-Vollard}, {Lasky},
  {Laxen}, {Lazzarini}, {Lazzaro}, {Leaci}, {Leavey}, {Lee}, {Lee}, {Lee}, {Lee}, {Lee}, {Lehmann}, {Lenon}, {Leon}, {Leonardi}, {Leroy}, {Letendre}, {Levin}, {Li}, {Linker}, {Littenberg}, {Liu}, {Liu}, {Lo}, {Lockerbie}, {London}, {Lord}, {Lorenzini}, {Loriette}, {Lormand}, {Losurdo}, {Lough}, {Lousto}, {Lovelace}, {L{\"u}ck}, {Lumaca}, {Lundgren}, {Lynch}, {Ma}, {Macas}, {Macfoy}, {Machenschalk}, {MacInnis}, {Macleod}, {Maga{\~n}a Hernandez}, {Maga{\~n}a-Sandoval}, {Maga{\~n}a Zertuche}, {Magee}, {Majorana}, {Maksimovic}, {Man}, {Mandic}, {Mangano}, {Mansell}, {Manske}, {Mantovani}, {Marchesoni}, {Marion}, {M{\'a}rka}, {M{\'a}rka}, {Markakis}, {Markosyan}, {Markowitz}, {Maros}, {Marquina}, {Marsh}, {Martelli}, {Martellini}, {Martin}, {Martin}, {Martynov}, {Marx}, {Mason}, {Massera}, {Masserot}, {Massinger}, {Masso-Reid}, {Mastrogiovanni}, {Matas}, {Matichard}, {Matone}, {Mavalvala}, {Mazumder}, {McCarthy}, {McClelland}, {McCormick}, {McCuller}, {McGuire}, {McIntyre}, {McIver}, {McManus}, {McNeill}, {McRae},
  {McWilliams}, {Meacher}, {Meadors}, {Mehmet}, {Meidam}, {Mejuto-Villa}, {Melatos}, {Mendell}, {Mercer}, {Merilh}, {Merzougui}, {Meshkov}, {Messenger}, {Messick}, {Metzdorff}, {Meyers}, {Miao}, {Michel}, {Middleton}, {Mikhailov}, {Milano}, {Miller}, {Miller}, {Miller}, {Millhouse}, {Milovich-Goff}, {Minazzoli}, {Minenkov}, {Ming}, {Mishra}, {Mitra}, {Mitrofanov}, {Mitselmakher}, {Mittleman}, {Moffa}, {Moggi}, {Mogushi}, {Mohan}, {Mohapatra}, {Molina}, {Montani}, {Moore}, {Moraru}, {Moreno}, {Morisaki}, {Morriss}, {Mours}, {Mow-Lowry}, {Mueller}, {Muir}, {Mukherjee}, {Mukherjee}, {Mukherjee}, {Mukund}, {Mullavey}, {Munch}, {Mu{\~n}iz}, {Muratore}, {Murray}, {Nagar}, {Napier}, {Nardecchia}, {Naticchioni}, {Nayak}, {Neilson}, {Nelemans}, {Nelson}, {Nery}, {Neunzert}, {Nevin}, {Newport}, {Newton}, {Ng}, {Nguyen}, {Nguyen}, {Nichols}, {Nielsen}, {Nissanke}, {Nitz}, {Noack}, {Nocera}, {Nolting}, {North}, {Nuttall}, {Oberling}, {O'Dea}, {Ogin}, {Oh}, {Oh}, {Ohme}, {Okada}, {Oliver}, {Oppermann}, {Oram}, {O'Reilly},
  {Ormiston}, {Ortega}, {O'Shaughnessy}, {Ossokine}, {Ottaway}, {Overmier}, {Owen}, {Pace}, {Page}, {Page}, {Pai}, {Pai}, {Palamos}, {Palashov}, {Palomba}, {Pal-Singh}, {Pan}, {Pan}, {Pang}, {Pang}, {Pankow}, {Pannarale}, {Pant}, {Paoletti}, {Paoli}, {Papa}, {Parida}, {Parker}, {Pascucci}, {Pasqualetti}, {Passaquieti}, {Passuello}, {Patil}, {Patricelli}, {Pearlstone}, {Pedraza}, {Pedurand}, {Pekowsky}, {Pele}, {Penn}, {Perez}, {Perreca}, {Perri}, {Pfeiffer}, {Phelps}, {Piccinni}, {Pichot}, {Piergiovanni}, {Pierro}, {Pillant}, {Pinard}, {Pinto}, {Pirello}, {Pitkin}, {Poe}, {Poggiani}, {Popolizio}, {Porter}, {Post}, {Powell}, {Prasad}, {Pratt}, {Pratten}, {Predoi}, {Prestegard}, {Prijatelj}, {Principe}, {Privitera}, {Prix}, {Prodi}, {Prokhorov}, {Puncken}, {Punturo}, {Puppo}, {P{\"u}rrer}, {Qi}, {Quetschke}, {Quintero}, {Quitzow-James}, {Raab}, {Rabeling}, {Radkins}, {Raffai}, {Raja}, {Rajan}, {Rajbhandari}, {Rakhmanov}, {Ramirez}, {Ramos-Buades}, {Rapagnani}, {Raymond}, {Razzano}, {Read}, {Regimbau}, {Rei},
  {Reid}, {Reitze}, {Ren}, {Reyes}, {Ricci}, {Ricker}, {Rieger}, {Riles}, {Rizzo}, {Robertson}, {Robie}, {Robinet}, {Rocchi}, {Rolland}, {Rollins}, {Roma}, {Romano}, {Romano}, {Romel}, {Romie}, {Rosi{\'n}ska}, {Ross}, {Rowan}, {R{\"u}diger}, {Ruggi}, {Rutins}, {Ryan}, {Sachdev}, {Sadecki}, {Sadeghian}, {Sakellariadou}, {Salconi}, {Saleem}, {Salemi}, {Samajdar}, {Sammut}, {Sampson}, {Sanchez}, {Sanchez}, {Sanchis-Gual}, {Sandberg}, {Sanders}, {Sassolas}, {Sathyaprakash}, {Saulson}, {Sauter}, {Savage}, {Sawadsky}, {Schale}, {Scheel}, {Scheuer}, {Schmidt}, {Schmidt}, {Schnabel}, {Schofield}, {Sch{\"o}nbeck}, {Schreiber}, {Schuette}, {Schulte}, {Schutz}, {Schwalbe}, {Scott}, {Scott}, {Seidel}, {Sellers}, {Sengupta}, {Sentenac}, {Sequino}, {Sergeev}, {Shaddock}, {Shaffer}, {Shah}, {Shahriar}, {Shaner}, {Shao}, {Shapiro}, {Shawhan}, {Sheperd}, {Shoemaker}, {Shoemaker}, {Siellez}, {Siemens}, {Sieniawska}, {Sigg}, {Silva}, {Singer}, {Singh}, {Singhal}, {Sintes}, {Slagmolen}, {Smith}, {Smith}, {Smith}, {Somala},
  {Son}, {Sonnenberg}, {Sorazu}, {Sorrentino}, {Souradeep}, {Spencer}, {Srivastava}, {Staats}, {Staley}, {Steinke}, {Steinlechner}, {Steinlechner}, {Steinmeyer}, {Stevenson}, {Stone}, {Stops}, {Strain}, {Stratta}, {Strigin}, {Strunk}, {Sturani}, {Stuver}, {Summerscales}, {Sun}, {Sunil}, {Suresh}, {Sutton}, {Swinkels}, {Szczepa{\'n}czyk}, {Tacca}, {Tait}, {Talbot}, {Talukder}, {Tanner}, {T{\'a}pai}, {Taracchini}, {Tasson}, {Taylor}, {Taylor}, {Tewari}, {Theeg}, {Thies}, {Thomas}, {Thomas}, {Thomas}, {Thorne}, {Thorne}, {Thrane}, {Tiwari}, {Tiwari}, {Tokmakov}, {Toland}, {Tonelli}, {Tornasi}, {Torres-Forn{\'e}}, {Torrie}, {T{\"o}yr{\"a}}, {Travasso}, {Traylor}, {Trinastic}, {Tringali}, {Trozzo}, {Tsang}, {Tse}, {Tso}, {Tsukada}, {Tsuna}, {Tuyenbayev}, {Ueno}, {Ugolini}, {Unnikrishnan}, {Urban}, {Usman}, {Vahlbruch}, {Vajente}, {Valdes}, {Vallisneri}, {van Bakel}, {van Beuzekom}, {van den Brand}, {Van Den Broeck}, {Vander-Hyde}, {van der Schaaf}, {van Heijningen}, {van Veggel}, {Vardaro}, {Varma}, {Vass},
  {Vas{\'u}th}, {Vecchio}, {Vedovato}, {Veitch}, {Veitch}, {Venkateswara}, {Venugopalan}, {Verkindt}, {Vetrano}, {Vicer{\'e}}, {Viets}, {Vinciguerra}, {Vine}, {Vinet}, {Vitale}, {Vo}, {Vocca}, {Vorvick}, {Vyatchanin}, {Wade}, {Wade}, {Wade}, {Walet}, {Walker}, {Wallace}, {Walsh}, {Wang}, {Wang}, {Wang}, {Wang}, {Wang}, {Ward}, {Warner}, {Was}, {Watchi}, {Weaver}, {Wei}, {Weinert}, {Weinstein}, {Weiss}, {Wen}, {Wessel}, {We{\ss}els}, {Westerweck}, {Westphal}, {Wette}, {Whelan}, {Whitcomb}, {Whiting}, {Whittle}, {Wilken}, {Williams}, {Williams}, {Williamson}, {Willis}, {Willke}, {Wimmer}, {Winkler}, {Wipf}, {Wittel}, {Woan}, {Woehler}, {Wofford}, {Wong}, {Worden}, {Wright}, {Wu}, {Wysocki}, {Xiao}, {Yamamoto}, {Yancey}, {Yang}, {Yap}, {Yazback}, {Yu}, {Yu}, {Yvert}, {Zadro{\.Z}ny}, {Zanolin}, {Zelenova}, {Zendri}, {Zevin}, {Zhang}, {Zhang}, {Zhang}, {Zhang}, {Zhao}, {Zhou}, {Zhou}, {Zhu}, {Zhu}, {Zimmerman}, {Zucker}, {Zweizig}, {LIGO Scientific Collaboration}, \& {Virgo Collaboration}}]{2017PhRvL.119p1101A}
{Abbott}, B.~P., {Abbott}, R., {Abbott}, T.~D., {et~al.} 2017{\natexlab{a}}, \prl, 119, 161101, \dodoi{10.1103/PhysRevLett.119.161101}

\bibitem[{{Abbott} {et~al.}(2017{\natexlab{b}}){Abbott}, {Abbott}, {Abbott}, {Acernese}, {Ackley}, {Adams}, {Adams}, {Addesso}, {Adhikari}, {Adya}, {Affeldt}, {Afrough}, {Agarwal}, {Agathos}, {Agatsuma}, {Aggarwal}, {Aguiar}, {Aiello}, {Ain}, {Ajith}, {Allen}, {Allen}, {Allocca}, {Altin}, {Amato}, {Ananyeva}, {Anderson}, {Anderson}, {Angelova}, {Antier}, {Appert}, {Arai}, {Araya}, {Areeda}, {Arnaud}, {Arun}, {Ascenzi}, {Ashton}, {Ast}, {Aston}, {Astone}, {Atallah}, {Aufmuth}, {Aulbert}, {AultONeal}, {Austin}, {Avila-Alvarez}, {Babak}, {Bacon}, {Bader}, {Bae}, {Baker}, {Baldaccini}, {Ballardin}, {Ballmer}, {Banagiri}, {Barayoga}, {Barclay}, {Barish}, {Barker}, {Barkett}, {Barone}, {Barr}, {Barsotti}, {Barsuglia}, {Barta}, {Barthelmy}, {Bartlett}, {Bartos}, {Bassiri}, {Basti}, {Batch}, {Bawaj}, {Bayley}, {Bazzan}, {B{\'e}csy}, {Beer}, {Bejger}, {Belahcene}, {Bell}, {Berger}, {Bergmann}, {Bero}, {Berry}, {Bersanetti}, {Bertolini}, {Betzwieser}, {Bhagwat}, {Bhandare}, {Bilenko}, {Billingsley}, {Billman}, {Birch},
  {Birney}, {Birnholtz}, {Biscans}, {Biscoveanu}, {Bisht}, {Bitossi}, {Biwer}, {Bizouard}, {Blackburn}, {Blackman}, {Blair}, {Blair}, {Blair}, {Bloemen}, {Bock}, {Bode}, {Boer}, {Bogaert}, {Bohe}, {Bondu}, {Bonilla}, {Bonnand}, {Boom}, {Bork}, {Boschi}, {Bose}, {Bossie}, {Bouffanais}, {Bozzi}, {Bradaschia}, {Brady}, {Branchesi}, {Brau}, {Briant}, {Brillet}, {Brinkmann}, {Brisson}, {Brockill}, {Broida}, {Brooks}, {Brown}, {Brown}, {Brunett}, {Buchanan}, {Buikema}, {Bulik}, {Bulten}, {Buonanno}, {Buskulic}, {Buy}, {Byer}, {Cabero}, {Cadonati}, {Cagnoli}, {Cahillane}, {Calder{\'o}n Bustillo}, {Callister}, {Calloni}, {Camp}, {Canepa}, {Canizares}, {Cannon}, {Cao}, {Cao}, {Capano}, {Capocasa}, {Carbognani}, {Caride}, {Carney}, {Casanueva Diaz}, {Casentini}, {Caudill}, {Cavagli{\`a}}, {Cavalier}, {Cavalieri}, {Cella}, {Cepeda}, {Cerd{\'a}-Dur{\'a}n}, {Cerretani}, {Cesarini}, {Chamberlin}, {Chan}, {Chao}, {Charlton}, {Chase}, {Chassande-Mottin}, {Chatterjee}, {Chatziioannou}, {Cheeseboro}, {Chen}, {Chen}, {Chen},
  {Cheng}, {Chia}, {Chincarini}, {Chiummo}, {Chmiel}, {Cho}, {Cho}, {Chow}, {Christensen}, {Chu}, {Chua}, {Chua}, {Chung}, {Chung}, {Ciani}, {Ciolfi}, {Cirelli}, {Cirone}, {Clara}, {Clark}, {Clearwater}, {Cleva}, {Cocchieri}, {Coccia}, {Cohadon}, {Cohen}, {Colla}, {Collette}, {Cominsky}, {Constancio}, {Conti}, {Cooper}, {Corban}, {Corbitt}, {Cordero-Carri{\'o}n}, {Corley}, {Cornish}, {Corsi}, {Cortese}, {Costa}, {Coughlin}, {Coughlin}, {Coulon}, {Countryman}, {Couvares}, {Covas}, {Cowan}, {Coward}, {Cowart}, {Coyne}, {Coyne}, {Creighton}, {Creighton}, {Cripe}, {Crowder}, {Cullen}, {Cumming}, {Cunningham}, {Cuoco}, {Dal Canton}, {D{\'a}lya}, {Danilishin}, {D'Antonio}, {Danzmann}, {Dasgupta}, {Da Silva Costa}, {Dattilo}, {Dave}, {Davier}, {Davis}, {Daw}, {Day}, {De}, {DeBra}, {Degallaix}, {De Laurentis}, {Del{\'e}glise}, {Del Pozzo}, {Demos}, {Denker}, {Dent}, {De Pietri}, {Dergachev}, {De Rosa}, {DeRosa}, {De Rossi}, {DeSalvo}, {de Varona}, {Devenson}, {Dhurandhar}, {D{\'\i}az}, {Di Fiore}, {Di Giovanni}, {Di
  Girolamo}, {Di Lieto}, {Di Pace}, {Di Palma}, {Di Renzo}, {Doctor}, {Dolique}, {Donovan}, {Dooley}, {Doravari}, {Dorrington}, {Douglas}, {Dovale {\'A}lvarez}, {Downes}, {Drago}, {Dreissigacker}, {Driggers}, {Du}, {Ducrot}, {Dupej}, {Dwyer}, {Edo}, {Edwards}, {Effler}, {Ehrens}, {Eichholz}, {Eikenberry}, {Eisenstein}, {Essick}, {Estevez}, {Etienne}, {Etzel}, {Evans}, {Evans}, {Factourovich}, {Fafone}, {Fair}, {Fairhurst}, {Fan}, {Farinon}, {Farr}, {Farr}, {Fauchon-Jones}, {Favata}, {Fays}, {Fee}, {Fehrmann}, {Feicht}, {Fejer}, {Fernandez-Galiana}, {Ferrante}, {Ferreira}, {Ferrini}, {Fidecaro}, {Finstad}, {Fiori}, {Fiorucci}, {Fishbach}, {Fisher}, {Fitz-Axen}, {Flaminio}, {Fletcher}, {Fong}, {Font}, {Forsyth}, {Forsyth}, {Fournier}, {Frasca}, {Frasconi}, {Frei}, {Freise}, {Frey}, {Frey}, {Fries}, {Fritschel}, {Frolov}, {Fulda}, {Fyffe}, {Gabbard}, {Gadre}, {Gaebel}, {Gair}, {Gammaitoni}, {Ganija}, {Gaonkar}, {Garcia-Quiros}, {Garufi}, {Gateley}, {Gaudio}, {Gaur}, {Gayathri}, {Gehrels}, {Gemme}, {Genin},
  {Gennai}, {George}, {George}, {Gergely}, {Germain}, {Ghonge}, {Ghosh}, {Ghosh}, {Ghosh}, {Giaime}, {Giardina}, {Giazotto}, {Gill}, {Glover}, {Goetz}, {Goetz}, {Gomes}, {Goncharov}, {Gonz{\'a}lez}, {Gonzalez Castro}, {Gopakumar}, {Gorodetsky}, {Gossan}, {Gosselin}, {Gouaty}, {Grado}, {Graef}, {Granata}, {Grant}, {Gras}, {Gray}, {Greco}, {Green}, {Gretarsson}, {Griswold}, {Groot}, {Grote}, {Grunewald}, {Gruning}, {Guidi}, {Guo}, {Gupta}, {Gupta}, {Gushwa}, {Gustafson}, {Gustafson}, {Halim}, {Hall}, {Hall}, {Hamilton}, {Hammond}, {Haney}, {Hanke}, {Hanks}, {Hanna}, {Hannam}, {Hannuksela}, {Hanson}, {Hardwick}, {Harms}, {Harry}, {Harry}, {Hart}, {Haster}, {Haughian}, {Healy}, {Heidmann}, {Heintze}, {Heitmann}, {Hello}, {Hemming}, {Hendry}, {Heng}, {Hennig}, {Heptonstall}, {Heurs}, {Hild}, {Hinderer}, {Hoak}, {Hofman}, {Holt}, {Holz}, {Hopkins}, {Horst}, {Hough}, {Houston}, {Howell}, {Hreibi}, {Hu}, {Huerta}, {Huet}, {Hughey}, {Husa}, {Huttner}, {Huynh-Dinh}, {Indik}, {Inta}, {Intini}, {Isa}, {Isac}, {Isi},
  {Iyer}, {Izumi}, {Jacqmin}, {Jani}, {Jaranowski}, {Jawahar}, {Jim{\'e}nez-Forteza}, {Johnson}, {Jones}, {Jones}, {Jonker}, {Ju}, {Junker}, {Kalaghatgi}, {Kalogera}, {Kamai}, {Kandhasamy}, {Kang}, {Kanner}, {Kapadia}, {Karki}, {Karvinen}, {Kasprzack}, {Katolik}, {Katsavounidis}, {Katzman}, {Kaufer}, {Kawabe}, {K{\'e}f{\'e}lian}, {Keitel}, {Kemball}, {Kennedy}, {Kent}, {Key}, {Khalili}, {Khan}, {Khan}, {Khan}, {Khazanov}, {Kijbunchoo}, {Kim}, {Kim}, {Kim}, {Kim}, {Kim}, {Kim}, {Kimbrell}, {King}, {King}, {Kinley-Hanlon}, {Kirchhoff}, {Kissel}, {Kleybolte}, {Klimenko}, {Knowles}, {Koch}, {Koehlenbeck}, {Koley}, {Kondrashov}, {Kontos}, {Korobko}, {Korth}, {Kowalska}, {Kozak}, {Kr{\"a}mer}, {Kringel}, {Krishnan}, {Kr{\'o}lak}, {Kuehn}, {Kumar}, {Kumar}, {Kumar}, {Kuo}, {Kutynia}, {Kwang}, {Lackey}, {Lai}, {Landry}, {Lang}, {Lange}, {Lantz}, {Lanza}, {Larson}, {Lartaux-Vollard}, {Lasky}, {Laxen}, {Lazzarini}, {Lazzaro}, {Leaci}, {Leavey}, {Lee}, {Lee}, {Lee}, {Lee}, {Lee}, {Lehmann}, {Lenon}, {Leonardi}, {Leroy},
  {Letendre}, {Levin}, {Li}, {Linker}, {Littenberg}, {Liu}, {Lo}, {Lockerbie}, {London}, {Lord}, {Lorenzini}, {Loriette}, {Lormand}, {Losurdo}, {Lough}, {Lousto}, {Lovelace}, {L{\"u}ck}, {Lumaca}, {Lundgren}, {Lynch}, {Ma}, {Macas}, {Macfoy}, {Machenschalk}, {MacInnis}, {Macleod}, {Maga{\~n}a Hernandez}, {Maga{\~n}a-Sandoval}, {Maga{\~n}a Zertuche}, {Magee}, {Majorana}, {Maksimovic}, {Man}, {Mandic}, {Mangano}, {Mansell}, {Manske}, {Mantovani}, {Marchesoni}, {Marion}, {M{\'a}rka}, {M{\'a}rka}, {Markakis}, {Markosyan}, {Markowitz}, {Maros}, {Marquina}, {Marsh}, {Martelli}, {Martellini}, {Martin}, {Martin}, {Martynov}, {Mason}, {Massera}, {Masserot}, {Massinger}, {Masso-Reid}, {Mastrogiovanni}, {Matas}, {Matichard}, {Matone}, {Mavalvala}, {Mazumder}, {McCarthy}, {McClelland}, {McCormick}, {McCuller}, {McGuire}, {McIntyre}, {McIver}, {McManus}, {McNeill}, {McRae}, {McWilliams}, {Meacher}, {Meadors}, {Mehmet}, {Meidam}, {Mejuto-Villa}, {Melatos}, {Mendell}, {Mercer}, {Merilh}, {Merzougui}, {Meshkov}, {Messenger},
  {Messick}, {Metzdorff}, {Meyers}, {Miao}, {Michel}, {Middleton}, {Mikhailov}, {Milano}, {Miller}, {Miller}, {Miller}, {Millhouse}, {Milovich-Goff}, {Minazzoli}, {Minenkov}, {Ming}, {Mishra}, {Mitra}, {Mitrofanov}, {Mitselmakher}, {Mittleman}, {Moffa}, {Moggi}, {Mogushi}, {Mohan}, {Mohapatra}, {Montani}, {Moore}, {Moraru}, {Moreno}, {Morriss}, {Mours}, {Mow-Lowry}, {Mueller}, {Muir}, {Mukherjee}, {Mukherjee}, {Mukherjee}, {Mukund}, {Mullavey}, {Munch}, {Mu{\~n}iz}, {Muratore}, {Murray}, {Napier}, {Nardecchia}, {Naticchioni}, {Nayak}, {Neilson}, {Nelemans}, {Nelson}, {Nery}, {Neunzert}, {Nevin}, {Newport}, {Newton}, {Ng}, {Nguyen}, {Nguyen}, {Nichols}, {Nielsen}, {Nissanke}, {Nitz}, {Noack}, {Nocera}, {Nolting}, {North}, {Nuttall}, {Oberling}, {O'Dea}, {Ogin}, {Oh}, {Oh}, {Ohme}, {Okada}, {Oliver}, {Oppermann}, {Oram}, {O'Reilly}, {Ormiston}, {Ortega}, {O'Shaughnessy}, {Ossokine}, {Ottaway}, {Overmier}, {Owen}, {Pace}, {Page}, {Page}, {Pai}, {Pai}, {Palamos}, {Palashov}, {Palomba}, {Pal-Singh}, {Pan}, {Pan},
  {Pang}, {Pang}, {Pankow}, {Pannarale}, {Pant}, {Paoletti}, {Paoli}, {Papa}, {Parida}, {Parker}, {Pascucci}, {Pasqualetti}, {Passaquieti}, {Passuello}, {Patil}, {Patricelli}, {Pearlstone}, {Pedraza}, {Pedurand}, {Pekowsky}, {Pele}, {Penn}, {Perez}, {Perreca}, {Perri}, {Pfeiffer}, {Phelps}, {Piccinni}, {Pichot}, {Piergiovanni}, {Pierro}, {Pillant}, {Pinard}, {Pinto}, {Pirello}, {Pitkin}, {Poe}, {Poggiani}, {Popolizio}, {Porter}, {Post}, {Powell}, {Prasad}, {Pratt}, {Pratten}, {Predoi}, {Prestegard}, {Price}, {Prijatelj}, {Principe}, {Privitera}, {Prodi}, {Prokhorov}, {Puncken}, {Punturo}, {Puppo}, {P{\"u}rrer}, {Qi}, {Quetschke}, {Quintero}, {Quitzow-James}, {Raab}, {Rabeling}, {Radkins}, {Raffai}, {Raja}, {Rajan}, {Rajbhandari}, {Rakhmanov}, {Ramirez}, {Ramos-Buades}, {Rapagnani}, {Raymond}, {Razzano}, {Read}, {Regimbau}, {Rei}, {Reid}, {Reitze}, {Ren}, {Reyes}, {Ricci}, {Ricker}, {Rieger}, {Riles}, {Rizzo}, {Robertson}, {Robie}, {Robinet}, {Rocchi}, {Rolland}, {Rollins}, {Roma}, {Romano}, {Romel}, {Romie},
  {Rosi{\'n}ska}, {Ross}, {Rowan}, {R{\"u}diger}, {Ruggi}, {Rutins}, {Ryan}, {Sachdev}, {Sadecki}, {Sadeghian}, {Sakellariadou}, {Salconi}, {Saleem}, {Salemi}, {Samajdar}, {Sammut}, {Sampson}, {Sanchez}, {Sanchez}, {Sanchis-Gual}, {Sandberg}, {Sanders}, {Sassolas}, {Sathyaprakash}, {Saulson}, {Sauter}, {Savage}, {Sawadsky}, {Schale}, {Scheel}, {Scheuer}, {Schmidt}, {Schmidt}, {Schnabel}, {Schofield}, {Sch{\"o}nbeck}, {Schreiber}, {Schuette}, {Schulte}, {Schutz}, {Schwalbe}, {Scott}, {Scott}, {Seidel}, {Sellers}, {Sengupta}, {Sentenac}, {Sequino}, {Sergeev}, {Shaddock}, {Shaffer}, {Shah}, {Shahriar}, {Shaner}, {Shao}, {Shapiro}, {Shawhan}, {Sheperd}, {Shoemaker}, {Shoemaker}, {Siellez}, {Siemens}, {Sieniawska}, {Sigg}, {Silva}, {Singer}, {Singh}, {Singhal}, {Sintes}, {Slagmolen}, {Smith}, {Smith}, {Smith}, {Somala}, {Son}, {Sonnenberg}, {Sorazu}, {Sorrentino}, {Souradeep}, {Spencer}, {Srivastava}, {Staats}, {Staley}, {Steinke}, {Steinlechner}, {Steinlechner}, {Steinmeyer}, {Stevenson}, {Stone}, {Stops},
  {Strain}, {Stratta}, {Strigin}, {Strunk}, {Sturani}, {Stuver}, {Summerscales}, {Sun}, {Sunil}, {Suresh}, {Sutton}, {Swinkels}, {Szczepa{\'n}czyk}, {Tacca}, {Tait}, {Talbot}, {Talukder}, {Tanner}, {T{\'a}pai}, {Taracchini}, {Tasson}, {Taylor}, {Taylor}, {Tewari}, {Theeg}, {Thies}, {Thomas}, {Thomas}, {Thomas}, {Thorne}, {Thorne}, {Thrane}, {Tiwari}, {Tiwari}, {Tokmakov}, {Toland}, {Tonelli}, {Tornasi}, {Torres-Forn{\'e}}, {Torrie}, {T{\"o}yr{\"a}}, {Travasso}, {Traylor}, {Trinastic}, {Tringali}, {Trozzo}, {Tsang}, {Tse}, {Tso}, {Tsukada}, {Tsuna}, {Tuyenbayev}, {Ueno}, {Ugolini}, {Unnikrishnan}, {Urban}, {Usman}, {Vahlbruch}, {Vajente}, {Valdes}, {van Bakel}, {van Beuzekom}, {van den Brand}, {Van Den Broeck}, {Vander-Hyde}, {van der Schaaf}, {van Heijningen}, {van Veggel}, {Vardaro}, {Varma}, {Vass}, {Vas{\'u}th}, {Vecchio}, {Vedovato}, {Veitch}, {Veitch}, {Venkateswara}, {Venugopalan}, {Verkindt}, {Vetrano}, {Vicer{\'e}}, {Viets}, {Vinciguerra}, {Vine}, {Vinet}, {Vitale}, {Vo}, {Vocca}, {Vorvick},
  {Vyatchanin}, {Wade}, {Wade}, {Wade}, {Walet}, {Walker}, {Wallace}, {Walsh}, {Wang}, {Wang}, {Wang}, {Wang}, {Wang}, {Ward}, {Warner}, {Was}, {Watchi}, {Weaver}, {Wei}, {Weinert}, {Weinstein}, {Weiss}, {Wen}, {Wessel}, {Wessels}, {Westerweck}, {Westphal}, {Wette}, {Whelan}, {Whitcomb}, {Whiting}, {Whittle}, {Wilken}, {Williams}, {Williams}, {Williamson}, {Willis}, {Willke}, {Wimmer}, {Winkler}, {Wipf}, {Wittel}, {Woan}, {Woehler}, {Wofford}, {Wong}, {Worden}, {Wright}, {Wu}, {Wysocki}, {Xiao}, {Yamamoto}, {Yancey}, {Yang}, {Yap}, {Yazback}, {Yu}, {Yu}, {Yvert}, {Zadro{\.z}ny}, {Zanolin}, {Zelenova}, {Zendri}, {Zevin}, {Zhang}, {Zhang}, {Zhang}, {Zhang}, {Zhao}, {Zhou}, {Zhou}, {Zhu}, {Zhu}, {Zimmerman}, {Zucker}, {Zweizig}, {LIGO Scientific Collaboration}, {Virgo Collaboration}, {Wilson-Hodge}, {Bissaldi}, {Blackburn}, {Briggs}, {Burns}, {Cleveland}, {Connaughton}, {Gibby}, {Giles}, {Goldstein}, {Hamburg}, {Jenke}, {Hui}, {Kippen}, {Kocevski}, {McBreen}, {Meegan}, {Paciesas}, {Poolakkil}, {Preece},
  {Racusin}, {Roberts}, {Stanbro}, {Veres}, {von Kienlin}, {GBM}, {Savchenko}, {Ferrigno}, {Kuulkers}, {Bazzano}, {Bozzo}, {Brandt}, {Chenevez}, {Courvoisier}, {Diehl}, {Domingo}, {Hanlon}, {Jourdain}, {Laurent}, {Lebrun}, {Lutovinov}, {Martin-Carrillo}, {Mereghetti}, {Natalucci}, {Rodi}, {Roques}, {Sunyaev}, {Ubertini}, {INTEGRAL}, {Aartsen}, {Ackermann}, {Adams}, {Aguilar}, {Ahlers}, {Ahrens}, {Samarai}, {Altmann}, {Andeen}, {Anderson}, {Ansseau}, {Anton}, {Arg{\"u}elles}, {Auffenberg}, {Axani}, {Bagherpour}, {Bai}, {Barron}, {Barwick}, {Baum}, {Bay}, {Beatty}, {Becker Tjus}, {Bernardini}, {Besson}, {Binder}, {Bindig}, {Blaufuss}, {Blot}, {Bohm}, {B{\"o}rner}, {Bos}, {Bose}, {B{\"o}ser}, {Botner}, {Bourbeau}, {Bourbeau}, {Bradascio}, {Braun}, {Brayeur}, {Brenzke}, {Bretz}, {Bron}, {Brostean-Kaiser}, {Burgman}, {Carver}, {Casey}, {Casier}, {Cheung}, {Chirkin}, {Christov}, {Clark}, {Classen}, {Coenders}, {Collin}, {Conrad}, {Cowen}, {Cross}, {Day}, {de Andr{\'e}}, {De Clercq}, {DeLaunay}, {Dembinski}, {De
  Ridder}, {Desiati}, {de Vries}, {de Wasseige}, {de With}, {DeYoung}, {D{\'\i}az-V{\'e}lez}, {di Lorenzo}, {Dujmovic}, {Dumm}, {Dunkman}, {Dvorak}, {Eberhardt}, {Ehrhardt}, {Eichmann}, {Eller}, {Evenson}, {Fahey}, {Fazely}, {Felde}, {Filimonov}, {Finley}, {Flis}, {Franckowiak}, {Friedman}, {Fuchs}, {Gaisser}, {Gallagher}, {Gerhardt}, {Ghorbani}, {Giang}, {Glauch}, {Gl{\"u}senkamp}, {Goldschmidt}, {Gonzalez}, {Grant}, {Griffith}, {Haack}, {Hallgren}, {Halzen}, {Hanson}, {Hebecker}, {Heereman}, {Helbing}, {Hellauer}, {Hickford}, {Hignight}, {Hill}, {Hoffman}, {Hoffmann}, {Hokanson-Fasig}, {Hoshina}, {Huang}, {Huber}, {Hultqvist}, {H{\"u}nnefeld}, {In}, {Ishihara}, {Jacobi}, {Japaridze}, {Jeong}, {Jero}, {Jones}, {Kalaczynski}, {Kang}, {Kappes}, {Karg}, {Karle}, {Kauer}, {Keivani}, {Kelley}, {Kheirandish}, {Kim}, {Kim}, {Kintscher}, {Kiryluk}, {Kittler}, {Klein}, {Kohnen}, {Koirala}, {Kolanoski}, {K{\"o}pke}, {Kopper}, {Kopper}, {Koschinsky}, {Koskinen}, {Kowalski}, {Krings}, {Kroll}, {Kr{\"u}ckl}, {Kunnen},
  {Kunwar}, {Kurahashi}, {Kuwabara}, {Kyriacou}, {Labare}, {Lanfranchi}, {Larson}, {Lauber}, {Lesiak-Bzdak}, {Leuermann}, {Liu}, {Lu}, {L{\"u}nemann}, {Luszczak}, {Madsen}, {Maggi}, {Mahn}, {Mancina}, {Maruyama}, {Mase}, {Maunu}, {McNally}, {Meagher}, {Medici}, {Meier}, {Menne}, {Merino}, {Meures}, {Miarecki}, {Micallef}, {Moment{\'e}}, {Montaruli}, {Moore}, {Moulai}, {Nahnhauer}, {Nakarmi}, {Naumann}, {Neer}, {Niederhausen}, {Nowicki}, {Nygren}, {Obertacke Pollmann}, {Olivas}, {O'Murchadha}, {Palczewski}, {Pandya}, {Pankova}, {Peiffer}, {Pepper}, {P{\'e}rez de los Heros}, {Pieloth}, {Pinat}, {Price}, {Przybylski}, {Raab}, {R{\"a}del}, {Rameez}, {Rawlins}, {Rea}, {Reimann}, {Relethford}, {Relich}, {Resconi}, {Rhode}, {Richman}, {Robertson}, {Rongen}, {Rott}, {Ruhe}, {Ryckbosch}, {Rysewyk}, {S{\"a}lzer}, {Sanchez Herrera}, {Sandrock}, {Sandroos}, {Santander}, {Sarkar}, {Sarkar}, {Satalecka}, {Schlunder}, {Schmidt}, {Schneider}, {Schoenen}, {Sch{\"o}neberg}, {Schumacher}, {Seckel}, {Seunarine}, {Soedingrekso},
  {Soldin}, {Song}, {Spiczak}, {Spiering}, {Stachurska}, {Stamatikos}, {Stanev}, {Stasik}, {Stettner}, {Steuer}, {Stezelberger}, {Stokstad}, {St{\"o}ssl}, {Strotjohann}, {Stuttard}, {Sullivan}, {Sutherland}, {Taboada}, {Tatar}, {Tenholt}, {Ter-Antonyan}, {Terliuk}, {Te{\v{s}}i{\'c}}, {Tilav}, {Toale}, {Tobin}, {Toscano}, {Tosi}, {Tselengidou}, {Tung}, {Turcati}, {Turley}, {Ty}, {Unger}, {Usner}, {Vandenbroucke}, {Van Driessche}, {van Eijndhoven}, {Vanheule}, {van Santen}, {Vehring}, {Vogel}, {Vraeghe}, {Walck}, {Wallace}, {Wallraff}, {Wandler}, {Wandkowsky}, {Waza}, {Weaver}, {Weiss}, {Wendt}, {Werthebach}, {Whelan}, {Wiebe}, {Wiebusch}, {Wille}, {Williams}, {Wills}, {Wolf}, {Wood}, {Woolsey}, {Woschnagg}, {Xu}, {Xu}, {Xu}, {Yanez}, {Yodh}, {Yoshida}, {Yuan}, {Zoll}, {IceCube Collaboration}, {Balasubramanian}, {Mate}, {Bhalerao}, {Bhattacharya}, {Vibhute}, {Dewangan}, {Rao}, {Vadawale}, {AstroSat Cadmium Zinc Telluride Imager Team}, {Svinkin}, {Hurley}, {Aptekar}, {Frederiks}, {Golenetskii}, {Kozlova},
  {Lysenko}, {Oleynik}, {Tsvetkova}, {Ulanov}, {Cline}, {IPN Collaboration}, {Li}, {Xiong}, {Zhang}, {Lu}, {Song}, {Cao}, {Chang}, {Chen}, {Chen}, {Chen}, {Chen}, {Chen}, {Chen}, {Cui}, {Cui}, {Deng}, {Dong}, {Du}, {Fu}, {Gao}, {Gao}, {Gao}, {Ge}, {Gu}, {Guan}, {Guo}, {Han}, {Hu}, {Huang}, {Huo}, {Jia}, {Jiang}, {Jiang}, {Jin}, {Jin}, {Li}, {Li}, {Li}, {Li}, {Li}, {Li}, {Li}, {Li}, {Li}, {Li}, {Li}, {Liang}, {Liao}, {Liu}, {Liu}, {Liu}, {Liu}, {Liu}, {Liu}, {Liu}, {Lu}, {Lu}, {Luo}, {Ma}, {Meng}, {Nang}, {Nie}, {Ou}, {Qu}, {Sai}, {Sun}, {Tan}, {Tao}, {Tao}, {Tuo}, {Wang}, {Wang}, {Wang}, {Wang}, {Wang}, {Wen}, {Wu}, {Wu}, {Xiao}, {Xu}, {Xu}, {Yan}, {Yang}, {Yang}, {Yang}, {Zhang}, {Zhang}, {Zhang}, {Zhang}, {Zhang}, {Zhang}, {Zhang}, {Zhang}, {Zhang}, {Zhang}, {Zhang}, {Zhang}, {Zhang}, {Zhang}, {Zhang}, {Zhang}, {Zhang}, {Zhang}, {Zhao}, {Zhao}, {Zhao}, {Zheng}, {Zhu}, {Zhu}, {Zou}, {Insight-HXMT Collaboration}, {Albert}, {Andr{\'e}}, {Anghinolfi}, {Ardid}, {Aubert}, {Aublin}, {Avgitas}, {Baret},
  {Barrios-Mart{\'\i}}, {Basa}, {Belhorma}, {Bertin}, {Biagi}, {Bormuth}, {Bourret}, {Bouwhuis}, {Br{\^a}nza{\c{s}}}, {Bruijn}, {Brunner}, {Busto}, {Capone}, {Caramete}, {Carr}, {Celli}, {Cherkaoui El Moursli}, {Chiarusi}, {Circella}, {Coelho}, {Coleiro}, {Coniglione}, {Costantini}, {Coyle}, {Creusot}, {D{\'\i}az}, {Deschamps}, {De Bonis}, {Distefano}, {Di Palma}, {Domi}, {Donzaud}, {Dornic}, {Drouhin}, {Eberl}, {El Bojaddaini}, {El Khayati}, {Els{\"a}sser}, {Enzenh{\"o}fer}, {Ettahiri}, {Fassi}, {Felis}, {Fusco}, {Gay}, {Giordano}, {Glotin}, {Gr{\'e}goire}, {Ruiz}, {Graf}, {Hallmann}, {van Haren}, {Heijboer}, {Hello}, {Hern{\'a}ndez-Rey}, {H{\"o}ssl}, {Hofest{\"a}dt}, {Hugon}, {Illuminati}, {James}, {de Jong}, {Jongen}, {Kadler}, {Kalekin}, {Katz}, {Kiessling}, {Kouchner}, {Kreter}, {Kreykenbohm}, {Kulikovskiy}, {Lachaud}, {Lahmann}, {Lef{\`e}vre}, {Leonora}, {Lotze}, {Loucatos}, {Marcelin}, {Margiotta}, {Marinelli}, {Mart{\'\i}nez-Mora}, {Mele}, {Melis}, {Michael}, {Migliozzi}, {Moussa}, {Navas}, {Nezri},
  {Organokov}, {P{\u{a}}v{\u{a}}la{\c{s}}}, {Pellegrino}, {Perrina}, {Piattelli}, {Popa}, {Pradier}, {Quinn}, {Racca}, {Riccobene}, {S{\'a}nchez-Losa}, {Salda{\~n}a}, {Salvadori}, {Samtleben}, {Sanguineti}, {Sapienza}, {Sieger}, {Spurio}, {Stolarczyk}, {Taiuti}, {Tayalati}, {Trovato}, {Turpin}, {T{\"o}nnis}, {Vallage}, {Van Elewyck}, {Versari}, {Vivolo}, {Vizzoca}, {Wilms}, {Zornoza}, {Z{\'u}{\~n}iga}, {ANTARES Collaboration}, {Beardmore}, {Breeveld}, {Burrows}, {Cenko}, {Cusumano}, {D'A{\`\i}}, {de Pasquale}, {Emery}, {Evans}, {Giommi}, {Gronwall}, {Kennea}, {Krimm}, {Kuin}, {Lien}, {Marshall}, {Melandri}, {Nousek}, {Oates}, {Osborne}, {Pagani}, {Page}, {Palmer}, {Perri}, {Siegel}, {Sbarufatti}, {Tagliaferri}, {Tohuvavohu}, {Swift Collaboration}, {Tavani}, {Verrecchia}, {Bulgarelli}, {Evangelista}, {Pacciani}, {Feroci}, {Pittori}, {Giuliani}, {Del Monte}, {Donnarumma}, {Argan}, {Trois}, {Ursi}, {Cardillo}, {Piano}, {Longo}, {Lucarelli}, {Munar-Adrover}, {Fuschino}, {Labanti}, {Marisaldi}, {Minervini},
  {Fioretti}, {Parmiggiani}, {Gianotti}, {Trifoglio}, {Di Persio}, {Antonelli}, {Barbiellini}, {Caraveo}, {Cattaneo}, {Costa}, {Colafrancesco}, {D'Amico}, {Ferrari}, {Morselli}, {Paoletti}, {Picozza}, {Pilia}, {Rappoldi}, {Soffitta}, {Vercellone}, {AGILE Team}, {Foley}, {Coulter}, {Kilpatrick}, {Drout}, {Piro}, {Shappee}, {Siebert}, {Simon}, {Ulloa}, {Kasen}, {Madore}, {Murguia-Berthier}, {Pan}, {Prochaska}, {Ramirez-Ruiz}, {Rest}, {Rojas-Bravo}, {1M2H Team}, {Berger}, {Soares-Santos}, {Annis}, {Alexander}, {Allam}, {Balbinot}, {Blanchard}, {Brout}, {Butler}, {Chornock}, {Cook}, {Cowperthwaite}, {Diehl}, {Drlica-Wagner}, {Drout}, {Durret}, {Eftekhari}, {Finley}, {Fong}, {Frieman}, {Fryer}, {Garc{\'\i}a-Bellido}, {Gruendl}, {Hartley}, {Herner}, {Kessler}, {Lin}, {Lopes}, {Louren{\c{c}}o}, {Margutti}, {Marshall}, {Matheson}, {Medina}, {Metzger}, {Mu{\~n}oz}, {Muir}, {Nicholl}, {Nugent}, {Palmese}, {Paz-Chinch{\'o}n}, {Quataert}, {Sako}, {Sauseda}, {Schlegel}, {Scolnic}, {Secco}, {Smith}, {Sobreira}, {Villar},
  {Vivas}, {Wester}, {Williams}, {Yanny}, {Zenteno}, {Zhang}, {Abbott}, {Banerji}, {Bechtol}, {Benoit-L{\'e}vy}, {Bertin}, {Brooks}, {Buckley-Geer}, {Burke}, {Capozzi}, {Carnero Rosell}, {Carrasco Kind}, {Castander}, {Crocce}, {Cunha}, {D'Andrea}, {da Costa}, {Davis}, {DePoy}, {Desai}, {Dietrich}, {Eifler}, {Fernandez}, {Flaugher}, {Fosalba}, {Gaztanaga}, {Gerdes}, {Giannantonio}, {Goldstein}, {Gruen}, {Gschwend}, {Gutierrez}, {Honscheid}, {James}, {Jeltema}, {Johnson}, {Johnson}, {Kent}, {Krause}, {Kron}, {Kuehn}, {Lahav}, {Lima}, {Maia}, {March}, {Martini}, {McMahon}, {Menanteau}, {Miller}, {Miquel}, {Mohr}, {Nichol}, {Ogando}, {Plazas}, {Romer}, {Roodman}, {Rykoff}, {Sanchez}, {Scarpine}, {Schindler}, {Schubnell}, {Sevilla-Noarbe}, {Sheldon}, {Smith}, {Smith}, {Stebbins}, {Suchyta}, {Swanson}, {Tarle}, {Thomas}, {Troxel}, {Tucker}, {Vikram}, {Walker}, {Wechsler}, {Weller}, {Carlin}, {Gill}, {Li}, {Marriner}, {Neilsen}, {Dark Energy Camera GW-EM Collaboration}, {DES Collaboration}, {Haislip}, {Kouprianov},
  {Reichart}, {Sand}, {Tartaglia}, {Valenti}, {Yang}, {DLT40 Collaboration}, {Benetti}, {Brocato}, {Campana}, {Cappellaro}, {Covino}, {D'Avanzo}, {D'Elia}, {Getman}, {Ghirlanda}, {Ghisellini}, {Limatola}, {Nicastro}, {Palazzi}, {Pian}, {Piranomonte}, {Possenti}, {Rossi}, {Salafia}, {Tomasella}, {Amati}, {Antonelli}, {Bernardini}, {Bufano}, {Capaccioli}, {Casella}, {Dadina}, {De Cesare}, {Di Paola}, {Giuffrida}, {Giunta}, {Israel}, {Lisi}, {Maiorano}, {Mapelli}, {Masetti}, {Pescalli}, {Pulone}, {Salvaterra}, {Schipani}, {Spera}, {Stamerra}, {Stella}, {Testa}, {Turatto}, {Vergani}, {Aresu}, {Bachetti}, {Buffa}, {Burgay}, {Buttu}, {Caria}, {Carretti}, {Casasola}, {Castangia}, {Carboni}, {Casu}, {Concu}, {Corongiu}, {Deiana}, {Egron}, {Fara}, {Gaudiomonte}, {Gusai}, {Ladu}, {Loru}, {Leurini}, {Marongiu}, {Melis}, {Melis}, {Migoni}, {Milia}, {Navarrini}, {Orlati}, {Ortu}, {Palmas}, {Pellizzoni}, {Perrodin}, {Pisanu}, {Poppi}, {Righini}, {Saba}, {Serra}, {Serrau}, {Stagni}, {Surcis}, {Vacca}, {Vargiu}, {Hunt},
  {Jin}, {Klose}, {Kouveliotou}, {Mazzali}, {M{\o}ller}, {Nava}, {Piran}, {Selsing}, {Vergani}, {Wiersema}, {Toma}, {Higgins}, {Mundell}, {di Serego Alighieri}, {G{\'o}tz}, {Gao}, {Gomboc}, {Kaper}, {Kobayashi}, {Kopac}, {Mao}, {Starling}, {Steele}, {van der Horst}, {GRAWITA: GRAvitational Wave Inaf TeAm}, {Acero}, {Atwood}, {Baldini}, {Barbiellini}, {Bastieri}, {Berenji}, {Bellazzini}, {Bissaldi}, {Blandford}, {Bloom}, {Bonino}, {Bottacini}, {Bregeon}, {Buehler}, {Buson}, {Cameron}, {Caputo}, {Caraveo}, {Cavazzuti}, {Chekhtman}, {Cheung}, {Chiang}, {Ciprini}, {Cohen-Tanugi}, {Cominsky}, {Costantin}, {Cuoco}, {D'Ammando}, {de Palma}, {Digel}, {Di Lalla}, {Di Mauro}, {Di Venere}, {Dubois}, {Fegan}, {Focke}, {Franckowiak}, {Fukazawa}, {Funk}, {Fusco}, {Gargano}, {Gasparrini}, {Giglietto}, {Giordano}, {Giroletti}, {Glanzman}, {Green}, {Grondin}, {Guillemot}, {Guiriec}, {Harding}, {Horan}, {J{\'o}hannesson}, {Kamae}, {Kensei}, {Kuss}, {La Mura}, {Latronico}, {Lemoine-Goumard}, {Longo}, {Loparco}, {Lovellette},
  {Lubrano}, {Magill}, {Maldera}, {Manfreda}, {Mazziotta}, {McEnery}, {Meyer}, {Michelson}, {Mirabal}, {Monzani}, {Moretti}, {Morselli}, {Moskalenko}, {Negro}, {Nuss}, {Ojha}, {Omodei}, {Orienti}, {Orlando}, {Palatiello}, {Paliya}, {Paneque}, {Pesce-Rollins}, {Piron}, {Porter}, {Principe}, {Rain{\`o}}, {Rando}, {Razzano}, {Razzaque}, {Reimer}, {Reimer}, {Reposeur}, {Rochester}, {Saz Parkinson}, {Sgr{\`o}}, {Siskind}, {Spada}, {Spandre}, {Suson}, {Takahashi}, {Tanaka}, {Thayer}, {Thayer}, {Thompson}, {Tibaldo}, {Torres}, {Torresi}, {Troja}, {Venters}, {Vianello}, {Zaharijas}, {Fermi Large Area Telescope Collaboration}, {Allison}, {Bannister}, {Dobie}, {Kaplan}, {Lenc}, {Lynch}, {Murphy}, {Sadler}, {Australia Telescope Compact Array}, {Hotan}, {James}, {Oslowski}, {Raja}, {Shannon}, {Whiting}, {Australian SKA Pathfinder}, {Arcavi}, {Howell}, {McCully}, {Hosseinzadeh}, {Hiramatsu}, {Poznanski}, {Barnes}, {Zaltzman}, {Vasylyev}, {Maoz}, {Las Cumbres Observatory Group}, {Cooke}, {Bailes}, {Wolf}, {Deller},
  {Lidman}, {Wang}, {Gendre}, {Andreoni}, {Ackley}, {Pritchard}, {Bessell}, {Chang}, {M{\"o}ller}, {Onken}, {Scalzo}, {Ridden-Harper}, {Sharp}, {Tucker}, {Farrell}, {Elmer}, {Johnston}, {Venkatraman Krishnan}, {Keane}, {Green}, {Jameson}, {Hu}, {Ma}, {Sun}, {Wu}, {Wang}, {Shang}, {Hu}, {Ashley}, {Yuan}, {Li}, {Tao}, {Zhu}, {Zhang}, {Suntzeff}, {Zhou}, {Yang}, {Orange}, {Morris}, {Cucchiara}, {Giblin}, {Klotz}, {Staff}, {Thierry}, {Schmidt}, {OzGrav}, {(Deeper}, {Wider}, {program}, {AST3}, {CAASTRO Collaborations}, {Tanvir}, {Levan}, {Cano}, {de Ugarte-Postigo}, {Gonz{\'a}lez-Fern{\'a}ndez}, {Greiner}, {Hjorth}, {Irwin}, {Kr{\"u}hler}, {Mandel}, {Milvang-Jensen}, {O'Brien}, {Rol}, {Rosetti}, {Rosswog}, {Rowlinson}, {Steeghs}, {Th{\"o}ne}, {Ulaczyk}, {Watson}, {Bruun}, {Cutter}, {Figuera Jaimes}, {Fujii}, {Fruchter}, {Gompertz}, {Jakobsson}, {Hodosan}, {J{\`e}rgensen}, {Kangas}, {Kann}, {Rabus}, {Schr{\o}der}, {Stanway}, {Wijers}, {VINROUGE Collaboration}, {Lipunov}, {Gorbovskoy}, {Kornilov}, {Tyurina},
  {Balanutsa}, {Kuznetsov}, {Vlasenko}, {Podesta}, {Lopez}, {Podesta}, {Levato}, {Saffe}, {Mallamaci}, {Budnev}, {Gress}, {Kuvshinov}, {Gorbunov}, {Vladimirov}, {Zimnukhov}, {Gabovich}, {Yurkov}, {Sergienko}, {Rebolo}, {Serra-Ricart}, {Tlatov}, {Ishmuhametova}, {MASTER Collaboration}, {Abe}, {Aoki}, {Aoki}, {Asakura}, {Baar}, {Barway}, {Bond}, {Doi}, {Finet}, {Fujiyoshi}, {Furusawa}, {Honda}, {Itoh}, {Kanda}, {Kawabata}, {Kawabata}, {Kim}, {Koshida}, {Kuroda}, {Lee}, {Liu}, {Matsubayashi}, {Miyazaki}, {Morihana}, {Morokuma}, {Motohara}, {Murata}, {Nagai}, {Nagashima}, {Nagayama}, {Nakaoka}, {Nakata}, {Ohsawa}, {Ohshima}, {Ohta}, {Okita}, {Saito}, {Saito}, {Sako}, {Sekiguchi}, {Sumi}, {Tajitsu}, {Takahashi}, {Takayama}, {Tamura}, {Tanaka}, {Tanaka}, {Terai}, {Tominaga}, {Tristram}, {Uemura}, {Utsumi}, {Yamaguchi}, {Yasuda}, {Yoshida}, {Zenko}, {J-GEM}, {Adams}, {Anupama}, {Bally}, {Barway}, {Bellm}, {Blagorodnova}, {Cannella}, {Chandra}, {Chatterjee}, {Clarke}, {Cobb}, {Cook}, {Copperwheat}, {De}, {Emery},
  {Feindt}, {Foster}, {Fox}, {Frail}, {Fremling}, {Frohmaier}, {Garcia}, {Ghosh}, {Giacintucci}, {Goobar}, {Gottlieb}, {Grefenstette}, {Hallinan}, {Harrison}, {Heida}, {Helou}, {Ho}, {Horesh}, {Hotokezaka}, {Ip}, {Itoh}, {Jacobs}, {Jencson}, {Kasen}, {Kasliwal}, {Kassim}, {Kim}, {Kiran}, {Kuin}, {Kulkarni}, {Kupfer}, {Lau}, {Madsen}, {Mazzali}, {Miller}, {Miyasaka}, {Mooley}, {Myers}, {Nakar}, {Ngeow}, {Nugent}, {Ofek}, {Palliyaguru}, {Pavana}, {Perley}, {Peters}, {Pike}, {Piran}, {Qi}, {Quimby}, {Rana}, {Rosswog}, {Rusu}, {Sadler}, {Van Sistine}, {Sollerman}, {Xu}, {Yan}, {Yatsu}, {Yu}, {Zhang}, {Zhao}, {GROWTH}, {JAGWAR}, {Caltech-NRAO}, {TTU-NRAO}, {NuSTAR Collaborations}, {Chambers}, {Huber}, {Schultz}, {Bulger}, {Flewelling}, {Magnier}, {Lowe}, {Wainscoat}, {Waters}, {Willman}, {Pan-STARRS}, {Ebisawa}, {Hanyu}, {Harita}, {Hashimoto}, {Hidaka}, {Hori}, {Ishikawa}, {Isobe}, {Iwakiri}, {Kawai}, {Kawai}, {Kawamuro}, {Kawase}, {Kitaoka}, {Makishima}, {Matsuoka}, {Mihara}, {Morita}, {Morita}, {Nakahira},
  {Nakajima}, {Nakamura}, {Negoro}, {Oda}, {Sakamaki}, {Sasaki}, {Serino}, {Shidatsu}, {Shimomukai}, {Sugawara}, {Sugita}, {Sugizaki}, {Tachibana}, {Takao}, {Tanimoto}, {Tomida}, {Tsuboi}, {Tsunemi}, {Ueda}, {Ueno}, {Yamada}, {Yamaoka}, {Yamauchi}, {Yatabe}, {Yoneyama}, {Yoshii}, {MAXI Team}, {Coward}, {Crisp}, {Macpherson}, {Andreoni}, {Laugier}, {Noysena}, {Klotz}, {Gendre}, {Thierry}, {Turpin}, {Consortium}, {Im}, {Choi}, {Kim}, {Yoon}, {Lim}, {Lee}, {Lee}, {Kim}, {Ko}, {Joe}, {Kwon}, {Kim}, {Lim}, {Choi}, {KU Collaboration}, {Fynbo}, {Malesani}, {Xu}, {Optical Telescope}, {Smartt}, {Jerkstrand}, {Kankare}, {Sim}, {Fraser}, {Inserra}, {Maguire}, {Leloudas}, {Magee}, {Shingles}, {Smith}, {Young}, {Kotak}, {Gal-Yam}, {Lyman}, {Homan}, {Agliozzo}, {Anderson}, {Angus}, {Ashall}, {Barbarino}, {Bauer}, {Berton}, {Botticella}, {Bulla}, {Cannizzaro}, {Cartier}, {Cikota}, {Clark}, {De Cia}, {Della Valle}, {Dennefeld}, {Dessart}, {Dimitriadis}, {Elias-Rosa}, {Firth}, {Fl{\"o}rs}, {Frohmaier}, {Galbany},
  {Gonz{\'a}lez-Gait{\'a}n}, {Gromadzki}, {Guti{\'e}rrez}, {Hamanowicz}, {Harmanen}, {Heintz}, {Hernandez}, {Hodgkin}, {Hook}, {Izzo}, {James}, {Jonker}, {Kerzendorf}, {Kostrzewa-Rutkowska}, {Kromer}, {Kuncarayakti}, {Lawrence}, {Manulis}, {Mattila}, {McBrien}, {M{\"u}ller}, {Nordin}, {O'Neill}, {Onori}, {Palmerio}, {Pastorello}, {Patat}, {Pignata}, {Podsiadlowski}, {Razza}, {Reynolds}, {Roy}, {Ruiter}, {Rybicki}, {Salmon}, {Pumo}, {Prentice}, {Seitenzahl}, {Smith}, {Sollerman}, {Sullivan}, {Szegedi}, {Taddia}, {Taubenberger}, {Terreran}, {Van Soelen}, {Vos}, {Walton}, {Wright}, {Wyrzykowski}, {Yaron}, {pre=''(''>ePESSTO}, {Chen}, {Kr{\"u}hler}, {Schady}, {Wiseman}, {Greiner}, {Rau}, {Schweyer}, {Klose}, {Nicuesa Guelbenzu}, {GROND}, {Palliyaguru}, {Tech University}, {Shara}, {Williams}, {Vaisanen}, {Potter}, {Romero Colmenero}, {Crawford}, {Buckley}, {Mao}, {SALT Group}, {D{\'\i}az}, {Macri}, {Garc{\'\i}a Lambas}, {Mendes de Oliveira}, {Nilo Castell{\'o}n}, {Ribeiro}, {S{\'a}nchez}, {Schoenell}, {Abramo},
  {Akras}, {Alcaniz}, {Artola}, {Beroiz}, {Bonoli}, {Cabral}, {Camuccio}, {Chavushyan}, {Coelho}, {Colazo}, {Costa-Duarte}, {Cuevas Larenas}, {Dom{\'\i}nguez Romero}, {Dultzin}, {Fern{\'a}ndez}, {Garc{\'\i}a}, {Girardini}, {Gon{\c{c}}alves}, {Gon{\c{c}}alves}, {Gurovich}, {Jim{\'e}nez-Teja}, {Kanaan}, {Lares}, {Lopes de Oliveira}, {L{\'o}pez-Cruz}, {Melia}, {Molino}, {Padilla}, {Pe{\~n}uela}, {Placco}, {Qui{\~n}ones}, {Ram{\'\i}rez Rivera}, {Renzi}, {Riguccini}, {R{\'\i}os-L{\'o}pez}, {Rodriguez}, {Sampedro}, {Schneiter}, {Sodr{\'e}}, {Starck}, {Torres-Flores}, {Tornatore}, {Zadro{\.z}ny}, {Castillo}, {TOROS: Transient Robotic Observatory of South Collaboration}, {Castro-Tirado}, {Tello}, {Hu}, {Zhang}, {Cunniffe}, {Castell{\'o}n}, {Hiriart}, {Caballero-Garc{\'\i}a}, {Jel{\'\i}nek}, {Kub{\'a}nek}, {P{\'e}rez del Pulgar}, {Park}, {Jeong}, {Castro Cer{\'o}n}, {Pandey}, {Yock}, {Querel}, {Fan}, {Wang}, {BOOTES Collaboration}, {Beardsley}, {Brown}, {Crosse}, {Emrich}, {Franzen}, {Gaensler}, {Horsley},
  {Johnston-Hollitt}, {Kenney}, {Morales}, {Pallot}, {Sokolowski}, {Steele}, {Tingay}, {Trott}, {Walker}, {Wayth}, {Williams}, {Wu}, {Murchison Widefield Array}, {Yoshida}, {Sakamoto}, {Kawakubo}, {Yamaoka}, {Takahashi}, {Asaoka}, {Ozawa}, {Torii}, {Shimizu}, {Tamura}, {Ishizaki}, {Cherry}, {Ricciarini}, {Penacchioni}, {Marrocchesi}, {CALET Collaboration}, {Pozanenko}, {Volnova}, {Mazaeva}, {Minaev}, {Krugov}, {Kusakin}, {Reva}, {Moskvitin}, {Rumyantsev}, {Inasaridze}, {Klunko}, {Tungalag}, {Schmalz}, {Burhonov}, {IKI-GW Follow-up Collaboration}, {Abdalla}, {Abramowski}, {Aharonian}, {Ait Benkhali}, {Ang{\"u}ner}, {Arakawa}, {Arrieta}, {Aubert}, {Backes}, {Balzer}, {Barnard}, {Becherini}, {Becker Tjus}, {Berge}, {Bernhard}, {Bernl{\"o}hr}, {Blackwell}, {B{\"o}ttcher}, {Boisson}, {Bolmont}, {Bonnefoy}, {Bordas}, {Bregeon}, {Brun}, {Brun}, {Bryan}, {B{\"u}chele}, {Bulik}, {Capasso}, {Caroff}, {Carosi}, {Casanova}, {Cerruti}, {Chakraborty}, {Chaves}, {Chen}, {Chevalier}, {Colafrancesco}, {Condon}, {Conrad},
  {Davids}, {Decock}, {Deil}, {Devin}, {deWilt}, {Dirson}, {Djannati-Ata{\"\i}}, {Donath}, {O'C. Drury}, {Dutson}, {Dyks}, {Edwards}, {Egberts}, {Emery}, {Ernenwein}, {Eschbach}, {Farnier}, {Fegan}, {Fernandes}, {Fiasson}, {Fontaine}, {Funk}, {F{\"u}ssling}, {Gabici}, {Gallant}, {Garrigoux}, {Gat{\'e}}, {Giavitto}, {Giebels}, {Glawion}, {Glicenstein}, {Gottschall}, {Grondin}, {Hahn}, {Haupt}, {Hawkes}, {Heinzelmann}, {Henri}, {Hermann}, {Hinton}, {Hofmann}, {Hoischen}, {Holch}, {Holler}, {Horns}, {Ivascenko}, {Iwasaki}, {Jacholkowska}, {Jamrozy}, {Jankowsky}, {Jankowsky}, {Jingo}, {Jouvin}, {Jung-Richardt}, {Kastendieck}, {Katarzy{\'n}ski}, {Katsuragawa}, {Kerszberg}, {Khangulyan}, {Kh{\'e}lifi}, {King}, {Klepser}, {Klochkov}, {Klu{\'z}niak}, {Komin}, {Kosack}, {Krakau}, {Kraus}, {Kr{\"u}ger}, {Laffon}, {Lamanna}, {Lau}, {Lees}, {Lefaucheur}, {Lemi{\`e}re}, {Lemoine-Goumard}, {Lenain}, {Leser}, {Lohse}, {Lorentz}, {Liu}, {Lypova}, {Malyshev}, {Marandon}, {Marcowith}, {Mariaud}, {Marx}, {Maurin}, {Maxted},
  {Mayer}, {Meintjes}, {Meyer}, {Mitchell}, {Moderski}, {Mohamed}, {Mohrmann}, {Mor{\r{a}}}, {Moulin}, {Murach}, {Nakashima}, {de Naurois}, {Ndiyavala}, {Niederwanger}, {Niemiec}, {Oakes}, {O'Brien}, {Odaka}, {Ohm}, {Ostrowski}, {Oya}, {Padovani}, {Panter}, {Parsons}, {Pekeur}, {Pelletier}, {Perennes}, {Petrucci}, {Peyaud}, {Piel}, {Pita}, {Poireau}, {Poon}, {Prokhorov}, {Prokoph}, {P{\"u}hlhofer}, {Punch}, {Quirrenbach}, {Raab}, {Rauth}, {Reimer}, {Reimer}, {Renaud}, {de los Reyes}, {Rieger}, {Rinchiuso}, {Romoli}, {Rowell}, {Rudak}, {Rulten}, {Sahakian}, {Saito}, {Sanchez}, {Santangelo}, {Sasaki}, {Schlickeiser}, {Sch{\"u}ssler}, {Schulz}, {Schwanke}, {Schwemmer}, {Seglar-Arroyo}, {Settimo}, {Seyffert}, {Shafi}, {Shilon}, {Shiningayamwe}, {Simoni}, {Sol}, {Spanier}, {Spir-Jacob}, {Stawarz}, {Steenkamp}, {Stegmann}, {Steppa}, {Sushch}, {Takahashi}, {Tavernet}, {Tavernier}, {Taylor}, {Terrier}, {Tibaldo}, {Tiziani}, {Tluczykont}, {Trichard}, {Tsirou}, {Tsuji}, {Tuffs}, {Uchiyama}, {van der Walt}, {van Eldik},
  {van Rensburg}, {van Soelen}, {Vasileiadis}, {Veh}, {Venter}, {Viana}, {Vincent}, {Vink}, {Voisin}, {V{\"o}lk}, {Vuillaume}, {Wadiasingh}, {Wagner}, {Wagner}, {Wagner}, {White}, {Wierzcholska}, {Willmann}, {W{\"o}rnlein}, {Wouters}, {Yang}, {Zaborov}, {Zacharias}, {Zanin}, {Zdziarski}, {Zech}, {Zefi}, {Ziegler}, {Zorn}, {{\.Z}ywucka}, {H.~E.~S.~S. Collaboration}, {Fender}, {Broderick}, {Rowlinson}, {Wijers}, {Stewart}, {ter Veen}, {Shulevski}, {LOFAR Collaboration}, {Kavic}, {Simonetti}, {League}, {Tsai}, {Obenberger}, {Nathaniel}, {Taylor}, {Dowell}, {Liebling}, {Estes}, {Lippert}, {Sharma}, {Vincent}, {Farella}, {Wavelength Array}, {Abeysekara}, {Albert}, {Alfaro}, {Alvarez}, {Arceo}, {Arteaga-Vel{\'a}zquez}, {Avila Rojas}, {Ayala Solares}, {Barber}, {Becerra Gonzalez}, {Becerril}, {Belmont-Moreno}, {BenZvi}, {Berley}, {Bernal}, {Braun}, {Brisbois}, {Caballero-Mora}, {Capistr{\'a}n}, {Carrami{\~n}ana}, {Casanova}, {Castillo}, {Cotti}, {Cotzomi}, {Couti{\~n}o de Le{\'o}n}, {De Le{\'o}n}, {De la Fuente},
  {Diaz Hernandez}, {Dichiara}, {Dingus}, {DuVernois}, {D{\'\i}az-V{\'e}lez}, {Ellsworth}, {Engel}, {Enr{\'\i}quez-Rivera}, {Fiorino}, {Fleischhack}, {Fraija}, {Garc{\'\i}a-Gonz{\'a}lez}, {Garfias}, {Gerhardt}, {Gonz{\~o}lez Mu{\~n}oz}, {Gonz{\'a}lez}, {Goodman}, {Hampel-Arias}, {Harding}, {Hernandez}, {Hernandez-Almada}, {Hona}, {H{\"u}ntemeyer}, {Iriarte}, {Jardin-Blicq}, {Joshi}, {Kaufmann}, {Kieda}, {Lara}, {Lauer}, {Lennarz}, {Le{\'o}n Vargas}, {Linnemann}, {Longinotti}, {Raya}, {Luna-Garc{\'\i}a}, {L{\'o}pez-Coto}, {Malone}, {Marinelli}, {Martinez}, {Martinez-Castellanos}, {Mart{\'\i}nez-Castro}, {Mart{\'\i}nez-Huerta}, {Matthews}, {Miranda-Romagnoli}, {Moreno}, {Mostaf{\'a}}, {Nellen}, {Newbold}, {Nisa}, {Noriega-Papaqui}, {Pelayo}, {Pretz}, {P{\'e}rez-P{\'e}rez}, {Ren}, {Rho}, {Rivi{\`e}re}, {Rosa-Gonz{\'a}lez}, {Rosenberg}, {Ruiz-Velasco}, {Salazar}, {Salesa Greus}, {Sandoval}, {Schneider}, {Schoorlemmer}, {Sinnis}, {Smith}, {Springer}, {Surajbali}, {Tibolla}, {Tollefson}, {Torres}, {Ukwatta},
  {Weisgarber}, {Westerhoff}, {Wisher}, {Wood}, {Yapici}, {Yodh}, {Younk}, {Zhou}, {{\'A}lvarez}, {HAWC Collaboration}, {Aab}, {Abreu}, {Aglietta}, {Albuquerque}, {Albury}, {Allekotte}, {Almela}, {Alvarez Castillo}, {Alvarez-Mu{\~n}iz}, {Anastasi}, {Anchordoqui}, {Andrada}, {Andringa}, {Aramo}, {Arsene}, {Asorey}, {Assis}, {Avila}, {Badescu}, {Balaceanu}, {Barbato}, {Barreira Luz}, {Becker}, {Bellido}, {Berat}, {Bertaina}, {Bertou}, {Biermann}, {Biteau}, {Blaess}, {Blanco}, {Blazek}, {Bleve}, {Boh{\'a}{\v{c}}ov{\'a}}, {Bonifazi}, {Borodai}, {Botti}, {Brack}, {Brancus}, {Bretz}, {Bridgeman}, {Briechle}, {Buchholz}, {Bueno}, {Buitink}, {Buscemi}, {Caballero-Mora}, {Caccianiga}, {Cancio}, {Canfora}, {Caruso}, {Castellina}, {Catalani}, {Cataldi}, {Cazon}, {Chavez}, {Chinellato}, {Chudoba}, {Clay}, {Cobos Cerutti}, {Colalillo}, {Coleman}, {Collica}, {Coluccia}, {Concei{\c{c}}{\~a}o}, {Consolati}, {Contreras}, {Cooper}, {Coutu}, {Covault}, {Cronin}, {D'Amico}, {Daniel}, {Dasso}, {Daumiller}, {Dawson}, {Day}, {de
  Almeida}, {de Jong}, {De Mauro}, {de Mello Neto}, {De Mitri}, {de Oliveira}, {de Souza}, {Debatin}, {Deligny}, {D{\'\i}az Castro}, {Diogo}, {Dobrigkeit}, {D'Olivo}, {Dorosti}, {Dos Anjos}, {Dova}, {Dundovic}, {Ebr}, {Engel}, {Erdmann}, {Erfani}, {Escobar}, {Espadanal}, {Etchegoyen}, {Falcke}, {Farmer}, {Farrar}, {Fauth}, {Fazzini}, {Feldbusch}, {Fenu}, {Fick}, {Figueira}, {Filip{\v{c}}i{\v{c}}}, {Freire}, {Fujii}, {Fuster}, {Ga{\"\i}or}, {Garc{\'\i}a}, {Gat{\'e}}, {Gemmeke}, {Gherghel-Lascu}, {Ghia}, {Giaccari}, {Giammarchi}, {Giller}, {G{\l}as}, {Glaser}, {Golup}, {G{\'o}mez Berisso}, {G{\'o}mez Vitale}, {Gonz{\'a}lez}, {Gorgi}, {Gottowik}, {Grillo}, {Grubb}, {Guarino}, {Guedes}, {Halliday}, {Hampel}, {Hansen}, {Harari}, {Harrison}, {Harvey}, {Haungs}, {Hebbeker}, {Heck}, {Heimann}, {Herve}, {Hill}, {Hojvat}, {Holt}, {Homola}, {H{\"o}randel}, {Horvath}, {Hrabovsk{\'y}}, {Huege}, {Hulsman}, {Insolia}, {Isar}, {Jandt}, {Johnsen}, {Josebachuili}, {Jurysek}, {K{\"a}{\"a}p{\"a}}, {Kampert}, {Keilhauer},
  {Kemmerich}, {Kemp}, {Kieckhafer}, {Klages}, {Kleifges}, {Kleinfeller}, {Krause}, {Krohm}, {Kuempel}, {Kukec Mezek}, {Kunka}, {Kuotb Awad}, {Lago}, {LaHurd}, {Lang}, {Lauscher}, {Legumina}, {Leigui de Oliveira}, {Letessier-Selvon}, {Lhenry-Yvon}, {Link}, {Lo Presti}, {Lopes}, {L{\'o}pez}, {L{\'o}pez Casado}, {Lorek}, {Luce}, {Lucero}, {Malacari}, {Mallamaci}, {Mandat}, {Mantsch}, {Mariazzi}, {Maris}, {Marsella}, {Martello}, {Martinez}, {Mart{\'\i}nez Bravo}, {Mas{\'\i}as Meza}, {Mathes}, {Mathys}, {Matthews}, {Matthiae}, {Mayotte}, {Mazur}, {Medina}, {Medina-Tanco}, {Melo}, {Menshikov}, {Merenda}, {Michal}, {Micheletti}, {Middendorf}, {Miramonti}, {Mitrica}, {Mockler}, {Mollerach}, {Montanet}, {Morello}, {Morlino}, {M{\"u}ller}, {M{\"u}ller}, {Muller}, {M{\"u}ller}, {Mussa}, {Naranjo}, {Nguyen}, {Niculescu-Oglinzanu}, {Niechciol}, {Niemietz}, {Niggemann}, {Nitz}, {Nosek}, {Novotny}, {No{\v{z}}ka}, {N{\'u}{\~n}ez}, {Oikonomou}, {Olinto}, {Palatka}, {Pallotta}, {Papenbreer}, {Parente}, {Parra}, {Paul},
  {Pech}, {Pedreira}, {P{\c{e}}kala}, {Pe{\~n}a-Rodriguez}, {Pereira}, {Perlin}, {Perrone}, {Peters}, {Petrera}, {Phuntsok}, {Pierog}, {Pimenta}, {Pirronello}, {Platino}, {Plum}, {Poh}, {Porowski}, {Prado}, {Privitera}, {Prouza}, {Quel}, {Querchfeld}, {Quinn}, {Ramos-Pollan}, {Rautenberg}, {Ravignani}, {Ridky}, {Riehn}, {Risse}, {Ristori}, {Rizi}, {Rodrigues de Carvalho}, {Rodriguez Fernandez}, {Rodriguez Rojo}, {Roncoroni}, {Roth}, {Roulet}, {Rovero}, {Ruehl}, {Saffi}, {Saftoiu}, {Salamida}, {Salazar}, {Saleh}, {Salina}, {S{\'a}nchez}, {Sanchez-Lucas}, {Santos}, {Santos}, {Sarazin}, {Sarmento}, {Sarmiento-Cano}, {Sato}, {Schauer}, {Scherini}, {Schieler}, {Schimp}, {Schmidt}, {Scholten}, {Schov{\'a}nek}, {Schr{\"o}der}, {Schr{\"o}der}, {Schulz}, {Schumacher}, {Sciutto}, {Segreto}, {Shadkam}, {Shellard}, {Sigl}, {Silli}, {{\v{S}}m{\'\i}da}, {Snow}, {Sommers}, {Sonntag}, {Soriano}, {Squartini}, {Stanca}, {Stani{\v{c}}}, {Stasielak}, {Stassi}, {Stolpovskiy}, {Strafella}, {Streich}, {Suarez}, {Suarez-Dur{\'a}n},
  {Sudholz}, {Suomij{\"a}rvi}, {Supanitsky}, {{\v{S}}up{\'\i}k}, {Swain}, {Szadkowski}, {Taboada}, {Taborda}, {Timmermans}, {Todero Peixoto}, {Tomankova}, {Tom{\'e}}, {Torralba Elipe}, {Travnicek}, {Trini}, {Tueros}, {Ulrich}, {Unger}, {Urban}, {Vald{\'e}s Galicia}, {Vali{\~n}o}, {Valore}, {van Aar}, {van Bodegom}, {van den Berg}, {van Vliet}, {Varela}, {Vargas C{\'a}rdenas}, {V{\'a}zquez}, {Veberi{\v{c}}}, {Ventura}, {Vergara Quispe}, {Verzi}, {Vicha}, {Villase{\~n}or}, {Vorobiov}, {Wahlberg}, {Wainberg}, {Walz}, {Watson}, {Weber}, {Weindl}, {Wiede{\'n}ski}, {Wiencke}, {Wilczy{\'n}ski}, {Wirtz}, {Wittkowski}, {Wundheiler}, {Yang}, {Yushkov}, {Zas}, {Zavrtanik}, {Zavrtanik}, {Zepeda}, {Zimmermann}, {Ziolkowski}, {Zong}, {Zuccarello}, {Pierre Auger Collaboration}, {Kim}, {Schulze}, {Bauer}, {Corral-Santana}, {de Gregorio-Monsalvo}, {Gonz{\'a}lez-L{\'o}pez}, {Hartmann}, {Ishwara-Chandra}, {Mart{\'\i}n}, {Mehner}, {Misra}, {Micha{\l}owski}, {Resmi}, {ALMA Collaboration}, {Paragi}, {Agudo}, {An}, {Beswick},
  {Casadio}, {Frey}, {Jonker}, {Kettenis}, {Marcote}, {Moldon}, {Szomoru}, {van Langevelde}, {Yang}, {Euro VLBI Team}, {Cwiek}, {Cwiok}, {Czyrkowski}, {Dabrowski}, {Kasprowicz}, {Mankiewicz}, {Nawrocki}, {Opiela}, {Piotrowski}, {Wrochna}, {Zaremba}, {{\.Z}arnecki}, {Pi of Sky Collaboration}, {Haggard}, {Nynka}, {Ruan}, {Chandra Team at McGill University}, {Bland}, {Booler}, {Devillepoix}, {de Gois}, {Hancock}, {Howie}, {Paxman}, {Sansom}, {Towner}, {Desert Fireball Network}, {Tonry}, {Coughlin}, {Stubbs}, {Denneau}, {Heinze}, {Stalder}, {Weiland}, {ATLAS}, {Eatough}, {Kramer}, {Kraus}, {Time Resolution Universe Survey}, {Troja}, {Piro}, {Becerra Gonz{\'a}lez}, {Butler}, {Fox}, {Khandrika}, {Kutyrev}, {Lee}, {Ricci}, {Ryan}, {S{\'a}nchez-Ram{\'\i}rez}, {Veilleux}, {Watson}, {Wieringa}, {Burgess}, {van Eerten}, {Fontes}, {Fryer}, {Korobkin}, {Wollaeger}, {RIMAS}, {RATIR}, {Camilo}, {Foley}, {Goedhart}, {Makhathini}, {Oozeer}, {Smirnov}, {Fender}, {Woudt}, \& {South Africa/MeerKAT}}]{2017ApJ...848L..12A}
---. 2017{\natexlab{b}}, \apjl, 848, L12, \dodoi{10.3847/2041-8213/aa91c9}

\bibitem[{{Abbott} {et~al.}(2018){Abbott}, {Abbott}, {Abbott}, {Acernese}, {Ackley}, {Adams}, {Adams}, {Addesso}, {Adhikari}, {Adya}, {Affeldt}, {Agarwal}, {Agathos}, {Agatsuma}, {Aggarwal}, {Aguiar}, {Aiello}, {Ain}, {Ajith}, {Allen}, {Allen}, {Allocca}, {Aloy}, {Altin}, {Amato}, {Ananyeva}, {Anderson}, {Anderson}, {Angelova}, {Antier}, {Appert}, {Arai}, {Araya}, {Areeda}, {Ar{\`e}ne}, {Arnaud}, {Arun}, {Ascenzi}, {Ashton}, {Ast}, {Aston}, {Astone}, {Atallah}, {Aubin}, {Aufmuth}, {Aulbert}, {AultONeal}, {Austin}, {Avila-Alvarez}, {Babak}, {Bacon}, {Badaracco}, {Bader}, {Bae}, {Baker}, {Baldaccini}, {Ballardin}, {Ballmer}, {Banagiri}, {Barayoga}, {Barclay}, {Barish}, {Barker}, {Barkett}, {Barnum}, {Barone}, {Barr}, {Barsotti}, {Barsuglia}, {Barta}, {Bartlett}, {Bartos}, {Bassiri}, {Basti}, {Batch}, {Bawaj}, {Bayley}, {Bazzan}, {B{\'e}csy}, {Beer}, {Bejger}, {Belahcene}, {Bell}, {Beniwal}, {Bensch}, {Berger}, {Bergmann}, {Bernuzzi}, {Bero}, {Berry}, {Bersanetti}, {Bertolini}, {Betzwieser}, {Bhandare},
  {Bilenko}, {Bilgili}, {Billingsley}, {Billman}, {Birch}, {Birney}, {Birnholtz}, {Biscans}, {Biscoveanu}, {Bisht}, {Bitossi}, {Bizouard}, {Blackburn}, {Blackman}, {Blair}, {Blair}, {Blair}, {Bloemen}, {Bock}, {Bode}, {Boer}, {Boetzel}, {Bogaert}, {Bohe}, {Bondu}, {Bonilla}, {Bonnand}, {Booker}, {Boom}, {Booth}, {Bork}, {Boschi}, {Bose}, {Bossie}, {Bossilkov}, {Bosveld}, {Bouffanais}, {Bozzi}, {Bradaschia}, {Brady}, {Bramley}, {Branchesi}, {Brau}, {Briant}, {Brighenti}, {Brillet}, {Brinkmann}, {Brisson}, {Brockill}, {Brooks}, {Brown}, {Brunett}, {Buchanan}, {Buikema}, {Bulik}, {Bulten}, {Buonanno}, {Buskulic}, {Buy}, {Byer}, {Cabero}, {Cadonati}, {Cagnoli}, {Cahillane}, {Calder{\'o}n Bustillo}, {Callister}, {Calloni}, {Camp}, {Canepa}, {Canizares}, {Cannon}, {Cao}, {Cao}, {Capano}, {Capocasa}, {Carbognani}, {Caride}, {Carney}, {Carullo}, {Casanueva Diaz}, {Casentini}, {Caudill}, {Cavagli{\`a}}, {Cavalier}, {Cavalieri}, {Cella}, {Cepeda}, {Cerd{\'a}-Dur{\'a}n}, {Cerretani}, {Cesarini}, {Chaibi}, {Chamberlin},
  {Chan}, {Chao}, {Charlton}, {Chase}, {Chassande-Mottin}, {Chatterjee}, {Chatziioannou}, {Cheeseboro}, {Chen}, {Chen}, {Chen}, {Cheng}, {Chia}, {Chincarini}, {Chiummo}, {Chmiel}, {Cho}, {Cho}, {Chow}, {Christensen}, {Chu}, {Chua}, {Chua}, {Chung}, {Chung}, {Ciani}, {Ciobanu}, {Ciolfi}, {Cipriano}, {Cirelli}, {Cirone}, {Clara}, {Clark}, {Clearwater}, {Cleva}, {Cocchieri}, {Coccia}, {Cohadon}, {Cohen}, {Colla}, {Collette}, {Collins}, {Cominsky}, {Constancio}, {Conti}, {Cooper}, {Corban}, {Corbitt}, {Cordero-Carri{\'o}n}, {Corley}, {Cornish}, {Corsi}, {Cortese}, {Costa}, {Cotesta}, {Coughlin}, {Coughlin}, {Coulon}, {Countryman}, {Couvares}, {Covas}, {Cowan}, {Coward}, {Cowart}, {Coyne}, {Coyne}, {Creighton}, {Creighton}, {Cripe}, {Crowder}, {Cullen}, {Cumming}, {Cunningham}, {Cuoco}, {Canton}, {D{\'a}lya}, {Danilishin}, {D'Antonio}, {Danzmann}, {Dasgupta}, {Da Silva Costa}, {Dattilo}, {Dave}, {Davier}, {Davis}, {Daw}, {Day}, {DeBra}, {Deenadayalan}, {Degallaix}, {De Laurentis}, {Del{\'e}glise}, {Del Pozzo},
  {Demos}, {Denker}, {Dent}, {De Pietri}, {Derby}, {Dergachev}, {De Rosa}, {De Rossi}, {DeSalvo}, {de Varona}, {Dhurandhar}, {D{\'\i}az}, {Dietrich}, {Di Fiore}, {Di Giovanni}, {Di Girolamo}, {Di Lieto}, {Ding}, {Di Pace}, {Di Palma}, {Di Renzo}, {Dmitriev}, {Doctor}, {Dolique}, {Donovan}, {Dooley}, {Doravari}, {Dorrington}, {Dovale {\'A}lvarez}, {Downes}, {Drago}, {Dreissigacker}, {Driggers}, {Du}, {Dupej}, {Dwyer}, {Easter}, {Edo}, {Edwards}, {Effler}, {Eggenstein}, {Ehrens}, {Eichholz}, {Eikenberry}, {Eisenmann}, {Eisenstein}, {Essick}, {Estelles}, {Estevez}, {Etienne}, {Etzel}, {Evans}, {Evans}, {Fafone}, {Fair}, {Fairhurst}, {Fan}, {Farinon}, {Farr}, {Farr}, {Fauchon-Jones}, {Favata}, {Fays}, {Fee}, {Fehrmann}, {Feicht}, {Fejer}, {Feng}, {Fernandez-Galiana}, {Ferrante}, {Ferreira}, {Ferrini}, {Fidecaro}, {Fiori}, {Fiorucci}, {Fishbach}, {Fisher}, {Fishner}, {Fitz-Axen}, {Flaminio}, {Fletcher}, {Fong}, {Font}, {Forsyth}, {Forsyth}, {Fournier}, {Frasca}, {Frasconi}, {Frei}, {Freise}, {Frey}, {Frey},
  {Fritschel}, {Frolov}, {Fulda}, {Fyffe}, {Gabbard}, {Gadre}, {Gaebel}, {Gair}, {Gammaitoni}, {Ganija}, {Gaonkar}, {Garcia}, {Garc{\'\i}a-Quir{\'o}s}, {Garufi}, {Gateley}, {Gaudio}, {Gaur}, {Gayathri}, {Gemme}, {Genin}, {Gennai}, {George}, {George}, {Gergely}, {Germain}, {Ghonge}, {Ghosh}, {Ghosh}, {Ghosh}, {Giacomazzo}, {Giaime}, {Giardina}, {Giazotto}, {Gill}, {Giordano}, {Glover}, {Goetz}, {Goetz}, {Goncharov}, {Gonz{\'a}lez}, {Gonzalez Castro}, {Gopakumar}, {Gorodetsky}, {Gossan}, {Gosselin}, {Gouaty}, {Grado}, {Graef}, {Granata}, {Grant}, {Gras}, {Gray}, {Greco}, {Green}, {Green}, {Gretarsson}, {Groot}, {Grote}, {Grunewald}, {Gruning}, {Guidi}, {Gulati}, {Guo}, {Gupta}, {Gupta}, {Gushwa}, {Gustafson}, {Gustafson}, {Halim}, {Hall}, {Hall}, {Hamilton}, {Hamilton}, {Hammond}, {Haney}, {Hanke}, {Hanks}, {Hanna}, {Hannam}, {Hannuksela}, {Hanson}, {Hardwick}, {Harms}, {Harry}, {Harry}, {Hart}, {Haster}, {Haughian}, {Healy}, {Heidmann}, {Heintze}, {Heitmann}, {Hello}, {Hemming}, {Hendry}, {Heng}, {Hennig},
  {Heptonstall}, {Hernandez}, {Heurs}, {Hild}, {Hinderer}, {Ho}, {Hoak}, {Hochheim}, {Hofman}, {Holland}, {Holt}, {Holz}, {Hopkins}, {Horst}, {Hough}, {Houston}, {Howell}, {Hreibi}, {Huerta}, {Huet}, {Hughey}, {Hulko}, {Husa}, {Huttner}, {Huynh-Dinh}, {Iess}, {Indik}, {Ingram}, {Inta}, {Intini}, {Irwin}, {Isa}, {Isac}, {Isi}, {Iyer}, {Izumi}, {Jacqmin}, {Jani}, {Jaranowski}, {Johnson}, {Johnson}, {Jones}, {Jones}, {Jonker}, {Ju}, {Junker}, {Kalaghatgi}, {Kalogera}, {Kamai}, {Kandhasamy}, {Kang}, {Kanner}, {Kapadia}, {Karki}, {Karvinen}, {Kasprzack}, {Katolik}, {Katsanevas}, {Katsavounidis}, {Katzman}, {Kaufer}, {Kawabe}, {Keerthana}, {K{\'e}f{\'e}lian}, {Keitel}, {Kemball}, {Kennedy}, {Key}, {Khalili}, {Khamesra}, {Khan}, {Khan}, {Khan}, {Khan}, {Khazanov}, {Kijbunchoo}, {Kim}, {Kim}, {Kim}, {Kim}, {Kim}, {Kim}, {King}, {King}, {Kinley-Hanlon}, {Kirchhoff}, {Kissel}, {Kleybolte}, {Klimenko}, {Knowles}, {Koch}, {Koehlenbeck}, {Koley}, {Kondrashov}, {Kontos}, {Korobko}, {Korth}, {Kowalska}, {Kozak},
  {Kr{\"a}mer}, {Kringel}, {Krishnan}, {Kr{\'o}lak}, {Kuehn}, {Kumar}, {Kumar}, {Kumar}, {Kuo}, {Kutynia}, {Kwang}, {Lackey}, {Lai}, {Landry}, {Landry}, {Lang}, {Lange}, {Lantz}, {Lanza}, {Lartaux-Vollard}, {Lasky}, {Laxen}, {Lazzarini}, {Lazzaro}, {Leaci}, {Leavey}, {Lee}, {Lee}, {Lee}, {Lee}, {Lee}, {Lehmann}, {Lenon}, {Leonardi}, {Leroy}, {Letendre}, {Levin}, {Li}, {Li}, {Li}, {Linker}, {Littenberg}, {Liu}, {Liu}, {Lo}, {Lockerbie}, {London}, {Longo}, {Lorenzini}, {Loriette}, {Lormand}, {Losurdo}, {Lough}, {Lousto}, {Lovelace}, {L{\"u}ck}, {Lumaca}, {Lundgren}, {Lynch}, {Ma}, {Macas}, {Macfoy}, {Machenschalk}, {MacInnis}, {Macleod}, {Maga{\~n}a Hernandez}, {Maga{\~n}a-Sandoval}, {Maga{\~n}a Zertuche}, {Magee}, {Majorana}, {Maksimovic}, {Man}, {Mandic}, {Mangano}, {Mansell}, {Manske}, {Mantovani}, {Marchesoni}, {Marion}, {M{\'a}rka}, {M{\'a}rka}, {Markakis}, {Markosyan}, {Markowitz}, {Maros}, {Marquina}, {Martelli}, {Martellini}, {Martin}, {Martin}, {Martynov}, {Mason}, {Massera}, {Masserot}, {Massinger},
  {Masso-Reid}, {Mastrogiovanni}, {Matas}, {Matichard}, {Matone}, {Mavalvala}, {Mazumder}, {McCann}, {McCarthy}, {McClelland}, {McCormick}, {McCuller}, {McGuire}, {McIver}, {McManus}, {McRae}, {McWilliams}, {Meacher}, {Meadors}, {Mehmet}, {Meidam}, {Mejuto-Villa}, {Melatos}, {Mendell}, {Mendoza-Gandara}, {Mercer}, {Mereni}, {Merilh}, {Merzougui}, {Meshkov}, {Messenger}, {Messick}, {Metzdorff}, {Meyers}, {Miao}, {Michel}, {Middleton}, {Mikhailov}, {Milano}, {Miller}, {Miller}, {Miller}, {Miller}, {Millhouse}, {Mills}, {Milovich-Goff}, {Minazzoli}, {Minenkov}, {Ming}, {Mishra}, {Mitra}, {Mitrofanov}, {Mitselmakher}, {Mittleman}, {Moffa}, {Mogushi}, {Mohan}, {Mohapatra}, {Montani}, {Moore}, {Moraru}, {Moreno}, {Morisaki}, {Mours}, {Mow-Lowry}, {Mueller}, {Muir}, {Mukherjee}, {Mukherjee}, {Mukherjee}, {Mukund}, {Mullavey}, {Munch}, {Mu{\~n}iz}, {Muratore}, {Murray}, {Nagar}, {Napier}, {Nardecchia}, {Naticchioni}, {Nayak}, {Neilson}, {Nelemans}, {Nelson}, {Nery}, {Neunzert}, {Nevin}, {Newport}, {Ng}, {Ng},
  {Nguyen}, {Nguyen}, {Nichols}, {Nielsen}, {Nissanke}, {Nitz}, {Nocera}, {Nolting}, {North}, {Nuttall}, {Obergaulinger}, {Oberling}, {O'Brien}, {O'Dea}, {Ogin}, {Oh}, {Oh}, {Ohme}, {Ohta}, {Okada}, {Oliver}, {Oppermann}, {Oram}, {O'Reilly}, {Ormiston}, {Ortega}, {O'Shaughnessy}, {Ossokine}, {Ottaway}, {Overmier}, {Owen}, {Pace}, {Pagano}, {Page}, {Page}, {Pai}, {Pai}, {Palamos}, {Palashov}, {Palomba}, {Pal-Singh}, {Pan}, {Pan}, {Pang}, {Pang}, {Pankow}, {Pannarale}, {Pant}, {Paoletti}, {Paoli}, {Papa}, {Parida}, {Parker}, {Pascucci}, {Pasqualetti}, {Passaquieti}, {Passuello}, {Patil}, {Patricelli}, {Pearlstone}, {Pedersen}, {Pedraza}, {Pedurand}, {Pekowsky}, {Pele}, {Penn}, {Perego}, {Perez}, {Perreca}, {Perri}, {Pfeiffer}, {Phelps}, {Phukon}, {Piccinni}, {Pichot}, {Piergiovanni}, {Pierro}, {Pillant}, {Pinard}, {Pinto}, {Pirello}, {Pitkin}, {Poggiani}, {Popolizio}, {Porter}, {Possenti}, {Post}, {Powell}, {Prasad}, {Pratt}, {Pratten}, {Predoi}, {Prestegard}, {Principe}, {Privitera}, {Prodi}, {Prokhorov},
  {Puncken}, {Punturo}, {Puppo}, {P{\"u}rrer}, {Qi}, {Quetschke}, {Quintero}, {Quitzow-James}, {Raab}, {Rabeling}, {Radkins}, {Raffai}, {Raja}, {Rajan}, {Rajbhandari}, {Rakhmanov}, {Ramirez}, {Ramos-Buades}, {Rana}, {Rapagnani}, {Raymond}, {Razzano}, {Read}, {Regimbau}, {Rei}, {Reid}, {Reitze}, {Ren}, {Ricci}, {Ricker}, {Riemenschneider}, {Riles}, {Rizzo}, {Robertson}, {Robie}, {Robinet}, {Robson}, {Rocchi}, {Rolland}, {Rollins}, {Roma}, {Romano}, {Romel}, {Romie}, {Rosi{\'n}ska}, {Ross}, {Rowan}, {R{\"u}diger}, {Ruggi}, {Rutins}, {Ryan}, {Sachdev}, {Sadecki}, {Sakellariadou}, {Salconi}, {Saleem}, {Salemi}, {Samajdar}, {Sammut}, {Sampson}, {Sanchez}, {Sanchez}, {Sanchis-Gual}, {Sandberg}, {Sanders}, {Sarin}, {Sassolas}, {Sathyaprakash}, {Saulson}, {Sauter}, {Savage}, {Sawadsky}, {Schale}, {Scheel}, {Scheuer}, {Schmidt}, {Schnabel}, {Schofield}, {Sch{\"o}nbeck}, {Schreiber}, {Schuette}, {Schulte}, {Schutz}, {Schwalbe}, {Scott}, {Scott}, {Seidel}, {Sellers}, {Sengupta}, {Sentenac}, {Sequino}, {Sergeev},
  {Setyawati}, {Shaddock}, {Shaffer}, {Shah}, {Shahriar}, {Shaner}, {Shao}, {Shapiro}, {Shawhan}, {Shen}, {Shoemaker}, {Shoemaker}, {Siellez}, {Siemens}, {Sieniawska}, {Sigg}, {Silva}, {Singer}, {Singh}, {Singhal}, {Sintes}, {Slagmolen}, {Slaven-Blair}, {Smith}, {Smith}, {Smith}, {Somala}, {Son}, {Sorazu}, {Sorrentino}, {Souradeep}, {Spencer}, {Srivastava}, {Staats}, {Steinke}, {Steinlechner}, {Steinlechner}, {Steinmeyer}, {Steltner}, {Stevenson}, {Stocks}, {Stone}, {Stops}, {Strain}, {Stratta}, {Strigin}, {Strunk}, {Sturani}, {Stuver}, {Summerscales}, {Sun}, {Sunil}, {Suresh}, {Sutton}, {Swinkels}, {Szczepa{\'n}czyk}, {Tacca}, {Tait}, {Talbot}, {Talukder}, {Tanner}, {T{\'a}pai}, {Taracchini}, {Tasson}, {Taylor}, {Taylor}, {Tewari}, {Theeg}, {Thies}, {Thomas}, {Thomas}, {Thomas}, {Thorne}, {Thrane}, {Tiwari}, {Tiwari}, {Tokmakov}, {Toland}, {Tonelli}, {Tornasi}, {Torres-Forn{\'e}}, {Torrie}, {T{\"o}yr{\"a}}, {Travasso}, {Traylor}, {Trinastic}, {Tringali}, {Trovato}, {Trozzo}, {Tsang}, {Tse}, {Tso}, {Tsuna},
  {Tsukada}, {Tuyenbayev}, {Ueno}, {Ugolini}, {Urban}, {Usman}, {Vahlbruch}, {Vajente}, {Valdes}, {van Bakel}, {van Beuzekom}, {van den Brand}, {Van Den Broeck}, {Vander-Hyde}, {van der Schaaf}, {van Heijningen}, {van Veggel}, {Vardaro}, {Varma}, {Vass}, {Vas{\'u}th}, {Vecchio}, {Vedovato}, {Veitch}, {Veitch}, {Venkateswara}, {Venugopalan}, {Verkindt}, {Vetrano}, {Vicer{\'e}}, {Viets}, {Vinciguerra}, {Vine}, {Vinet}, {Vitale}, {Vo}, {Vocca}, {Vorvick}, {Vyatchanin}, {Wade}, {Wade}, {Wade}, {Walet}, {Walker}, {Wallace}, {Walsh}, {Wang}, {Wang}, {Wang}, {Wang}, {Wang}, {Ward}, {Warner}, {Was}, {Watchi}, {Weaver}, {Wei}, {Weinert}, {Weinstein}, {Weiss}, {Wellmann}, {Wen}, {Wessel}, {We{\ss}els}, {Westerweck}, {Wette}, {Whelan}, {Whiting}, {Whittle}, {Wilken}, {Williams}, {Williams}, {Williamson}, {Willis}, {Willke}, {Wimmer}, {Winkler}, {Wipf}, {Wittel}, {Woan}, {Woehler}, {Wofford}, {Wong}, {Worden}, {Wright}, {Wu}, {Wysocki}, {Xiao}, {Yam}, {Yamamoto}, {Yancey}, {Yang}, {Yap}, {Yazback}, {Yu}, {Yu}, {Yvert},
  {Zadro{\.Z}ny}, {Zanolin}, {Zelenova}, {Zendri}, {Zevin}, {Zhang}, {Zhang}, {Zhang}, {Zhang}, {Zhang}, {Zhao}, {Zhou}, {Zhou}, {Zhu}, {Zhu}, {Zimmerman}, {Zlochower}, {Zucker}, {Zweizig}, {LIGO Scientific Collaboration}, \& {Virgo Collaboration}}]{2018PhRvL.121p1101A}
---. 2018, \prl, 121, 161101, \dodoi{10.1103/PhysRevLett.121.161101}

\bibitem[{{Abbott} {et~al.}(2020){Abbott}, {Abbott}, {Abbott}, {Abraham}, {Acernese}, {Ackley}, {Adams}, {Adhikari}, {Adya}, {Affeldt}, {Agathos}, {Agatsuma}, {Aggarwal}, {Aguiar}, {Aiello}, {Ain}, {Ajith}, {Allen}, {Allocca}, {Aloy}, {Altin}, {Amato}, {Anand}, {Ananyeva}, {Anderson}, {Anderson}, {Angelova}, {Antier}, {Appert}, {Arai}, {Araya}, {Areeda}, {Ar{\`e}ne}, {Arnaud}, {Aronson}, {Arun}, {Ascenzi}, {Ashton}, {Aston}, {Astone}, {Aubin}, {Aufmuth}, {AultONeal}, {Austin}, {Avendano}, {Avila-Alvarez}, {Babak}, {Bacon}, {Badaracco}, {Bader}, {Bae}, {Baird}, {Baker}, {Baldaccini}, {Ballardin}, {Ballmer}, {Bals}, {Banagiri}, {Barayoga}, {Barbieri}, {Barclay}, {Barish}, {Barker}, {Barkett}, {Barnum}, {Barone}, {Barr}, {Barsotti}, {Barsuglia}, {Barta}, {Bartlett}, {Bartos}, {Bassiri}, {Basti}, {Bawaj}, {Bayley}, {Baylor}, {Bazzan}, {B{\'e}csy}, {Bejger}, {Belahcene}, {Bell}, {Beniwal}, {Benjamin}, {Berger}, {Bergmann}, {Bernuzzi}, {Berry}, {Bersanetti}, {Bertolini}, {Betzwieser}, {Bhandare}, {Bidler}, {Biggs},
  {Bilenko}, {Bilgili}, {Billingsley}, {Birney}, {Birnholtz}, {Biscans}, {Bischi}, {Biscoveanu}, {Bisht}, {Bitossi}, {Bizouard}, {Blackburn}, {Blackman}, {Blair}, {Blair}, {Blair}, {Bloemen}, {Bobba}, {Bode}, {Boer}, {Boetzel}, {Bogaert}, {Bondu}, {Bonnand}, {Booker}, {Boom}, {Bork}, {Boschi}, {Bose}, {Bossilkov}, {Bosveld}, {Bouffanais}, {Bozzi}, {Bradaschia}, {Brady}, {Bramley}, {Branchesi}, {Brau}, {Breschi}, {Briant}, {Briggs}, {Brighenti}, {Brillet}, {Brinkmann}, {Brockill}, {Brooks}, {Brooks}, {Brown}, {Brunett}, {Buikema}, {Bulik}, {Bulten}, {Buonanno}, {Buskulic}, {Buy}, {Byer}, {Cabero}, {Cadonati}, {Cagnoli}, {Cahillane}, {Calder{\'o}n Bustillo}, {Callister}, {Calloni}, {Camp}, {Campbell}, {Canepa}, {Cannon}, {Cao}, {Cao}, {Carapella}, {Carbognani}, {Caride}, {Carney}, {Carullo}, {Casanueva Diaz}, {Casentini}, {Caudill}, {Cavagli{\`a}}, {Cavalier}, {Cavalieri}, {Cella}, {Cerd{\'a}-Dur{\'a}n}, {Cesarini}, {Chaibi}, {Chakravarti}, {Chamberlin}, {Chan}, {Chao}, {Charlton}, {Chase}, {Chassande-Mottin},
  {Chatterjee}, {Chaturvedi}, {Chatziioannou}, {Cheeseboro}, {Chen}, {Chen}, {Chen}, {Cheng}, {Cheong}, {Chia}, {Chiadini}, {Chincarini}, {Chiummo}, {Cho}, {Cho}, {Cho}, {Christensen}, {Chu}, {Chua}, {Chung}, {Chung}, {Ciani}, {Cie{\'s}lar}, {Ciobanu}, {Ciolfi}, {Cipriano}, {Cirone}, {Clara}, {Clark}, {Clearwater}, {Cleva}, {Coccia}, {Cohadon}, {Cohen}, {Colleoni}, {Collette}, {Collins}, {Colpi}, {Cominsky}, {Constancio}, {Conti}, {Cooper}, {Corban}, {Corbitt}, {Cordero-Carri{\'o}n}, {Corezzi}, {Corley}, {Cornish}, {Corre}, {Corsi}, {Cortese}, {Costa}, {Cotesta}, {Coughlin}, {Coughlin}, {Coulon}, {Countryman}, {Couvares}, {Covas}, {Cowan}, {Coward}, {Cowart}, {Coyne}, {Coyne}, {Creighton}, {Creighton}, {Cripe}, {Croquette}, {Crowder}, {Cullen}, {Cumming}, {Cunningham}, {Cuoco}, {Dal Canton}, {D{\'a}lya}, {D'Angelo}, {Danilishin}, {D'Antonio}, {Danzmann}, {Dasgupta}, {Da Silva Costa}, {Datrier}, {Dattilo}, {Dave}, {Davier}, {Davis}, {Daw}, {DeBra}, {Deenadayalan}, {Degallaix}, {De Laurentis}, {Del{\'e}glise},
  {De Lillo}, {Del Pozzo}, {DeMarchi}, {Demos}, {Dent}, {De Pietri}, {De Rosa}, {De Rossi}, {DeSalvo}, {de Varona}, {Dhurandhar}, {D{\'\i}az}, {Dietrich}, {Di Fiore}, {DiFronzo}, {Di Giorgio}, {Di Giovanni}, {Di Giovanni}, {Di Girolamo}, {Di Lieto}, {Ding}, {Di Pace}, {Di Palma}, {Di Renzo}, {Divakarla}, {Dmitriev}, {Doctor}, {Donovan}, {Dooley}, {Doravari}, {Dorrington}, {Downes}, {Drago}, {Driggers}, {Du}, {Ducoin}, {Dudi}, {Dupej}, {Durante}, {Dwyer}, {Easter}, {Eddolls}, {Edo}, {Effler}, {Ehrens}, {Eichholz}, {Eikenberry}, {Eisenmann}, {Eisenstein}, {Errico}, {Essick}, {Estelles}, {Estevez}, {Etienne}, {Etzel}, {Evans}, {Evans}, {Fafone}, {Fairhurst}, {Fan}, {Farinon}, {Farr}, {Farr}, {Fauchon-Jones}, {Favata}, {Fays}, {Fazio}, {Fee}, {Feicht}, {Fejer}, {Feng}, {Fernandez-Galiana}, {Ferrante}, {Ferreira}, {Ferreira}, {Fidecaro}, {Fiori}, {Fiorucci}, {Fishbach}, {Fisher}, {Fishner}, {Fittipaldi}, {Fitz-Axen}, {Fiumara}, {Flaminio}, {Fletcher}, {Floden}, {Flynn}, {Fong}, {Font}, {Forsyth}, {Fournier},
  {Vivanco}, {Frasca}, {Frasconi}, {Frei}, {Freise}, {Frey}, {Frey}, {Fritschel}, {Frolov}, {Fronz{\`e}}, {Fulda}, {Fyffe}, {Gabbard}, {Gadre}, {Gaebel}, {Gair}, {Gamba}, {Gammaitoni}, {Gaonkar}, {Garc{\'\i}a-Quir{\'o}s}, {Garufi}, {Gateley}, {Gaudio}, {Gaur}, {Gayathri}, {Gemme}, {Genin}, {Gennai}, {George}, {George}, {George}, {Gergely}, {Ghonge}, {Ghosh}, {Ghosh}, {Ghosh}, {Giacomazzo}, {Giaime}, {Giardina}, {Gibson}, {Gill}, {Glover}, {Gniesmer}, {Godwin}, {Goetz}, {Goetz}, {Goncharov}, {Gonz{\'a}lez}, {Castro}, {Gopakumar}, {Gossan}, {Gosselin}, {Gouaty}, {Grace}, {Grado}, {Granata}, {Grant}, {Gras}, {Grassia}, {Gray}, {Gray}, {Greco}, {Green}, {Green}, {Gretarsson}, {Grimaldi}, {Grimm}, {Groot}, {Grote}, {Grunewald}, {Gruning}, {Guidi}, {Gulati}, {Guo}, {Gupta}, {Gupta}, {Gupta}, {Gustafson}, {Gustafson}, {Haegel}, {Halim}, {Hall}, {Hall}, {Hamilton}, {Hammond}, {Haney}, {Hanke}, {Hanks}, {Hanna}, {Hannam}, {Hannuksela}, {Hansen}, {Hanson}, {Harder}, {Hardwick}, {Haris}, {Harms}, {Harry}, {Harry},
  {Hasskew}, {Haster}, {Haughian}, {Hayes}, {Healy}, {Heidmann}, {Heintze}, {Heitmann}, {Hellman}, {Hello}, {Hemming}, {Hendry}, {Heng}, {Hennig}, {Heurs}, {Hild}, {Hinderer}, {Ho}, {Hochheim}, {Hofman}, {Holgado}, {Holland}, {Holt}, {Holz}, {Hopkins}, {Horst}, {Hough}, {Howell}, {Hoy}, {Huang}, {H{\"u}bner}, {Huerta}, {Huet}, {Hughey}, {Hui}, {Husa}, {Huttner}, {Huynh-Dinh}, {Idzkowski}, {Iess}, {Inchauspe}, {Ingram}, {Inta}, {Intini}, {Irwin}, {Isa}, {Isac}, {Isi}, {Iyer}, {Jacqmin}, {Jadhav}, {Jani}, {Janthalur}, {Jaranowski}, {Jariwala}, {Jenkins}, {Jiang}, {Johnson}, {Johnson-McDaniel}, {Jones}, {Jones}, {Jones}, {Jones}, {Jonker}, {Ju}, {Junker}, {Kalaghatgi}, {Kalogera}, {Kamai}, {Kandhasamy}, {Kang}, {Kanner}, {Kapadia}, {Karki}, {Kashyap}, {Kasprzack}, {Kastaun}, {Katsanevas}, {Katsavounidis}, {Katzman}, {Kaufer}, {Kawabe}, {Keerthana}, {K{\'e}f{\'e}lian}, {Keitel}, {Kennedy}, {Key}, {Khalili}, {Khan}, {Khan}, {Khazanov}, {Khetan}, {Khursheed}, {Kijbunchoo}, {Kim}, {Kim}, {Kim}, {Kim}, {Kim}, {Kim},
  {Kimball}, {King}, {Kinley-Hanlon}, {Kirchhoff}, {Kissel}, {Kleybolte}, {Klika}, {Klimenko}, {Knowles}, {Koch}, {Koehlenbeck}, {Koekoek}, {Koley}, {Kondrashov}, {Kontos}, {Koper}, {Korobko}, {Korth}, {Kovalam}, {Kozak}, {Kr{\"a}mer}, {Kringel}, {Krishnendu}, {Kr{\'o}lak}, {Krupinski}, {Kuehn}, {Kumar}, {Kumar}, {Kumar}, {Kumar}, {Kuo}, {Kutynia}, {Kwang}, {Lackey}, {Laghi}, {Lai}, {Lam}, {Landry}, {Landry}, {Lane}, {Lang}, {Lange}, {Lantz}, {Lanza}, {Lartaux-Vollard}, {Lasky}, {Laxen}, {Lazzarini}, {Lazzaro}, {Leaci}, {Leavey}, {Lecoeuche}, {Lee}, {Lee}, {Lee}, {Lee}, {Lee}, {Lee}, {Lehmann}, {Lenon}, {Leroy}, {Letendre}, {Levin}, {Li}, {Li}, {Li}, {Li}, {Li}, {Lin}, {Linde}, {Linker}, {Littenberg}, {Liu}, {Liu}, {Llorens-Monteagudo}, {Lo}, {London}, {Longo}, {Lorenzini}, {Loriette}, {Lormand}, {Losurdo}, {Lough}, {Lousto}, {Lovelace}, {Lower}, {Lucaccioni}, {L{\"u}ck}, {Lumaca}, {Lundgren}, {Lynch}, {Ma}, {Macas}, {Macfoy}, {MacInnis}, {Macleod}, {Macquet}, {Maga{\~n}a Hernandez}, {Maga{\~n}a-Sandoval},
  {Magee}, {Majorana}, {Maksimovic}, {Malik}, {Man}, {Mandic}, {Mangano}, {Mansell}, {Manske}, {Mantovani}, {Mapelli}, {Marchesoni}, {Marion}, {M{\'a}rka}, {M{\'a}rka}, {Markakis}, {Markosyan}, {Markowitz}, {Maros}, {Marquina}, {Marsat}, {Martelli}, {Martin}, {Martin}, {Martinez}, {Martynov}, {Masalehdan}, {Mason}, {Massera}, {Masserot}, {Massinger}, {Masso-Reid}, {Mastrogiovanni}, {Matas}, {Matichard}, {Matone}, {Mavalvala}, {McCann}, {McCarthy}, {McClelland}, {McCormick}, {McCuller}, {McGuire}, {McIsaac}, {McIver}, {McManus}, {McRae}, {McWilliams}, {Meacher}, {Meadors}, {Mehmet}, {Mehta}, {Meidam}, {Mejuto Villa}, {Melatos}, {Mendell}, {Mercer}, {Mereni}, {Merfeld}, {Merilh}, {Merzougui}, {Meshkov}, {Messenger}, {Messick}, {Messina}, {Metzdorff}, {Meyers}, {Meylahn}, {Miani}, {Miao}, {Michel}, {Middleton}, {Milano}, {Miller}, {Millhouse}, {Mills}, {Milovich-Goff}, {Minazzoli}, {Minenkov}, {Mishkin}, {Mishra}, {Mistry}, {Mitra}, {Mitrofanov}, {Mitselmakher}, {Mittleman}, {Mo}, {Moffa}, {Mogushi},
  {Mohapatra}, {Molina-Ruiz}, {Mondin}, {Montani}, {Moore}, {Moraru}, {Morawski}, {Moreno}, {Morisaki}, {Mours}, {Mow-Lowry}, {Muciaccia}, {Mukherjee}, {Mukherjee}, {Mukherjee}, {Mukherjee}, {Mukund}, {Mullavey}, {Munch}, {Mu{\~n}iz}, {Muratore}, {Murray}, {Nagar}, {Nardecchia}, {Naticchioni}, {Nayak}, {Neil}, {Neilson}, {Nelemans}, {Nelson}, {Nery}, {Neunzert}, {Nevin}, {Ng}, {Ng}, {Nguyen}, {Nguyen}, {Nichols}, {Nichols}, {Nissanke}, {Nocera}, {North}, {Nuttall}, {Obergaulinger}, {Oberling}, {O'Brien}, {Oganesyan}, {Ogin}, {Oh}, {Oh}, {Ohme}, {Ohta}, {Okada}, {Oliver}, {Oppermann}, {Oram}, {O'Reilly}, {Ormiston}, {Ortega}, {O'Shaughnessy}, {Ossokine}, {Ottaway}, {Overmier}, {Owen}, {Pace}, {Pagano}, {Page}, {Pagliaroli}, {Pai}, {Pai}, {Palamos}, {Palashov}, {Palomba}, {Pan}, {Panda}, {Pang}, {Pankow}, {Pannarale}, {Pant}, {Paoletti}, {Paoli}, {Parida}, {Parker}, {Pascucci}, {Pasqualetti}, {Passaquieti}, {Passuello}, {Patil}, {Patricelli}, {Payne}, {Pearlstone}, {Pechsiri}, {Pedersen}, {Pedraza}, {Pedurand},
  {Pele}, {Penn}, {Perego}, {Perez}, {P{\'e}rigois}, {Perreca}, {Petermann}, {Pfeiffer}, {Phelps}, {Phukon}, {Piccinni}, {Pichot}, {Piergiovanni}, {Pierro}, {Pillant}, {Pinard}, {Pinto}, {Pirello}, {Pitkin}, {Plastino}, {Poggiani}, {Pong}, {Ponrathnam}, {Popolizio}, {Porter}, {Powell}, {Prajapati}, {Prasad}, {Prasai}, {Prasanna}, {Pratten}, {Prestegard}, {Principe}, {Prodi}, {Prokhorov}, {Punturo}, {Puppo}, {P{\"u}rrer}, {Qi}, {Quetschke}, {Quinonez}, {Raab}, {Raaijmakers}, {Radkins}, {Radulesco}, {Raffai}, {Raja}, {Rajan}, {Rajbhandari}, {Rakhmanov}, {Ramirez}, {Ramos-Buades}, {Rana}, {Rao}, {Rapagnani}, {Raymond}, {Razzano}, {Read}, {Regimbau}, {Rei}, {Reid}, {Reitze}, {Rettegno}, {Ricci}, {Richardson}, {Richardson}, {Ricker}, {Riemenschneider}, {Riles}, {Rizzo}, {Robertson}, {Robinet}, {Rocchi}, {Rolland}, {Rollins}, {Roma}, {Romanelli}, {Romano}, {Romel}, {Romie}, {Rose}, {Rose}, {Rose}, {Rosell}, {Rosi{\'n}ska}, {Rosofsky}, {Ross}, {Rowan}, {Roy}, {R{\"u}diger}, {Ruggi}, {Rutins}, {Ryan}, {Sachdev},
  {Sadecki}, {Sakellariadou}, {Salafia}, {Salconi}, {Saleem}, {Samajdar}, {Sammut}, {Sanchez}, {Sanchez}, {Sanchis-Gual}, {Sanders}, {Santiago}, {Santos}, {Sarin}, {Sassolas}, {Sathyaprakash}, {Sauter}, {Savage}, {Schale}, {Scheel}, {Scheuer}, {Schmidt}, {Schnabel}, {Schofield}, {Sch{\"o}nbeck}, {Schreiber}, {Schulte}, {Schutz}, {Scott}, {Scott}, {Seidel}, {Sellers}, {Sengupta}, {Sennett}, {Sentenac}, {Sequino}, {Sergeev}, {Setyawati}, {Shaddock}, {Shaffer}, {Shahriar}, {Shaner}, {Sharma}, {Sharma}, {Shawhan}, {Shen}, {Shink}, {Shoemaker}, {Shoemaker}, {Shukla}, {ShyamSundar}, {Siellez}, {Sieniawska}, {Sigg}, {Singer}, {Singh}, {Singh}, {Singhal}, {Sintes}, {Sitmukhambetov}, {Skliris}, {Slagmolen}, {Slaven-Blair}, {Smith}, {Smith}, {Somala}, {Son}, {Soni}, {Sorazu}, {Sorrentino}, {Souradeep}, {Sowell}, {Spencer}, {Spera}, {Srivastava}, {Srivastava}, {Staats}, {Stachie}, {Standke}, {Steer}, {Steinke}, {Steinlechner}, {Steinlechner}, {Steinmeyer}, {Stevenson}, {Stocks}, {Stone}, {Stops}, {Strain}, {Stratta},
  {Strigin}, {Strunk}, {Sturani}, {Stuver}, {Sudhir}, {Summerscales}, {Sun}, {Sunil}, {Sur}, {Suresh}, {Sutton}, {Swinkels}, {Szczepa{\'n}czyk}, {Tacca}, {Tait}, {Talbot}, {Tanner}, {Tao}, {T{\'a}pai}, {Tapia}, {Tasson}, {Taylor}, {Tenorio}, {Terkowski}, {Thomas}, {Thomas}, {Thondapu}, {Thorne}, {Thrane}, {Tiwari}, {Tiwari}, {Tiwari}, {Toland}, {Tonelli}, {Tornasi}, {Torres-Forn{\'e}}, {Torrie}, {T{\"o}yr{\"a}}, {Travasso}, {Traylor}, {Tringali}, {Tripathee}, {Trovato}, {Trozzo}, {Tsang}, {Tse}, {Tso}, {Tsukada}, {Tsuna}, {Tsutsui}, {Tuyenbayev}, {Ueno}, {Ugolini}, {Unnikrishnan}, {Urban}, {Usman}, {Vahlbruch}, {Vajente}, {Valdes}, {Valentini}, {van Bakel}, {van Beuzekom}, {van den Brand}, {Van Den Broeck}, {Vander-Hyde}, {van der Schaaf}, {VanHeijningen}, {van Veggel}, {Vardaro}, {Varma}, {Vass}, {Vas{\'u}th}, {Vecchio}, {Vedovato}, {Veitch}, {Veitch}, {Venkateswara}, {Venugopalan}, {Verkindt}, {Vetrano}, {Vicer{\'e}}, {Viets}, {Vinciguerra}, {Vine}, {Vinet}, {Vitale}, {Vo}, {Vocca}, {Vorvick}, {Vyatchanin},
  {Wade}, {Wade}, {Wade}, {Walet}, {Walker}, {Wallace}, {Walsh}, {Wang}, {Wang}, {Wang}, {Wang}, {Ward}, {Warden}, {Warner}, {Was}, {Watchi}, {Weaver}, {Wei}, {Weinert}, {Weinstein}, {Weiss}, {Wellmann}, {Wen}, {Wessel}, {We{\ss}els}, {Westhouse}, {Wette}, {Whelan}, {White}, {Whiting}, {Whittle}, {Wilken}, {Williams}, {Williamson}, {Willis}, {Willke}, {Winkler}, {Wipf}, {Wittel}, {Woan}, {Woehler}, {Wofford}, {Wright}, {Wu}, {Wysocki}, {Xiao}, {Xu}, {Yamamoto}, {Yancey}, {Yang}, {Yang}, {Yang}, {Yap}, {Yazback}, {Yeeles}, {Yu}, {Yu}, {Yuen}, {Zadro{\.z}ny}, {Zadro{\.z}ny}, {Zanolin}, {Zelenova}, {Zendri}, {Zevin}, {Zhang}, {Zhang}, {Zhang}, {Zhao}, {Zhao}, {Zhou}, {Zhou}, {Zhu}, {Zimmerman}, {Zucker}, \& {Zweizig}}]{2020ApJ...892L...3A}
---. 2020, \apjl, 892, L3, \dodoi{10.3847/2041-8213/ab75f5}

\bibitem[{{Banik} {et~al.}(2014){Banik}, {Hempel}, \& {Bandyopadhyay}}]{2014ApJS..214...22B}
{Banik}, S., {Hempel}, M., \& {Bandyopadhyay}, D. 2014, \apjs, 214, 22, \dodoi{10.1088/0067-0049/214/2/22}

\bibitem[{{Baumgarte} {et~al.}(2000){Baumgarte}, {Shapiro}, \& {Shibata}}]{2000ApJ...528L..29B}
{Baumgarte}, T.~W., {Shapiro}, S.~L., \& {Shibata}, M. 2000, \apjl, 528, L29, \dodoi{10.1086/312425}

\bibitem[{{Belczynski} {et~al.}(2016){Belczynski}, {Repetto}, {Holz}, {O'Shaughnessy}, {Bulik}, {Berti}, {Fryer}, \& {Dominik}}]{2016ApJ...819..108B}
{Belczynski}, K., {Repetto}, S., {Holz}, D.~E., {et~al.} 2016, \apj, 819, 108, \dodoi{10.3847/0004-637X/819/2/108}

\bibitem[{{Belczynski} {et~al.}(2018){Belczynski}, {Askar}, {Arca-Sedda}, {Chruslinska}, {Donnari}, {Giersz}, {Benacquista}, {Spurzem}, {Jin}, {Wiktorowicz}, \& {Belloni}}]{2018A&A...615A..91B}
{Belczynski}, K., {Askar}, A., {Arca-Sedda}, M., {et~al.} 2018, \aap, 615, A91, \dodoi{10.1051/0004-6361/201732428}

\bibitem[{{Bhattacharya} \& {van den Heuvel}(1991)}]{1991PhR...203....1B}
{Bhattacharya}, D., \& {van den Heuvel}, E.~P.~J. 1991, \physrep, 203, 1, \dodoi{10.1016/0370-1573(91)90064-S}

\bibitem[{{Bildsten} {et~al.}(1997){Bildsten}, {Chakrabarty}, {Chiu}, {Finger}, {Koh}, {Nelson}, {Prince}, {Rubin}, {Scott}, {Stollberg}, {Vaughan}, {Wilson}, \& {Wilson}}]{1997ApJS..113..367B}
{Bildsten}, L., {Chakrabarty}, D., {Chiu}, J., {et~al.} 1997, \apjs, 113, 367, \dodoi{10.1086/313060}

\bibitem[{{Boldin} \& {Popov}(2010)}]{2010MNRAS.407.1090B}
{Boldin}, P.~A., \& {Popov}, S.~B. 2010, \mnras, 407, 1090, \dodoi{10.1111/j.1365-2966.2010.16910.x}

\bibitem[{{Bondi}(1952)}]{1952MNRAS.112..195B}
{Bondi}, H. 1952, \mnras, 112, 195, \dodoi{10.1093/mnras/112.2.195}

\bibitem[{{Bozzo} {et~al.}(2008){Bozzo}, {Falanga}, \& {Stella}}]{2008ApJ...683.1031B}
{Bozzo}, E., {Falanga}, M., \& {Stella}, L. 2008, \apj, 683, 1031, \dodoi{10.1086/589990}

\bibitem[{{Cameron} \& {Mock}(1967)}]{1967Natur.215..464C}
{Cameron}, A.~G.~W., \& {Mock}, M. 1967, \nat, 215, 464, \dodoi{10.1038/215464a0}

\bibitem[{{Chattopadhyay} {et~al.}(2020){Chattopadhyay}, {Stevenson}, {Hurley}, {Rossi}, \& {Flynn}}]{2020MNRAS.494.1587C}
{Chattopadhyay}, D., {Stevenson}, S., {Hurley}, J.~R., {Rossi}, L.~J., \& {Flynn}, C. 2020, \mnras, 494, 1587, \dodoi{10.1093/mnras/staa756}

\bibitem[{{Chruslinska} {et~al.}(2018){Chruslinska}, {Belczynski}, {Klencki}, \& {Benacquista}}]{2018MNRAS.474.2937C}
{Chruslinska}, M., {Belczynski}, K., {Klencki}, J., \& {Benacquista}, M. 2018, \mnras, 474, 2937, \dodoi{10.1093/mnras/stx2923}

\bibitem[{{Chu} {et~al.}(2022){Chu}, {Yu}, \& {Lu}}]{2022MNRAS.509.1557C}
{Chu}, Q., {Yu}, S., \& {Lu}, Y. 2022, \mnras, 509, 1557, \dodoi{10.1093/mnras/stab2882}

\bibitem[{{Claeys} {et~al.}(2014){Claeys}, {Pols}, {Izzard}, {Vink}, \& {Verbunt}}]{2014A&A...563A..83C}
{Claeys}, J.~S.~W., {Pols}, O.~R., {Izzard}, R.~G., {Vink}, J., \& {Verbunt}, F.~W.~M. 2014, \aap, 563, A83, \dodoi{10.1051/0004-6361/201322714}

\bibitem[{{Cook} {et~al.}(1992){Cook}, {Shapiro}, \& {Teukolsky}}]{1992ApJ...398..203C}
{Cook}, G.~B., {Shapiro}, S.~L., \& {Teukolsky}, S.~A. 1992, \apj, 398, 203, \dodoi{10.1086/171849}

\bibitem[{{Cook} {et~al.}(1994){Cook}, {Shapiro}, \& {Teukolsky}}]{1994ApJ...424..823C}
---. 1994, \apj, 424, 823, \dodoi{10.1086/173934}

\bibitem[{{Corbet}(1984)}]{1984A&A...141...91C}
{Corbet}, R.~H.~D. 1984, \aap, 141, 91

\bibitem[{{Corbet}(1986)}]{1986MNRAS.220.1047C}
---. 1986, \mnras, 220, 1047, \dodoi{10.1093/mnras/220.4.1047}

\bibitem[{{Davies} \& {Pringle}(1981)}]{1981MNRAS.196..209D}
{Davies}, R.~E., \& {Pringle}, J.~E. 1981, \mnras, 196, 209, \dodoi{10.1093/mnras/196.2.209}

\bibitem[{{De Marco} {et~al.}(2011){De Marco}, {Passy}, {Moe}, {Herwig}, {Mac Low}, \& {Paxton}}]{2011MNRAS.411.2277D}
{De Marco}, O., {Passy}, J.-C., {Moe}, M., {et~al.} 2011, \mnras, 411, 2277, \dodoi{10.1111/j.1365-2966.2010.17891.x}

\bibitem[{{Deng} {et~al.}(2024){Deng}, {Li}, {Shao}, \& {Xu}}]{2024arXiv240204658D}
{Deng}, Z.-L., {Li}, X.-D., {Shao}, Y., \& {Xu}, K. 2024, arXiv e-prints, arXiv:2402.04658, \dodoi{10.48550/arXiv.2402.04658}

\bibitem[{{Dewdney} {et~al.}(2009){Dewdney}, {Hall}, {Schilizzi}, \& {Lazio}}]{2009IEEEP..97.1482D}
{Dewdney}, P.~E., {Hall}, P.~J., {Schilizzi}, R.~T., \& {Lazio}, T.~J.~L.~W. 2009, IEEE Proceedings, 97, 1482, \dodoi{10.1109/JPROC.2009.2021005}

\bibitem[{{Di Stefano} {et~al.}(2023){Di Stefano}, {Kruckow}, {Gao}, {Neunteufel}, \& {Kobayashi}}]{2023ApJ...944...87D}
{Di Stefano}, R., {Kruckow}, M.~U., {Gao}, Y., {Neunteufel}, P.~G., \& {Kobayashi}, C. 2023, \apj, 944, 87, \dodoi{10.3847/1538-4357/acae9b}

\bibitem[{{Dominik} {et~al.}(2012){Dominik}, {Belczynski}, {Fryer}, {Holz}, {Berti}, {Bulik}, {Mandel}, \& {O'Shaughnessy}}]{2012ApJ...759...52D}
{Dominik}, M., {Belczynski}, K., {Fryer}, C., {et~al.} 2012, \apj, 759, 52, \dodoi{10.1088/0004-637X/759/1/52}

\bibitem[{{Dominik} {et~al.}(2013){Dominik}, {Belczynski}, {Fryer}, {Holz}, {Berti}, {Bulik}, {Mandel}, \& {O'Shaughnessy}}]{2013ApJ...779...72D}
---. 2013, \apj, 779, 72, \dodoi{10.1088/0004-637X/779/1/72}

\bibitem[{{Douchin} \& {Haensel}(2001)}]{2001A&A...380..151D}
{Douchin}, F., \& {Haensel}, P. 2001, \aap, 380, 151, \dodoi{10.1051/0004-6361:20011402}

\bibitem[{{Eggleton}(1983)}]{1983ApJ...268..368E}
{Eggleton}, P.~P. 1983, \apj, 268, 368, \dodoi{10.1086/160960}

\bibitem[{{Eggleton} {et~al.}(1989){Eggleton}, {Fitchett}, \& {Tout}}]{1989ApJ...347..998E}
{Eggleton}, P.~P., {Fitchett}, M.~J., \& {Tout}, C.~A. 1989, \apj, 347, 998, \dodoi{10.1086/168190}

\bibitem[{{Elsner} \& {Lamb}(1984)}]{1984ApJ...278..326E}
{Elsner}, R.~F., \& {Lamb}, F.~K. 1984, \apj, 278, 326, \dodoi{10.1086/161797}

\bibitem[{{Esposito} {et~al.}(2021){Esposito}, {Rea}, \& {Israel}}]{2021ASSL..461...97E}
{Esposito}, P., {Rea}, N., \& {Israel}, G.~L. 2021, in Astrophysics and Space Science Library, Vol. 461, Timing Neutron Stars: Pulsations, Oscillations and Explosions, ed. T.~M. {Belloni}, M.~{M{\'e}ndez}, \& C.~{Zhang}, 97--142, \dodoi{10.1007/978-3-662-62110-3_3}

\bibitem[{{Farrow} {et~al.}(2019){Farrow}, {Zhu}, \& {Thrane}}]{2019ApJ...876...18F}
{Farrow}, N., {Zhu}, X.-J., \& {Thrane}, E. 2019, \apj, 876, 18, \dodoi{10.3847/1538-4357/ab12e3}

\bibitem[{{Faucher-Gigu{\`e}re} \& {Kaspi}(2006)}]{2006ApJ...643..332F}
{Faucher-Gigu{\`e}re}, C.-A., \& {Kaspi}, V.~M. 2006, \apj, 643, 332, \dodoi{10.1086/501516}

\bibitem[{{Ferreras} {et~al.}(2003){Ferreras}, {Wyse}, \& {Silk}}]{2003MNRAS.345.1381F}
{Ferreras}, I., {Wyse}, R. F.~G., \& {Silk}, J. 2003, \mnras, 345, 1381, \dodoi{10.1046/j.1365-2966.2003.07056.x}

\bibitem[{{Fryer} {et~al.}(2012){Fryer}, {Belczynski}, {Wiktorowicz}, {Dominik}, {Kalogera}, \& {Holz}}]{2012ApJ...749...91F}
{Fryer}, C.~L., {Belczynski}, K., {Wiktorowicz}, G., {et~al.} 2012, \apj, 749, 91, \dodoi{10.1088/0004-637X/749/1/91}

\bibitem[{{Galaudage} {et~al.}(2021){Galaudage}, {Adamcewicz}, {Zhu}, {Stevenson}, \& {Thrane}}]{2021ApJ...909L..19G}
{Galaudage}, S., {Adamcewicz}, C., {Zhu}, X.-J., {Stevenson}, S., \& {Thrane}, E. 2021, \apjl, 909, L19, \dodoi{10.3847/2041-8213/abe7f6}

\bibitem[{{Ghosh}(2007)}]{2007rapp.book.....G}
{Ghosh}, P. 2007, {Rotation and Accretion Powered Pulsars}, Vol.~10, \dodoi{10.1142/4806}

\bibitem[{{Giacobbo} \& {Mapelli}(2018)}]{2018MNRAS.480.2011G}
{Giacobbo}, N., \& {Mapelli}, M. 2018, \mnras, 480, 2011, \dodoi{10.1093/mnras/sty1999}

\bibitem[{{Han}(1998)}]{1998MNRAS.296.1019H}
{Han}, Z. 1998, \mnras, 296, 1019, \dodoi{10.1046/j.1365-8711.1998.01475.x}

\bibitem[{{Han} {et~al.}(2003){Han}, {Podsiadlowski}, {Maxted}, \& {Marsh}}]{2003MNRAS.341..669H}
{Han}, Z., {Podsiadlowski}, P., {Maxted}, P.~F.~L., \& {Marsh}, T.~R. 2003, \mnras, 341, 669, \dodoi{10.1046/j.1365-8711.2003.06451.x}

\bibitem[{{Harding} \& {Leventhal}(1992)}]{1992Natur.357..388H}
{Harding}, A.~K., \& {Leventhal}, M. 1992, \nat, 357, 388, \dodoi{10.1038/357388a0}

\bibitem[{{Hirai} \& {Mandel}(2022)}]{2022ApJ...937L..42H}
{Hirai}, R., \& {Mandel}, I. 2022, \apjl, 937, L42, \dodoi{10.3847/2041-8213/ac9519}

\bibitem[{{Hobbs} {et~al.}(2005){Hobbs}, {Lorimer}, {Lyne}, \& {Kramer}}]{2005MNRAS.360..974H}
{Hobbs}, G., {Lorimer}, D.~R., {Lyne}, A.~G., \& {Kramer}, M. 2005, \mnras, 360, 974, \dodoi{10.1111/j.1365-2966.2005.09087.x}

\bibitem[{{Hoyle} \& {Lyttleton}(1939)}]{1939PCPS...35..405H}
{Hoyle}, F., \& {Lyttleton}, R.~A. 1939, Proceedings of the Cambridge Philosophical Society, 35, 405, \dodoi{10.1017/S0305004100021150}

\bibitem[{{Huang} {et~al.}(2018){Huang}, {Jiang}, {Li}, {Jin}, {Fan}, \& {Wei}}]{2018arXiv180403101H}
{Huang}, Y.-J., {Jiang}, J.-L., {Li}, X., {et~al.} 2018, arXiv e-prints, arXiv:1804.03101.
\newblock \doarXiv{1804.03101}

\bibitem[{{Hurley} {et~al.}(2000){Hurley}, {Pols}, \& {Tout}}]{2000MNRAS.315..543H}
{Hurley}, J.~R., {Pols}, O.~R., \& {Tout}, C.~A. 2000, \mnras, 315, 543, \dodoi{10.1046/j.1365-8711.2000.03426.x}

\bibitem[{{Hurley} {et~al.}(2002){Hurley}, {Tout}, \& {Pols}}]{2002MNRAS.329..897H}
{Hurley}, J.~R., {Tout}, C.~A., \& {Pols}, O.~R. 2002, \mnras, 329, 897, \dodoi{10.1046/j.1365-8711.2002.05038.x}

\bibitem[{{Igoshev} \& {Popov}(2013)}]{2013MNRAS.432..967I}
{Igoshev}, A.~P., \& {Popov}, S.~B. 2013, \mnras, 432, 967, \dodoi{10.1093/mnras/stt519}

\bibitem[{{Ikhsanov}(2001)}]{2001A&A...368L...5I}
{Ikhsanov}, N.~R. 2001, \aap, 368, L5, \dodoi{10.1051/0004-6361:20010140}

\bibitem[{{Illarionov} \& {Sunyaev}(1975)}]{1975A&A....39..185I}
{Illarionov}, A.~F., \& {Sunyaev}, R.~A. 1975, \aap, 39, 185

\bibitem[{{Iorio} {et~al.}(2023){Iorio}, {Mapelli}, {Costa}, {Spera}, {Escobar}, {Sgalletta}, {Trani}, {Korb}, {Santoliquido}, {Dall'Amico}, {Gaspari}, \& {Bressan}}]{2023MNRAS.524..426I}
{Iorio}, G., {Mapelli}, M., {Costa}, G., {et~al.} 2023, \mnras, 524, 426, \dodoi{10.1093/mnras/stad1630}

\bibitem[{{Kalogera} {et~al.}(2019){Kalogera}, {Berry}, {Colpi}, {Fairhurst}, {Justham}, {Mandel}, {Mangiagli}, {Mapelli}, {Mills}, {Sathyaprakash}, {Schneider}, {Tauris}, \& {Valiante}}]{2019BAAS...51c.242K}
{Kalogera}, V., {Berry}, C. P.~L., {Colpi}, M., {et~al.} 2019, \baas, 51, 242.
\newblock \doarXiv{1903.09220}

\bibitem[{{Karino}(2020)}]{2020PASJ...72...95K}
{Karino}, S. 2020, \pasj, 72, 95, \dodoi{10.1093/pasj/psaa087}

\bibitem[{{Keitel}(2019)}]{2019MNRAS.485.1665K}
{Keitel}, D. 2019, \mnras, 485, 1665, \dodoi{10.1093/mnras/stz358}

\bibitem[{{Kiziltan} {et~al.}(2013){Kiziltan}, {Kottas}, {De Yoreo}, \& {Thorsett}}]{2013ApJ...778...66K}
{Kiziltan}, B., {Kottas}, A., {De Yoreo}, M., \& {Thorsett}, S.~E. 2013, \apj, 778, 66, \dodoi{10.1088/0004-637X/778/1/66}

\bibitem[{{Kroupa}(2001)}]{2001MNRAS.322..231K}
{Kroupa}, P. 2001, \mnras, 322, 231, \dodoi{10.1046/j.1365-8711.2001.04022.x}

\bibitem[{{Kruckow}(2020)}]{2020A&A...639A.123K}
{Kruckow}, M.~U. 2020, \aap, 639, A123, \dodoi{10.1051/0004-6361/202037519}

\bibitem[{{Kruckow} {et~al.}(2018){Kruckow}, {Tauris}, {Langer}, {Kramer}, \& {Izzard}}]{2018MNRAS.481.1908K}
{Kruckow}, M.~U., {Tauris}, T.~M., {Langer}, N., {Kramer}, M., \& {Izzard}, R.~G. 2018, \mnras, 481, 1908, \dodoi{10.1093/mnras/sty2190}

\bibitem[{{Li} {et~al.}(2016){Li}, {Shao}, \& {Li}}]{2016ApJ...824..143L}
{Li}, T., {Shao}, Y., \& {Li}, X.-D. 2016, \apj, 824, 143, \dodoi{10.3847/0004-637X/824/2/143}

\bibitem[{{Licquia} \& {Newman}(2015)}]{2015ApJ...806...96L}
{Licquia}, T.~C., \& {Newman}, J.~A. 2015, \apj, 806, 96, \dodoi{10.1088/0004-637X/806/1/96}

\bibitem[{{Lipunov}(1992)}]{1992ans..book.....L}
{Lipunov}, V.~M. 1992, {Astrophysics of Neutron Stars}

\bibitem[{{Lorimer}(2008)}]{2008LRR....11....8L}
{Lorimer}, D.~R. 2008, Living Reviews in Relativity, 11, 8, \dodoi{10.12942/lrr-2008-8}

\bibitem[{{Maggiore} {et~al.}(2020){Maggiore}, {Van Den Broeck}, {Bartolo}, {Belgacem}, {Bertacca}, {Bizouard}, {Branchesi}, {Clesse}, {Foffa}, {Garc{\'\i}a-Bellido}, {Grimm}, {Harms}, {Hinderer}, {Matarrese}, {Palomba}, {Peloso}, {Ricciardone}, \& {Sakellariadou}}]{2020JCAP...03..050M}
{Maggiore}, M., {Van Den Broeck}, C., {Bartolo}, N., {et~al.} 2020, \jcap, 2020, 050, \dodoi{10.1088/1475-7516/2020/03/050}

\bibitem[{{Mandel} {et~al.}(2021){Mandel}, {M{\"u}ller}, {Riley}, {de Mink}, {Vigna-G{\'o}mez}, \& {Chattopadhyay}}]{2021MNRAS.500.1380M}
{Mandel}, I., {M{\"u}ller}, B., {Riley}, J., {et~al.} 2021, \mnras, 500, 1380, \dodoi{10.1093/mnras/staa3390}

\bibitem[{{Marchant} {et~al.}(2021){Marchant}, {Pappas}, {Gallegos-Garcia}, {Berry}, {Taam}, {Kalogera}, \& {Podsiadlowski}}]{2021A&A...650A.107M}
{Marchant}, P., {Pappas}, K. M.~W., {Gallegos-Garcia}, M., {et~al.} 2021, \aap, 650, A107, \dodoi{10.1051/0004-6361/202039992}

\bibitem[{{Miyaji} {et~al.}(1980){Miyaji}, {Nomoto}, {Yokoi}, \& {Sugimoto}}]{1980PASJ...32..303M}
{Miyaji}, S., {Nomoto}, K., {Yokoi}, K., \& {Sugimoto}, D. 1980, \pasj, 32, 303

\bibitem[{{Nan} {et~al.}(2011){Nan}, {Li}, {Jin}, {Wang}, {Zhu}, {Zhu}, {Zhang}, {Yue}, \& {Qian}}]{2011IJMPD..20..989N}
{Nan}, R., {Li}, D., {Jin}, C., {et~al.} 2011, International Journal of Modern Physics D, 20, 989, \dodoi{10.1142/S0218271811019335}

\bibitem[{{Nomoto}(1984)}]{1984ApJ...277..791N}
{Nomoto}, K. 1984, \apj, 277, 791, \dodoi{10.1086/161749}

\bibitem[{{Nomoto}(1987)}]{1987ApJ...322..206N}
---. 1987, \apj, 322, 206, \dodoi{10.1086/165716}

\bibitem[{{Olausen} \& {Kaspi}(2014)}]{2014ApJS..212....6O}
{Olausen}, S.~A., \& {Kaspi}, V.~M. 2014, \apjs, 212, 6, \dodoi{10.1088/0067-0049/212/1/6}

\bibitem[{{Os{\l}owski} {et~al.}(2011){Os{\l}owski}, {Bulik}, {Gondek-Rosi{\'n}ska}, \& {Belczy{\'n}ski}}]{2011MNRAS.413..461O}
{Os{\l}owski}, S., {Bulik}, T., {Gondek-Rosi{\'n}ska}, D., \& {Belczy{\'n}ski}, K. 2011, \mnras, 413, 461, \dodoi{10.1111/j.1365-2966.2010.18147.x}

\bibitem[{{{\"O}zel} \& {Freire}(2016)}]{2016ARA&A..54..401O}
{{\"O}zel}, F., \& {Freire}, P. 2016, \araa, 54, 401, \dodoi{10.1146/annurev-astro-081915-023322}

\bibitem[{{{\"O}zel} {et~al.}(2012){{\"O}zel}, {Psaltis}, {Narayan}, \& {Santos Villarreal}}]{2012ApJ...757...55O}
{{\"O}zel}, F., {Psaltis}, D., {Narayan}, R., \& {Santos Villarreal}, A. 2012, \apj, 757, 55, \dodoi{10.1088/0004-637X/757/1/55}

\bibitem[{{Peters}(1964)}]{Peters1964}
{Peters}, P.~C. 1964, Physical Review, 136, 1224, \dodoi{10.1103/PhysRev.136.B1224}

\bibitem[{{Popov} \& {Turolla}(2012)}]{2012MNRAS.421L.127P}
{Popov}, S.~B., \& {Turolla}, R. 2012, \mnras, 421, L127, \dodoi{10.1111/j.1745-3933.2012.01220.x}

\bibitem[{{Pringle} \& {Rees}(1972)}]{1972A&A....21....1P}
{Pringle}, J.~E., \& {Rees}, M.~J. 1972, \aap, 21, 1

\bibitem[{{Punturo} {et~al.}(2010){Punturo}, {Abernathy}, {Acernese}, {Allen}, {Andersson}, {Arun}, {Barone}, {Barr}, {Barsuglia}, {Beker}, {Beveridge}, {Birindelli}, {Bose}, {Bosi}, {Braccini}, {Bradaschia}, {Bulik}, {Calloni}, {Cella}, {Chassande Mottin}, {Chelkowski}, {Chincarini}, {Clark}, {Coccia}, {Colacino}, {Colas}, {Cumming}, {Cunningham}, {Cuoco}, {Danilishin}, {Danzmann}, {De Luca}, {De Salvo}, {Dent}, {De Rosa}, {Di Fiore}, {Di Virgilio}, {Doets}, {Fafone}, {Falferi}, {Flaminio}, {Franc}, {Frasconi}, {Freise}, {Fulda}, {Gair}, {Gemme}, {Gennai}, {Giazotto}, {Glampedakis}, {Granata}, {Grote}, {Guidi}, {Hammond}, {Hannam}, {Harms}, {Heinert}, {Hendry}, {Heng}, {Hennes}, {Hild}, {Hough}, {Husa}, {Huttner}, {Jones}, {Khalili}, {Kokeyama}, {Kokkotas}, {Krishnan}, {Lorenzini}, {L{\"u}ck}, {Majorana}, {Mandel}, {Mandic}, {Martin}, {Michel}, {Minenkov}, {Morgado}, {Mosca}, {Mours}, {M{\"u}ller{\textendash}Ebhardt}, {Murray}, {Nawrodt}, {Nelson}, {Oshaughnessy}, {Ott}, {Palomba}, {Paoli}, {Parguez},
  {Pasqualetti}, {Passaquieti}, {Passuello}, {Pinard}, {Poggiani}, {Popolizio}, {Prato}, {Puppo}, {Rabeling}, {Rapagnani}, {Read}, {Regimbau}, {Rehbein}, {Reid}, {Rezzolla}, {Ricci}, {Richard}, {Rocchi}, {Rowan}, {R{\"u}diger}, {Sassolas}, {Sathyaprakash}, {Schnabel}, {Schwarz}, {Seidel}, {Sintes}, {Somiya}, {Speirits}, {Strain}, {Strigin}, {Sutton}, {Tarabrin}, {Th{\"u}ring}, {van den Brand}, {van Leewen}, {van Veggel}, {van den Broeck}, {Vecchio}, {Veitch}, {Vetrano}, {Vicere}, {Vyatchanin}, {Willke}, {Woan}, {Wolfango}, \& {Yamamoto}}]{2010CQGra..27s4002P}
{Punturo}, M., {Abernathy}, M., {Acernese}, F., {et~al.} 2010, Classical and Quantum Gravity, 27, 194002, \dodoi{10.1088/0264-9381/27/19/194002}

\bibitem[{{Reitze} {et~al.}(2019){Reitze}, {Adhikari}, {Ballmer}, {Barish}, {Barsotti}, {Billingsley}, {Brown}, {Chen}, {Coyne}, {Eisenstein}, {Evans}, {Fritschel}, {Hall}, {Lazzarini}, {Lovelace}, {Read}, {Sathyaprakash}, {Shoemaker}, {Smith}, {Torrie}, {Vitale}, {Weiss}, {Wipf}, \& {Zucker}}]{2019BAAS...51g..35R}
{Reitze}, D., {Adhikari}, R.~X., {Ballmer}, S., {et~al.} 2019, in Bulletin of the American Astronomical Society, Vol.~51, 35.
\newblock \doarXiv{1907.04833}

\bibitem[{{Ridley} \& {Lorimer}(2010)}]{2010MNRAS.404.1081R}
{Ridley}, J.~P., \& {Lorimer}, D.~R. 2010, \mnras, 404, 1081, \dodoi{10.1111/j.1365-2966.2010.16342.x}

\bibitem[{{Riley} {et~al.}(2022){Riley}, {Agrawal}, {Barrett}, {Boyett}, {Broekgaarden}, {Chattopadhyay}, {Gaebel}, {Gittins}, {Hirai}, {Howitt}, {Justham}, {Khandelwal}, {Kummer}, {Lau}, {Mandel}, {de Mink}, {Neijssel}, {Riley}, {van Son}, {Stevenson}, {Vigna-G{\'o}mez}, {Vinciguerra}, {Wagg}, {Willcox}, \& {Team Compas}}]{2022ApJS..258...34R}
{Riley}, J., {Agrawal}, P., {Barrett}, J.~W., {et~al.} 2022, \apjs, 258, 34, \dodoi{10.3847/1538-4365/ac416c}

\bibitem[{{Romero-Shaw} {et~al.}(2020){Romero-Shaw}, {Farrow}, {Stevenson}, {Thrane}, \& {Zhu}}]{2020MNRAS.496L..64R}
{Romero-Shaw}, I.~M., {Farrow}, N., {Stevenson}, S., {Thrane}, E., \& {Zhu}, X.-J. 2020, \mnras, 496, L64, \dodoi{10.1093/mnrasl/slaa084}

\bibitem[{{R{\"o}pke} \& {De Marco}(2023)}]{2023LRCA....9....2R}
{R{\"o}pke}, F.~K., \& {De Marco}, O. 2023, Living Reviews in Computational Astrophysics, 9, 2, \dodoi{10.1007/s41115-023-00017-x}

\bibitem[{{Safarzadeh} {et~al.}(2020){Safarzadeh}, {Ramirez-Ruiz}, \& {Berger}}]{2020ApJ...900...13S}
{Safarzadeh}, M., {Ramirez-Ruiz}, E., \& {Berger}, E. 2020, \apj, 900, 13, \dodoi{10.3847/1538-4357/aba596}

\bibitem[{{Sgalletta} {et~al.}(2023){Sgalletta}, {Iorio}, {Mapelli}, {Artale}, {Boco}, {Chattopadhyay}, {Lapi}, {Possenti}, {Rinaldi}, \& {Spera}}]{2023MNRAS.526.2210S}
{Sgalletta}, C., {Iorio}, G., {Mapelli}, M., {et~al.} 2023, \mnras, 526, 2210, \dodoi{10.1093/mnras/stad2768}

\bibitem[{{Shakura} {et~al.}(2012){Shakura}, {Postnov}, {Kochetkova}, \& {Hjalmarsdotter}}]{2012MNRAS.420..216S}
{Shakura}, N., {Postnov}, K., {Kochetkova}, A., \& {Hjalmarsdotter}, L. 2012, \mnras, 420, 216, \dodoi{10.1111/j.1365-2966.2011.20026.x}

\bibitem[{{Snaith} {et~al.}(2014){Snaith}, {Haywood}, {Di Matteo}, {Lehnert}, {Combes}, {Katz}, \& {G{\'o}mez}}]{2014ApJ...781L..31S}
{Snaith}, O.~N., {Haywood}, M., {Di Matteo}, P., {et~al.} 2014, \apjl, 781, L31, \dodoi{10.1088/2041-8205/781/2/L31}

\bibitem[{{Tauris} {et~al.}(2013){Tauris}, {Langer}, {Moriya}, {Podsiadlowski}, {Yoon}, \& {Blinnikov}}]{2013ApJ...778L..23T}
{Tauris}, T.~M., {Langer}, N., {Moriya}, T.~J., {et~al.} 2013, \apjl, 778, L23, \dodoi{10.1088/2041-8205/778/2/L23}

\bibitem[{{Tauris} {et~al.}(2015){Tauris}, {Langer}, \& {Podsiadlowski}}]{2015MNRAS.451.2123T}
{Tauris}, T.~M., {Langer}, N., \& {Podsiadlowski}, P. 2015, \mnras, 451, 2123, \dodoi{10.1093/mnras/stv990}

\bibitem[{{Tauris} {et~al.}(2017){Tauris}, {Kramer}, {Freire}, {Wex}, {Janka}, {Langer}, {Podsiadlowski}, {Bozzo}, {Chaty}, {Kruckow}, {van den Heuvel}, {Antoniadis}, {Breton}, \& {Champion}}]{2017ApJ...846..170T}
{Tauris}, T.~M., {Kramer}, M., {Freire}, P.~C.~C., {et~al.} 2017, \apj, 846, 170, \dodoi{10.3847/1538-4357/aa7e89}

\bibitem[{{Timmes} {et~al.}(1996){Timmes}, {Woosley}, \& {Weaver}}]{1996ApJ...457..834T}
{Timmes}, F.~X., {Woosley}, S.~E., \& {Weaver}, T.~A. 1996, \apj, 457, 834, \dodoi{10.1086/176778}

\bibitem[{{Vigna-G{\'o}mez} {et~al.}(2021){Vigna-G{\'o}mez}, {Schr{\o}der}, {Ramirez-Ruiz}, {Aguilera-Dena}, {Batta}, {Langer}, \& {Willcox}}]{2021ApJ...920L..17V}
{Vigna-G{\'o}mez}, A., {Schr{\o}der}, S.~L., {Ramirez-Ruiz}, E., {et~al.} 2021, \apjl, 920, L17, \dodoi{10.3847/2041-8213/ac2903}

\bibitem[{{Vigna-G{\'o}mez} {et~al.}(2018){Vigna-G{\'o}mez}, {Neijssel}, {Stevenson}, {Barrett}, {Belczynski}, {Justham}, {de Mink}, {M{\"u}ller}, {Podsiadlowski}, {Renzo}, {Sz{\'e}csi}, \& {Mandel}}]{2018MNRAS.481.4009V}
{Vigna-G{\'o}mez}, A., {Neijssel}, C.~J., {Stevenson}, S., {et~al.} 2018, \mnras, 481, 4009, \dodoi{10.1093/mnras/sty2463}

\bibitem[{{Vink} {et~al.}(2001){Vink}, {de Koter}, \& {Lamers}}]{2001A&A...369..574V}
{Vink}, J.~S., {de Koter}, A., \& {Lamers}, H.~J.~G.~L.~M. 2001, \aap, 369, 574, \dodoi{10.1051/0004-6361:20010127}

\bibitem[{{Webbink}(1984)}]{1984ApJ...277..355W}
{Webbink}, R.~F. 1984, \apj, 277, 355, \dodoi{10.1086/161701}

\bibitem[{{Young} {et~al.}(1999){Young}, {Manchester}, \& {Johnston}}]{1999Natur.400..848Y}
{Young}, M.~D., {Manchester}, R.~N., \& {Johnston}, S. 1999, \nat, 400, 848, \dodoi{10.1038/23650}

\bibitem[{{Yu} \& {Jeffery}(2010)}]{2010A&A...521A..85Y}
{Yu}, S., \& {Jeffery}, C.~S. 2010, \aap, 521, A85, \dodoi{10.1051/0004-6361/201014827}

\bibitem[{{Yu} \& {Jeffery}(2011)}]{2011MNRAS.417.1392Y}
---. 2011, \mnras, 417, 1392, \dodoi{10.1111/j.1365-2966.2011.19352.x}

\bibitem[{{Yu} \& {Jeffery}(2015)}]{2015MNRAS.448.1078Y}
---. 2015, \mnras, 448, 1078, \dodoi{10.1093/mnras/stv059}

\bibitem[{{Zhang} \& {Kojima}(2006)}]{2006MNRAS.366..137Z}
{Zhang}, C.~M., \& {Kojima}, Y. 2006, \mnras, 366, 137, \dodoi{10.1111/j.1365-2966.2005.09802.x}

\bibitem[{{Zhang} {et~al.}(2004){Zhang}, {Li}, \& {Wang}}]{2004ChJAA...4..320Z}
{Zhang}, F., {Li}, X.-D., \& {Wang}, Z.-R. 2004, \cjaa, 4, 320, \dodoi{10.1088/1009-9271/4/4/320}

\end{thebibliography}
\bibliographystyle{aasjournal}

\end{document}